\documentclass[useAMS,usenatbib,fleqn]{mn2e}

\usepackage{url} 
\usepackage{graphicx}

\usepackage{hyperref}

\usepackage{amssymb} 

\usepackage{amsmath}

\usepackage{mathptmx} 

\bibpunct{(}{)}{,}{a}{}{,} 

\DeclareMathOperator{\logit}{logit}

\title[]{Machine learning search for variable stars}
\author[]{\parbox{\textwidth}{Ilya~N.~Pashchenko$^{1}$\thanks{E-mail: in4pashchenko@gmail.com},
Kirill~V.~Sokolovsky$^{2,3,1}$\thanks{kirx@kirx.net},
Panagiotis~Gavras$^2$\thanks{pgavras@noa.gr}
}\vspace{0.4cm}\\
\parbox{\textwidth}{$^{1}$Astro Space Center of Lebedev Physical Institute, Profsoyuznaya~St.~84/32, 117997~Moscow, Russia\\
$^{2}$IAASARS, National Observatory of Athens, Vas.~Pavlou \& I.~Metaxa, 15236~Penteli, Greece\\
$^{3}$Sternberg Astronomical Institute, Moscow State University, Universitetskii~pr.~13, 119992~Moscow, Russia\\
}}

\begin{document}

\date{Accepted XXXX Month XX. Received 2017 Month XX; in original form 2017 Month XX}

\pagerange{\pageref{firstpage}--\pageref{lastpage}} \pubyear{2017}

\maketitle

\label{firstpage}

\begin{abstract}
Photometric variability detection is often considered as a hypothesis testing
problem: an object is variable if the null-hypothesis that its brightness
is constant can be ruled out given the measurements and their uncertainties.
The practical applicability of this approach is limited by uncorrected systematic errors.
We propose a new variability detection technique sensitive to a wide range of variability types while being robust to outliers and underestimated measurement uncertainties.
We consider variability detection as a classification problem
that can be approached with machine learning.  
Logistic Regression ($LR$), Support Vector Machines ($SVM$),
$k$~Nearest Neighbors ($kNN$) Neural Nets ($NN$), Random
Forests ($RF$) and Stochastic Gradient Boosting classifier ($SGB$) are applied to 18
features (variability indices) quantifying scatter and/or correlation between points in a light curve. We use a subset of OGLE-II Large Magellanic Cloud (LMC) photometry (30265 light curves) that was searched for variability using traditional methods (168 known variable objects)
as the training set and then apply the $NN$ to a new test set of 31798 OGLE-II LMC light curves.
Among 205 candidates selected in the test set, 178 are real variables, while 13 low-amplitude variables are new discoveries.
The machine learning classifiers considered are 
found to be more efficient (select more variables and fewer false candidates)
compared to traditional techniques using individual variability indices or their linear combination. The $NN$, $SGB$, $SVM$ and $RF$ show a higher efficiency compared to $LR$ and $kNN$.
\end{abstract}

\begin{keywords}
methods: data analysis, methods: statistical, stars: variables: general
\end{keywords}

\section{Introduction}
\label{sec:intro}

A variety of astrophysical phenomena manifest themselves with optical
variability. The incomplete list includes accretion, ejection, explosions,
gravitational lensing, stellar magnetic activity, pulsations and eclipses.
Historically, variable objects were mostly identified by comparing their
brightness recorded on a pair of images \citep{1990vest.book.....H}. 
The photographic images were compared
with a blink-comparator or by placing a positive image of a photographic plate
taken at one epoch on top of the negative plate taken at a different epoch.
Difference image analysis (DIA;
\citealt{1998ApJ...503..325A,2016MNRAS.457..542B}) can be thought of as a
modern software implementation of this idea. The pairwise image comparison has
the obvious drawback that it can detect only high-amplitude variability: the
object's brightness difference between the two images must be a few times
larger than measurement errors associated with individual images.
To detect low-amplitude variability one needs to construct and analyze 
a light curve containing multiple measurements in order to effectively average-out
individual measurement errors. The multiple measurements may be performed
using DIA, point spread function (PSF) fitting, or aperture photometry.  The
effect of considering multiple measurements altogether instead of 
pairs
is illustrated by the large number of high-amplitude $\delta$~Scuti/SX~Phoenicis
stars (HADS) found using digitized photographic plates by
\cite{2010ARep...54.1000K}. These plates were earlier searched for variability by comparing pairs of images, but this search failed
to identify the majority of HADS variables despite having a comparable accuracy of individual measurements.

Detection of variability in a light curve may be considered a hypothesis testing
problem \citep{2005ESASP.576..513E,2006AJ....132..633H,2010AJ....139.1269D,2001A&A...373..576P}: 
an object is variable if the null-hypothesis that its brightness is
constant can be ruled out.
Uncorrected systematic errors and
corrupted measurements limit the practical applicability of this approach to
well-behaved data sets. Tests that take into account not only the
measurements themselves, but also the order
\citep{2006MNRAS.367.1521T,2013A&A...556A..20F,2016A&A...586A..36F} and times
\citep{1996PASP..108..851S,2003ChJAA...3..151Z} at which
the measurements were taken were also proposed. The variability detection
threshold for these tests often has to be determined empirically for a given
data set.
%
%
\cite{2017MNRAS.464..274S} investigated 24 ``variability indices'' (also
referred to as ``light curve features'') -- statistical characteristics
quantifying scatter and correlation between points in a light curve. 
The ability of these indices used individually or in a linear combination to
discriminate variable objects from non-variable ones was compared using multiple real
and simulated data sets.  

In this paper we explore a new variable star selection technique that
outperforms all the individual (or linearly combined) indices considered by
\cite{2017MNRAS.464..274S}. This is achieved by finding useful non-linear
combinations of these indices. We consider variability detection not as a
hypothesis testing problem, but as a binary classification problem (variable vs.
non-variable objects) and apply machine learning techniques to solve it.
The proposed technique does not critically depend on accurate photometric error estimates 
and is not sensitive to individual outlier measurements\footnote{We use light curve features (MAD, IQR, $1/\eta$; Table~\ref{tab:indexsummary}) that are by design not sensitive to outliers \citep{2017MNRAS.464..274S} and do not depend on the estimated photometric errors. While this is not the case for other features ($\sigma$, $\chi_{\rm red}^2$,...) these features will end up having less predictive power compared to the robust features if the sensitivity to outliers or the incorrectly estimated errors constitute a problem in the given data set. ML techniques described in Section~\ref{sec:classifiers} include procedures (bagging, dropout, appropriate choice of loss function) designed to minimize dependency on individual objects with outlier lightcurve feature values (that may result from corrupted photometry). The OGLE-II and $TF1$ data we use for tests are plagued with outlier measurements which do not end up having a critical impact on our ability to identify variable objects (Sections~\ref{sec:cv_comparison} and \ref{sec:blind}).}.
It can be applied to {\em any} large set of light curves given a representative subset of these light curves has been manually classified as variable or non-variable. 
This subset is used to train a machine learning (ML) classifier that will process the rest of the data. 

While preparing this paper we
learned
that the General Variability Detection module of the Gaia Variability Analysis
pipeline \citep{2017arXiv170203295E} is using a Random Forest classifier
trained on multiple variability indices computed for variables
identified in the OGLE-IV Gaia south ecliptic pole field by
\cite{2012AcA....62..219S}.
Earlier, \cite{2009MNRAS.400.1897S,2012AJ....143...65S} proposed to use
multiple variability indices together combining them with an infinite Gaussian
mixture model.
The method of \cite{2016AcA....66..421P}, while focusing
solely on eclipsing binaries, is similar in spirit to the method proposed here.
The authors use a set of features computed by the BLS period-finding algorithm
\citep{2002A&A...391..369K} as an input for the Random Forest classifier
trained on OGLE-III eclipsing binaries \citep{2011AcA....61..103G} in one of
the OGLE-IV \citep{2015AcA....65....1U} fields.
\cite{2016A&A...595A..82E} used machine learning to identify RRab stars in the
VVV survey data \citep{2010NewA...15..433M}.
\cite{2017A&A...605A.123P} propose a set of light curve features robust to individual outlier measurements and use them to compare multiple machine learning algorithms on classified OGLE-III light curves.
\cite{2017MNRAS.468.2189Z} consider an original set of features suitable for characterizing non-periodic and quasi-periodic light curves: parameters of the damped random walk and quasi-periodic oscillation models.

Taking into account the experience of authors listed above, we suggest the following points that we try to justify in this work:
\begin{itemize}
\item Machine learning can be used for variability detection \textit{in general}, not only for extracting variable objects of specific types, one type at a time.
\item This general variability detection problem is tractable for many different supervised learning algorithms.
\item Systematic search for optimal hyperparameters of a learning algorithm is needed to achieve its best performance.
\item Variability search with machine learning is effective even with a modest training sample size containing hundreds of known variables. The sample may be highly imbalanced (few variables and many constant stars).
\end{itemize}
The paper is structured as follows: Section~\ref{sec:data} describes the test
data, Section~\ref{sec:technique} describes the proposed variable object
selection technique, Section~\ref{sec:disc} discusses the results of its application to the
test data, while Section~\ref{sec:sum} summarizes our findings.

\section{Input data}
\label{sec:data}

The primary input for variability search is a set of time-series brightness measurements collected for a number of sources -- a set of light curves.
The light curves may be quite diverse even within one data set. 
They may have different number of measurements as not all sources are detected and successfully measured in each image. 
Measurements of different sources may be influenced to a different extent by systematic effects that depend on source color or the source position on an image. 
Some measurements get corrupted by random events such as cosmic ray hits or the object's image falling on a bad pixel. 
Light curves of variable sources may show a variety of patterns depending on the variability type and period (or typical timescale for non-periodic variables). To characterize such diverse light curves in a uniform way, we extract a set of light curve features (or ``variability indices''). 
We use the {\scshape VaST} code \citep{2017arXiv170207715S} to extract the features. Other feature extraction codes are also publicly available (\citealt{2015arXiv150600010N}, \citealt{2016A&A...587A..18K}, \citealt{2016arXiv161007717C}). 
The features computed by {\scshape VaST} are meant to be used for variability {\it detection} while the other codes are mainly concerned with features useful for {\it classification} of detected variables, but there is a great deal of overlap between the features useful for these two tasks.
In Section~\ref{sec:lightcurves} we describe the photometric data set used for our tests and discuss its inherent biases in Section~\ref{sec:bias}. 
In Section~\ref{sec:prepocessing} we present the utilized set of light curve features.

\subsection{Light curves}
\label{sec:lightcurves}

%
We use a small subset of publicly available Optical Gravitational
Lensing Experiment phase two (OGLE-II) PSF fitting $I$-band photometry of the
field LMC\_SC20 towards the Large Magellanic Cloud
\citep[LMC;][]{2005AcA....55...43S}.  OGLE-II observations are conducted with
the 1.3\,m Warsaw telescope at the Las Campanas Observatory, Chile
\citep{1997AcA....47..319U}.
Public OGLE-II photometry was used earlier to test new variability detection
techniques by \cite{2007ASPC..362..255S}.
OGLE-II LMC data were searched for various specific types of variable objects
including
microlensing events \citep{2009MNRAS.397.1228W},
variable red giants \citep{2004AcA....54..347S,2003MNRAS.343L..79K,2005AcA....55..331S},
RR~Lyrae stars \citep{2003AcA....53...93S},
eclipsing binaries \citep{2003AcA....53....1W},
cataclysmic variables \citep{2003PASP..115..193C},
quasars \citep{2002AcA....52..241E},
Cepheids \citep{1999AcA....49..223U}.
\cite{2001AcA....51..317Z} constructed a comprehensive catalog of candidate
variables (of all types) detected with DIA.
The field was also covered by later phases of the OGLE project
\citep{2008AcA....58...69U,2015AcA....65....1U}
as well as other time-domain surveys including
VMC \citep{2011A&A...527A.116C},
EROS \citep{2007A&A...469..387T},
MACHO \citep{2000ApJ...542..281A,2005IAUS..225..357B}.
Overall, the test field is well-studied for variability.

The LMC\_SC20 data set was manually searched for variable objects 
by \cite{2017MNRAS.464..274S}.
The authors identified 20 new variable stars hinting that
variability detectable in OGLE-II data is still not fully explored.
These findings also highlight the fact that in practice, one cannot expect to have 
a complete sample of variable stars by just searching catalogs of known variables, 
even in such a well-studied region of the as the LMC.

The use of the LMC\_SC20 data set allows us to
directly compare the effectiveness of the variability detection technique
proposed here to the techniques discussed by \cite{2017MNRAS.464..274S}.
Specifically, we want to compare the results obtained with machine
learning to the results of variability search by visual inspection of
light curves, which is the most reliable, but labor-intensive way of identifying
variable objects (hence the relatively small size of our training sample).
The sample consists of 30265 sources with high-quality (percentage of good
measurements ${\rm Pgood} \ge 98$; see Section~4.1 in \citealt{2005AcA....55...43S}) light curves each having 262 to 268 points;
among them are 168 variable sources of various types.  This data set is further
split into subsets randomly and multiple times in order to find the most
promising variable object selection technique as described in
Section~\ref{sec:technique}. The full LMC\_SC20 data set from
\cite{2017MNRAS.464..274S} is used to train the selected best classifier before
applying it to a new data set that was not previously searched for
variability by us.
The new data set consists of 31798 OGLE-II PSF $I$-band light curves from the
adjacent field LMC\_SC19 selected by applying the same quality cut (${\rm
Pgood} \ge 98$ resulting in 262--268 light curve points).
Three variable objects (and 893 non-variable ones) located in the overlapping sky region are present in both LMC\_SC19 and LMC\_SC20 data sets.

Table~\ref{tab:vartypes} presents the distribution of variability types available in the training (LMC\_SC20) data set 
and that recovered from the blind test data set (LMC\_SC19, Section~\ref{sec:blind}).
We adopted a published classification of variable objects whenever possible:
eclipsing binaries from \cite{2003AcA....53....1W,2011AcA....61..103G,2014A&A...566A..43K},
red giant variables from \cite{2005AcA....55..331S,2008AJ....136.1242F,2009AcA....59..239S,2011A&A...536A..60S},
RR~Lyrae variables from \cite{2003AcA....53...93S},
Cepheids from \cite{1999AcA....49..223U},
candidate Be stars from \cite{2005MNRAS.361.1055S},
QSO candidates from \cite{2002AcA....52..241E,2012ApJ...747..107K,2013ApJ...775...92K},
$\delta$\,Scuti stars from \cite{2010AcA....60....1P}.
\cite{2004AcA....54..347S} classified 1546 periodic red giants in the LMC as candidate ellipsoidal variables following the suggestion by \cite{1999IAUS..191..151W,2000PASA...17...18W} that one of the five period-luminosity sequences observed in LMC red giants may represent binary systems rather than a mode of pulsations. Considering that {\it i)}\,physical interpretation of this sequence as binary systems is not unambiguous; {\it ii)}\,in practice, the light curve shapes of these objects are indistinguishable from light curves of some semiregular variables; {\it iii)}\,eclipsing variables with periods $>10$\,d showing strong ellipsoidal variations often have bluer colors than the candidate ellipsoidal variables with no eclipses; for the purpose of this work we group the candidate ellipsoidal variables with other variable red giants in Table~\ref{tab:vartypes}. 
The lists of candiate Be stars of \cite{2005MNRAS.361.1055S} and QSO candidates of \cite{2002AcA....52..241E} have 99 common objects 11 of which are among the variable objects in our data sets. In Table~\ref{tab:vartypes} we combine them under the label ``blue irregular variables''.

\begin{table}
 \caption{Types of variable objects in the blind test (LMC\_SC19) and training (LMC\_SC20) data sets.}
 \label{tab:vartypes}
 \begin{tabular}{r@{~~~~}c@{~~~~}c}
    \hline
Type & LMC\_SC19 & LMC\_SC20 \\
    \hline
                    eclipsing binaries &  36 &  54 \\
      variable red giants (L/M/SR/ELL) &  54 &  52 \\
                    RR~Lyrae-type variables &  56 &  26 \\
     Cepheids (classical and Type\,II) &  17 &  20 \\
blue irregular variables (GCAS/BE/QSO) &  22 &  13 \\
                 $\delta$\,Scuti stars &   1 &   3 \\
\hline
                                 total & 186 & 168 \\
    \hline
    \end{tabular}
\end{table}

While the discussion below is based on the OGLE-II data, we also performed a similar analysis of two other data sets investigated by \cite{2017MNRAS.464..274S} that were collected with different telescopes and processed using different source extraction and photometry software: $Kr$ \citep{2013PZP....13...12L,2016PZP....16....4L} and 
$TF1$ \citep{2016MNRAS.461.3854B,2015PZP....15....7P,2014AstBu..69..368B}.
The results obtained with $Kr$ and $TF1$ are consistent with the ones presented in Sections~\ref{sec:disc} and \ref{sec:sum}. The main focus of our investigation was the OGLE-II LMC\_SC20 data set as many other OGLE-II light curves are readily available for variability search with the technique described here.

\subsection{Sources of bias in the training sample}
\label{sec:bias}

The training sample may be biased as the list of known variables in the LMC\_SC20 data set may not be exhaustive (and therefore some variable objects may be incorrectly labeled as non-variable). We try to minimize this by conducting our own variability search  \citep[used also in our previous work,][]{2017MNRAS.464..274S} based on visual inspection of light curves instead of relying on published lists of variables (Sec.~\ref{sec:lightcurves}), however this is still likely an approximation to the complete list of (detectable) variables in the 
selected 
set of light curves.

Another source of bias is the limited size of our training sample (Table~\ref{tab:vartypes}) that does not nearly represent all variability types and all possible variations in amplitude and period or variability time scale within each type. This translates in a non-trivial way to an incomplete coverage of the variability feature (introduced in Sec.~\ref{sec:prepocessing}) parameter space occupied by variable objects (see the discussion of learning curves in Sec.~\ref{sec:learningcurvesandprecondition}). The severity of this problem is hard to quantify a~priori. Positive results of variability search in the unseen data described in Sec.~\ref{sec:blind} indicate that this is not a critical issue.
This may partly be attributed to the fact that (while relying on the assumption that variable objects are rare) the section of the variability feature parameter space occupied by non-variable objects should be sampled well with $\sim30000$ 
non-variable sources in the training set.

\subsection{Variability features}
\label{sec:prepocessing}

\begin{figure*}
	\centering
	\includegraphics[width=1.0\textwidth]{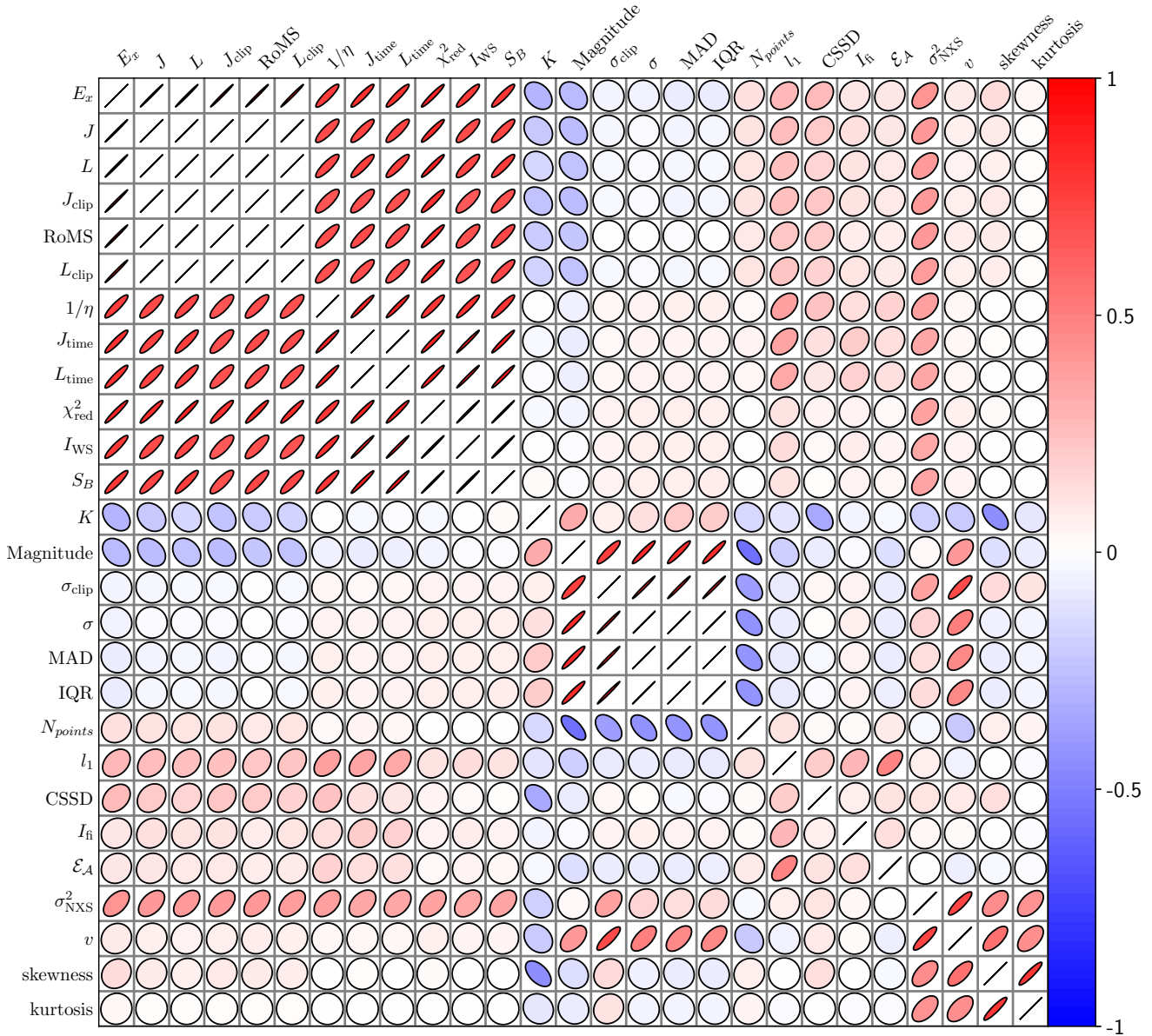}
        \caption{Correlation between the light curve features. Color and orientation of each ellipse represent the sign (red and rotated 45 degrees clockwise from vertical - positive while blue and rotated counterclockwise - negative) and eccentricity with color depth code the value of the Pearson correlation coefficient between the corresponding features (see the color bar). A nearly circular shape and white color indicate close to zero correlation between a pair of features while a narrow red (blue) ellipse indicates high positive (negative) correlation between features.}
	\label{fig:featurecorr}
\end{figure*}

\begin{figure}
	\centering
	\includegraphics[width=0.48\textwidth,clip=true,trim=0.5cm 0.3cm 0.8cm 0.7cm]{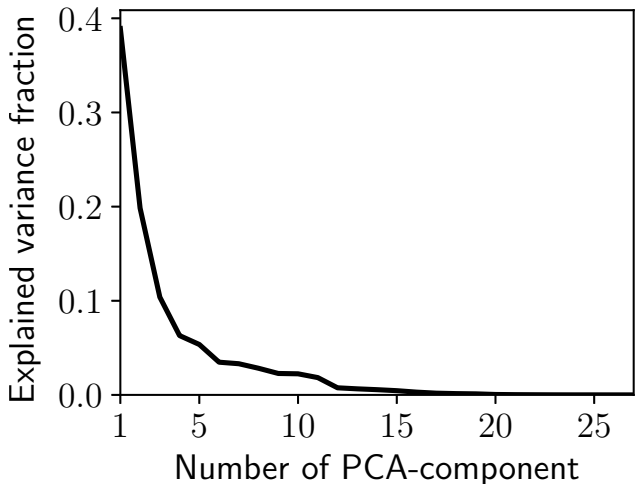}
        \caption{Fractional variance explained by each of the PCA-component. Also known as \textit{scree plot} \citep{screeplot}. Most of the variance can be explained by 10 PCA components confirming that many light curve features listed in Table~\ref{tab:indexsummary} are correlated (see also Figure~\ref{fig:featurecorr} and Section~\ref{sec:moretests}).}
	\label{fig:pca}
\end{figure}

\begin{table}
    \caption{Light curve features (variability indices). Features correlated with other features with $r > 0.995$ for the LMC\_SC20 data set (and excluded from the final analysis) are marked in italics}
    \label{tab:indexsummary}
    \begin{tabular}{r@{~~~}l}
    \hline
Index                                               & Reference  \\
    \hline
weighted standard deviation -- $\sigma$             & \text{\cite{2008AcA....58..279K}}   \\
\textit{clipped $\sigma$} -- $\sigma_{\rm clip}$             & \text{Section~\ref{sec:sigmaclip}}   \\
\textit{median abs. deviation} -- ${\rm MAD}$                & \text{\cite{2016PASP..128c5001Z}}   \\
interquartile range -- ${\rm IQR}$                  & \text{\cite{2017MNRAS.464..274S}}   \\
reduced $\chi^2$ statistic -- $\chi_{\rm red}^2$    & \text{\cite{2010AJ....139.1269D}}   \\
robust median statistic -- ${\rm RoMS}$             & \text{\cite{2007AJ....134.2067R}}   \\
norm. excess variance -- $\sigma_{\rm NXS}^2$       & \text{\cite{1997ApJ...476...70N}}   \\
norm. peak-to-peak amp. -- $v$                      & \text{\cite{2009AN....330..199S}}   \\
\textit{autocorrelation} -- $l_1$                            & \text{\cite{2011ASPC..442..447K}}   \\
inv. von~Neumann ratio -- $1/\eta$                  & \text{\cite{2009MNRAS.400.1897S}}   \\
Welch-Stetson index -- $I_{\rm WS}$                 & \text{\cite{1993AJ....105.1813W}}   \\
flux-independent index -- $I_{\rm fi}$              & \text{\cite{2015A&A...573A.100F}}   \\
Stetson's~$J$ index                                 & \text{\cite{1996PASP..108..851S}}   \\
time-weighted Stetson's~$J_{\rm time}$              & \text{\cite{2012AJ....143..140F}}   \\
\textit{clipped Stetson's}~$J_{\rm clip}$                    & \text{Section~\ref{sec:clipstetson}}   \\
\textit{Stetson's~$L$} index                                 & \text{\cite{1996PASP..108..851S}}   \\
\textit{time-weighted Stetson's}~$L_{\rm time}$              & \text{\cite{2012AJ....143..140F}}   \\
\textit{clipped Stetson's}~$L_{\rm clip}$                    & \text{Section~\ref{sec:clipstetson}}   \\
\textit{consec. same-sign dev. -- CSSD}   & \text{\cite{2009MNRAS.400.1897S}}   \\
$S_B$ statistic                                     & \text{\cite{2013A&A...556A..20F}}   \\
excursions -- $E_x$                                 & \text{\cite{2014ApJS..211....3P}}   \\
excess Abbe value -- $\mathcal{E}_\mathcal{A}$      & \text{\cite{2014A&A...568A..78M}}   \\
Stetson's~$K$ index                                 & \text{\cite{1996PASP..108..851S}}   \\
kurtosis                                            & \text{\cite{1997ESASP.402..441F}}   \\
skewness                                            & \text{\cite{1997ESASP.402..441F}}   \\
    \hline
    \end{tabular}
\end{table}

We initially considered 24 features listed in Table~\ref{tab:indexsummary} (a detailed discussion of these features is presented by \citealt{2017MNRAS.464..274S}).
Many of them are highly correlated (see Figure~\ref{fig:featurecorr})\footnote{\textit{biokit} Python package was used to generate the plot \url{https://github.com/biokit/biokit}}, 
in fact some represent the same quantity computed using different weighting or
clipping schemes ($\sigma$-$\sigma_{\rm clip}$, Stetson's~$K$-kurtosis,
$J$-$J_{\rm time}$-$J_{\rm clip}$, $L$-$L_{\rm time}$-$L_{\rm clip}$; see Section~\ref{sec:clip}). We
dropped the features
$\sigma_{\rm clip}$, $L$, $L_{\rm clip}$, $J_{\rm clip}$, ${\rm MAD}$,
$L_{\rm time}$ which are highly correlated to other corresponding features with $r >
0.995$ (Figure~\ref{fig:featurecorr}). 
The choice of which feature to keep among a few highly correlated ones was done in a quasi-random fashion. When processing a really large set of light curves, it would be wise to consider computational costs of features and keep the one that requires less time to calculate.
We checked that the number of remaining features is reasonable using \textit{Principal Component Analysis} \citep[PCA;][]{pearson1901}.
Most (95\%) of the variance in features can be explained by 10 PCA-components 
(Figure~\ref{fig:pca}).
This suggests that at least 10 original features are needed to describe most of the variance in the data.
We also dropped the features $CSSD$ and $l_1$ which in the implementation 
of \cite{2017MNRAS.464..274S} appeared to be less-informative for variability search.
We tried to log-transform positive features (such as $\sigma$ or $IQR$) to make their
distribution closer to the normal but it did not resulted in higher performance for any of the tested algorithms. 
Ensemble tree methods used in our work, Random Forests ($RF$; Section~\ref{sec:RF}) and Stochastic Gradient Boosting 
($SGB$; Section~\ref{sec:SGB}), are invariant to one-to-one transformations of the input feature data. 
Our pre-processing procedure includes scaling features by centering and
standardization for all methods except $RF$ and $SGB$. 
We note that to prevent overestimation of the classification performance, the 
data pre-processing and the feature selection should be done in a way that prevents 
any information leakage from the sample used to evaluate performance to the one used 
to build the classifier \citep[e.g.][]{doi:10.1093/bioinformatics/btp621}.

\section{Variable star identification as a classification problem}
\label{sec:technique}

We tackle the problem of variable star identification as a classification
problem. Classification is a \textit{supervised} learning problem
where one has a set of objects $X$, a set of responses $Y$ and some unknown
dependence $f: X \mapsto Y$ (\textit{target function}, e.g. \citealt{esl}, \citealt{ml}). 
The problem is to find (\textit{learn}) an algorithm (decision function) $f^{\star}$ that approximates the target function $f$ for all $X$ given only the subsample of all objects - $X_{{\rm train,} i}$ with known responses $Y_{{\rm train,} i}$ (called the \textit{training sample}). 
After being trained on ($X_{{\rm train,} i}$, $Y_{{\rm train,} i}$) the algorithm $f^{\star}$ can be used to predict the values of $Y$ for the {\it new data}: $X_{{\rm new}, i} \subset X$ and $X_{{\rm new}, i} \not \subset X_{{\rm train,} i}$.
Depending on the nature of $Y$, the problem can be formulated as regression ($Y = \mathbb{R}$), binary ($Y = \{0, 1\}$) or $K$-class classification ($Y = \{0, 1, ..., K\}$). The objects are characterized by a set of \textit{features} $\phi_{j}: X \mapsto D_j$, where $D_j$ could be $\{0, 1\}$ (binary feature), $|D_j| < \infty $ (nominal or ordinal feature if finite $D_j$ could be ordered) or $D_j = \mathbb{R}$ (qualitative feature).
The choice of features that capture properties related to the object's class
is crucial for a reliable classification. The chosen set of features determines the maximum classification performance that could be achieved for a given problem\footnote{A classifier with such performance is called \textit{Bayes classifier} \citep{esl} and its (maximum achievable) error rate is called \textit{Bayesian rate}. It 
is 
a 
theoretical construction as it uses generally unknown posterior probability of class membership $P(Y|X)$ for making 
predictions.}. 

When building a classifier $f^{\star}$ that provides high quality predictions on new unclassified data, given a set of features (that describe the data and constrain the maximum achievable quality of classification) one has to decide what family of algorithms $f^{\star}(\theta)$, parameterized by some parameter vector $\theta$, to use. 
If the chosen algorithm is not sufficiently flexible to approximate $f$, 
when presented with new data it will not be able to approach 
the highest classification performance allowed by the used set of features,
no matter how large the training sample is.
In this case, the algorithm prediction has high bias and it is said that the algorithm is \textit{underfitting}.
If the algorithm is too complex ($f^{\star}$ has many unconstrained parameters $\theta$) it can spend some of its degrees of freedom on learning noisy patterns specific to a given finite training sample. Thus algorithm's predictions on new data 
will be
unstable, sensitive to small changes in the training data. In that case, the predictions have high dispersion and it is said that the algorithm is \textit{overfitting.}\footnote{Overfitting could also be the result of the training sample being unrepresentative of the parent population, e.g. when the training data set is small or 
includes incorrectly
classified objects.} 
In both cases the algorithm's ability to generalize (that is to provide good quality classification of new data) decreases. 
This trade-off governed by the algorithm's complexity is called the \textit{Bias-Variance trade-off} \citep{esl}.

One can constrain the complexity of $f^{\star}$ or tune some other high-level algorithm property (e.g. algorithm behavior during training) to reduce the dependence of its predictions on the finite training sample used (i.e. algorithm dispersion\footnote{High-bias algorithms could also have significant dispersion, e.g. multivariate linear regression with highly correlated independent variables (features).}). 
To see how much the algorithm is overfitting one has to apply it to some classified data that are not part of the training sample.
Parameters that determine the algorithm performance on new data but cannot be learned using training data alone are called \textit{hyperparameters} (HP).

In summary, each algorithm has a set of conventional parameters $\theta$ (e.g.~coefficients of features in regression, Section~\ref{sec:LR}, weights of neurons in $NN$, Section~\ref{sec:NN}) that are learned from the training sample and hyperparameters that have to be set before training (e.g. number of trees in $RF$,
Sec.~\ref{sec:RF} or number of hidden layers or number of neurons in each hidden layer in $NN$, Section~\ref{sec:NN}). 
The hyperparameters include not only the complexity parameters (capacity to learn determined by the 
depth of a decision tree, number of hidden layers and neurons in each layer in a neural network, number of basis learners in an ensemble, value of regularization that penalizes models that are too complex), but also parameters that control the process of algorithm training, e.g. speed of learning (the learning rate in gradient descent methods of learning neural networks). The optimal set of hyperparameters for a given algorithm largely depends on the data set and might differ even between training samples of different sizes.

\subsection{Performance metric}
\label{sec:performancemetric}

To decide which variability detection technique works best, we need to define what exactly do we mean by ``best'', in other words -- adopt an appropriate performance metric.
As we deal with a highly imbalanced data set (non-variable stars outnumber variable ones by a factor of $\sim100$, Section~\ref{sec:lightcurves}), \textit{accuracy}, defined as the ratio
of correct predictions to the total number of cases evaluated, despite being
the most intuitive performance metric is not a proper measure of classification
algorithm performance. A high accuracy score could be obtained by just labeling all target objects with the majority
class \citep{Kononenko1991,Valverde-Albacete}. To avoid
this, one considers \textit{Precision}, $P = {\rm TP}/({\rm TP}+{\rm FP})$ and \textit{Recall}, $R = {\rm TP}/({\rm TP}+{\rm FN})$,
as well as their harmonic mean known as $F_1$-score
$$F_1 = 2 P R/(P+R),$$ 
where TP is the number of true positives (i.e. true variables classified as variables), FP is the number of false positives (non-variables classified as variables) and FN is the number of false negatives (true variables classified as non-variables; \citealt{evaluation}).

Suppose we test a classifier using it to select candidate variables from a set of light curves for which we already know the right answer: which light curve shows variability and which does not. Then $P$ is the probability that a randomly chosen object from the list of candidates is a true variable, while $R$ is the probability that a randomly chosen true variable is in the list of candidates.
%
There is a trade-off between high values of $R$ and $P$, i.e. recovery of all positive objects (true variables) and contamination by false positives (objects that algorithm wrongly classifies as variables).
$F_1$ is a useful compromise: it has a high value when both $R$ and $P$ are high, that is when the classifier does not miss many true variables \textit{and} the majority of objects classified as variables are actual variables.

Most classification algorithms instead of class labels (e.g. variable/non-variable) return probabilities $p_i$ of the $i$-th object representing a certain class\footnote{Actually they return some proxy of probability. To 
derive actual probabilities 
one has to \textit{calibrate} \citep{calibration} the classifier by comparing predicted and true frequency of classes for some independent data set.}. 
To assign class membership to objects being classified, one has to choose a threshold value $p_{\rm threshold}$ such that objects with probability $p_i$ of belonging to the class
$Y$ are assigned to that class if $p_i > p_{\rm threshold}$. 
$P$, $R$ and $F_1$ 
depend on the adopted threshold value. 
This can be utilized if the ``cost'' of false positives and false negatives is different. 
For example, if when visually inspecting a list of candidate variables we are willing to look at ten false candidates for every true variable, then for us the cost of false positives is ten times lower than the cost of false negatives. If the cost of false positives is high (e.g. if we want to obtain a list of candidates with the majority of the objects representing true variability) then $P$ is a suitable performance metric, if the cost of false negatives is high (if we want to recover as many true variables as possible) than $R$ can be used.
Alternatively, one may use
$$F_\beta = (1+\beta^2) R P / (R + \beta^2 P),$$
a score that attaches $\beta$ times as much importance to $R$ as $P$ \citep{evaluation}. 
In the case of equal costs $F_1$ works best.

To characterize the model's performance over all possible thresholds
(i.e.~under different values of FP/FN cost ratios), the Area Under \textit{ROC}
Curve ($AUC$; \citealt{FAWCETT2006861}) may be used as a performance metric. ROC-curve is a plot of $R$ against \textit{False Positive Rate} $FPR = {\rm FP}/({\rm FP}+{\rm TN})$, where ${\rm TN}$ is the number of true negatives (true non-variable stars correctly classified as non-variables). For binary classification, $AUC$ is the probability that given one positive and one negative example at random, the classifier ranks the positive example above the negative one.

As shown by \cite{Takaya}, in the case of highly imbalanced data $AUC$ weakly depends on the
algorithm performance (mainly because it considers the number of ${\rm TN}$) and other metrics
(such as Area Under Precision-Recall Curve - $AUPRC$) should be used instead. 
To compare methods in a similar to \cite{2017MNRAS.464..274S} manner we decided to search hyperparameters that
maximize the $F_1$-score using the default threshold value of 0.5.

\subsection{Classifiers}
\label{sec:classifiers}

We 
test the following
classifiers: Logistic Regression ($LR$), Support Vector
Machines with Radial Basis Functions ($SVM$), $k$~Nearest Neighbors
($kNN$), Neural Nets ($NN$), Random Forests ($RF$) and Stochastic Gradient
Boosting classifier ($SGB$).
These algorithms make different assumptions about classes 
and the target function and use different heuristics
and methods to tackle the problem of classification.
We use \texttt{scikit-learn} Python package \citep{scikit-learn} implementation of $SVM$, $RF$, $kNN$,
\texttt{XGBoost}\footnote{https://xgboost.readthedocs.io/en/latest/} \citep{2016arXiv160302754C}
implementation of $SGB$ and \texttt{Keras}\footnote{https://keras.io/}
library for $NN$--classification. 
In this subsection, we briefly describe these 
classifiers and their hyperparameters. More information may be found in the official documentation of \texttt{scikit-learn}, \texttt{XGBoost} and \texttt{Keras}.

\subsubsection{$k$~Nearest Neighbors ($kNN$)}
\label{sec:kNN}
The $kNN$ method is based on the hypothesis that similar objects usually share the same class.
The notion of ``similarity'' is defined in terms of a distance between objects in feature space. 
The object class predicted by $kNN$ is the class chosen by the majority of $k$ closest neighbors
of that object. The method differs from the other tested classification algorithms as no model is built during its training phase. Learning (i.e. approximation of the decision function, Section~\ref{sec:technique}) occurs only when new data are presented to the classifier.
Despite being quite simple, 
this method is very effective, especially in the situation where the hypothesis holds and the number of samples is relatively high.
The algorithm is nonparametric, i.e. its decision surface (boundary between classes in feature space) can be arbitrary complex and approximate any underlying dependence $f$ (Section~\ref{sec:technique}) given enough training data. The optimized hyperparameters are the number of neighbors $k$ and \textit{weights}
- the type of weighting being used. We tried uniform weights and weights inversely
proportional to the euclidean distance of a neighbor\footnote{We also experimented with some non-euclidean metrics supported by \texttt{scikit-learn} 
including \textit{chebyshev} and \textit{manhattan} distances, but their use resulted in a degraded performance.}. 

\subsubsection{Logistic Regression ($LR$)}
\label{sec:LR}
$LR$ is a generalized regression model used in cases of binary (or categorical, i.e. belonging to one of a limited number of classes) 
response variable. It differs from the standard linear regression with continuous response by the use of the \textit{link function} that transforms linear combinations of features to a binary response variable. $LR$ models the $\logit(p) = \log[p/(1-p)]$ of posterior class probability membership $p$ as a linear combination of features. 
Setting a threshold value of $p$ allows one to make the response binary.
We optimized two hyperparameters: 
\begin{itemize}
\item $C$ that defines the level of regularization used 
(i.e. the default $L2$-regularization, which penalizes complexity by adding a term to the objective function being minimized that consists of the sum of squares of feature coefficients) and 
\item relative weights of classes.
\end{itemize}

\subsubsection{Support Vector Machine ($SVM$)}
\label{sec:SVM}
Linear $SVM$ is searching for the \textit{optimal separating hyperplane} in the feature space that separates classes best in terms of maximum distance from closest objects of both classes to the hyperplane \citep{vapnik1996} thus maximizing the \textit{margin} between classes. This hyperplane is defined by a (usually) small number of objects in feature space that are close to the decision surface (\textit{support vectors}) and that are the hardest to classify. For classification problems with classes that cannot be separated using a linear surface, the use of special kernels reduces the problem to finding the optimal separating hyperplane in an enlarged (even infinite-dimensional for some kernels) transformed feature space without explicitly transforming features \citep{Boser:1992:TAO:130385.130401}. We optimized: 
\begin{itemize}
\item the kernel type - linear ($linear$), polynomial ($poly$) and Radial Basis Function kernel ($rbf$), 
\item degree of polynomial kernel for kernel $poly$, 
\item $C$ - ``soft margin'' regularization penalty parameter (it determines the relative influence of wrongly classified points - points on the ``wrong'' side of the optimal hyperplane), 
\item $gamma$ - kernel coefficient, 
\item the relative weights of the classes.
\end{itemize}

\subsubsection{Random Forest ($RF$)}
\label{sec:RF}
\begin{figure}
	\centering
	\includegraphics[width=0.47\textwidth]{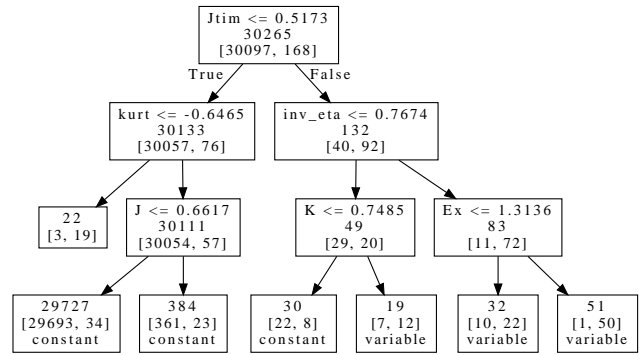}
        \caption{An example decision tree for the LMC\_SC20 data set. Nodes of the tree show the cuts on individual variability features (Table~\ref{tab:indexsummary}) values used to make a decision at each node. The numbers in each node are the number of all objects considered in this node, the number of non-variable and variable objects.}
	\label{fig:tree}
\end{figure}

$RF$ is an ensemble method. The ensemble methods use 
predictions of several weak learners\footnote{Weak learner is an algorithm performing not much better than random guessing.} and combine them all at once or sequentially to make more efficient predictions than would be possible with individual learners. 
$RF$ uses \textit{bagging} (bootstrap aggregation; \citealt{bagging}) which combines many weak learners with high variance 
trained on bootstrap samples\footnote{Bootstrap sample is a sample of the same size as the original one drawn with replacement from it.} of training data 
to reduce the 
variance of the final estimator. It usually uses a deep decision tree (tree with many branches) as a weak learner. An example decision tree classifier is presented in Figure~\ref{fig:tree}. We use a shallow tree with an easy-to-visualize structure. Hyperparameters of this tree were 
optimized for maximum performance as measured by $F_1 = 0.69$ (see \ref{sec:hyperpartuning} for details on measuring $F_1$). $RF$ 
relies on 
the idea of \textit{random subspace selection} \citep{subspace}, also known as attribute or feature bagging. The procedure is similar to bagging but instead of subsampling training objects it consists of using random subsets of features for creating and growing individual decision trees. This prevents $RF$ from being focused on a small number of highly informative features that may loose their predictive power on unseen data.
The optimized $RF$ hyperparameters are: 
\begin{itemize}
\item $n\_estimators$ - the number of decision trees to use in the ensemble, 
\item $max\_features$ - the number of features to use in search of best split of the node,  
\item $max\_depth$ - the maximum depth of the individual trees,
\item $min\_samples\_split$ - the minimum number of samples in the node of the
decision tree required to make a split, 
\item $min\_samples\_leaf$ - the minimum number
of samples required to be in the leaf (that is terminal) node of each tree and 
\item the relative weights of classes.
\end{itemize}

\subsubsection{Stochastic Gradient Boosting ($SGB$)}
\label{sec:SGB}
The idea of \textit{boosting} \citep{Schapire1990} is to incrementally build a classifier by re-weighting training examples giving more weight to misclassified objects.
Boosting sequentially combines multiple weak learners with high bias to reduce bias of the final estimator. Individual weak learners are not flexible/complex enough to approximate the underlying relation themselves (i.e. underfitting). We use shallow decision trees as the weak learners.
Gradient boosting treats boosting as an optimization algorithm and generalizes the boosting method to arbitrary differentiable objective functions \citep{friedman2001,Mason:1999:BAG:3009657.3009730}. 
%
Boosting can be combined with bagging and random subset selection (\textit{stochastic gradient boosting}) to prevent overfitting.
This is achieved by using only a subsample of training data on each iteration \citep{Friedman2002367} and a subset of features to decide which data and features should be used for splitting a tree node or creating another tree.
We optimize the following hyperparameters: 
\begin{itemize}
\item $learning\_rate$ - the scale value for the prediction of each tree (shrinkage); 
\item model complexity parameters: $max\_depth$ - the maximum depth of the individual trees, $gamma$ - the minimum objective function reduction required to make a further partition on a leaf node of the tree, $min\_child\_weigth$ - the minimum sum of weights of all examples in a child of a split required to make further splits, $max\_delta\_step$
- the maximum delta step allowed for each tree's weight estimation to be; 
\item parameters that make predictions to be more robust to noise: $subsample$ - subsample ratio of the training instances, that is the fraction of the training data set drawn at random without replacement at each iteration, $col\_sample\_bytree$ - the subsample ratio of columns (features) when constructing each tree, $col\_sample\_bylevel$ - subsample ratio of columns (features) for each split, in each level, $scale\_pos\_weigth$ - relative weights of classes, 
\item parameter that controls the model complexity through regularization: $reg\_lambda$ - $L2$-regularization term on weights. The parameter $n\_estimators$ - number of decision trees in model - was determined as the iteration after which the performance measure ($F_1$; see Section~\ref{sec:hyperpartuning}) have not
improved in the following 30 iterations (\textit{early stopping} rule).
\end{itemize}

\subsubsection{Neural Net ($NN$)}
\label{sec:NN}
We used a fully connected neural network topology and checked one and two
hidden layers. Though we did not expect complex decision surface geometry for
our problem, we decided to try two hidden layers, but include regularization
by means of constrains on neuron weights and the \textit{dropout}\footnote{Dropout
is a regularization method for $NN$ where a randomly selected fraction of neurons
do not participate in updating weights. That helps to avoid overfitting as shown by \cite{dropout}.}
technique to prevent overfitting. The input and the hidden layer(s) both had
\textit{rectified linear units} \citep{relu} activation functions \citep{haykin1999neural} and the output layer had a
sigmoid activation function for probabilistic predictions. The neuron weights were
initialized using the normal distribution.
The weight updates used the \textit{Stochastic Gradient Decent} ($SGD$) method on
subsets (minibatches) of training data \citep{2016arXiv160904747R}. We optimized the following hyperparameters: 
\begin{itemize}
\item network architecture parameters -- the number of hidden layers and neurons in each hidden
layer (size of the input layer was determined by the number of features);
\item regularization parameters -- the value of the dropout at each layer (except
output) and the maximum sum of weights for each layer;
\item parameters of $SGD$ (not specific to $NN$) -- the initial learning rate $lr$, the decay rate $decay$, rate of decreasing learning rate (learning rate schedule),
$momentum$ - parameter that determines the ``inertia'' of neuron weight updates with $SGD$, 
$batch\_size$ - the number of data points to use for calculating updates of neuron weights; 
\item $class\_weight$ - the relative weights of classes. $nb\_epochs$ - number of epochs, that is the number of times all training data were used for updating network weights - was determined by the early stopping rule.
\end{itemize}

\begin{table}
    \centering
		\caption{Variability selection algorithms and their hyperparameter values that maximize the $F_1^{CV}$ for the test data set LMC\_SC20.}
		\label{table:algos}
		\begin{center}
			\begin{tabular}{r@{~~~}l@{~~~~}r@{~~~}l@{~~~~}c}
				\hline
				Algorithm & Secion & Hyperparameter & Value & $F_1^{CV}$\\
				\hline
				\multicolumn{5}{c}{Machine learning algorithms}\\
				$kNN$   & \ref{sec:kNN} & $n\_neighbors$ & 6 & 0.68\\
                        &               & $weights$ & $distance$\\
				$LR$    & \ref{sec:LR}  & C & 50.78 & 0.68\\
				        &               & $class\_weight$ & 2.65\\
				$SVM$   & \ref{sec:SVM} & $kernel$ & $rbf$ & 0.80\\
				        &               & $C$ & 25.05\\
				        &               & $gamma$ & 0.017\\
				        &               & $class\_weight$ & 2.93\\
				$RF$    & \ref{sec:RF}  & $n\_estimators$ & 1400 & 0.77\\
				        &               & $max\_depth$ & 16\\
				        &               & $max\_features$ & 5\\
				        &               & $min\_samples\_split$ & 16\\
				        &               & $min\_samples\_leaf$ & 2\\
				        &               & $class\_weight$ & 28\\
				$SGB$   & \ref{sec:SGB} & $learning\_rate$ & 0.085 & 0.79\\
				        &               & $max\_depth$ & 6\\
				        &               & $min\_child\_weigth$ & 2.36\\
				        &               & $subsample$ & 0.44\\
				        &               & $colsample\_bytree$ & 0.35\\
				        &               & $colsample\_bylevel$ & 0.76\\
				        &               & $gamma$ & 4.16\\
				        &               & $scale\_pos\_weight$ & 4.09\\
				        &               & $max\_delta\_step$ & 2\\
				        &               & $reg\_lambda$ & 0.09\\
				$NN$	& \ref{sec:NN}  & num. of hidden layers & 1 & 0.81\\
        		        &               & num. neurons in hidden layer & 13\\
				        &               & $dropout$ on input layer & 0.00\\
				        &               & $dropout$ on hidden layer & 0.17\\
                        &               & sum of weights, input layer & 9.04\\
				        &               & sum of weights, hidden layer & 5.62\\
				        &               & $learning\_rate$ & 0.20\\
				        &               & $decay\_rate$ & 0.001\\
				        &               & $momentum$ & 0.95\\
				        &               & $class\_weight$ & 2.03\\
				        &               & $batch\_size$ & 1024\\
				\hline
				\multicolumn{5}{c}{Traditional methods}\\
				$J_{\rm time}^a$  & & selection threshold & $5.3\sigma$ & 0.59\\
				$L^b$             & & selection threshold & $6.5\sigma$ & 0.53\\
				PCA$^c_1$         & & selection threshold & $7.4\sigma$ & 0.49\\
				median$^d$        & &                     &             & 0.43\\

				\hline
			\end{tabular}
		\end{center}
\flushleft $^a$\,$J_{\rm time}$ is the variability index (Table~\ref{tab:indexsummary}) with the highest $F_1$-score for LMC\_SC20, but some short-period variables cannot be recovered with this index.
$^b$\,$L$ index has the highest $F_1$-score in this data set among the indices that may recover all known variables.
$^c$\,Admixture coefficient of the first PCA component used as a composite variability index (a linear combination of individual indices, see \citealt{2017MNRAS.464..274S} for details).
$^d$\,The last line in the table presents the median $F_1$-score of all variability indices 
compared by \cite{2017MNRAS.464..274S}.
\end{table}

\subsection{Hyperarameter tuning}
\label{sec:hyperpartuning}

Each algorithm's hyperparameters (listed in Table~\ref{table:algos}) were tuned using 
the \textit{Tree of Parzen Estimators (TPE)} algorithm \citep{TPE} 
implemented in \texttt{hyperopt}\footnote{\url{http://hyperopt.github.io/hyperopt/}}.
$TPE$ is a Bayesian approach to optimization, which models conditional
probability $p(\lambda | c)$, where $\lambda$ represents the values of hyperparameters and $c$
is some loss function (criterion one desires to minimize) by two Gaussian
Mixture Models\footnote{Gaussian Mixture Model is a probabilistic model that assumes data are generated from a mixture of a finite number of Gaussian distributions with unknown parameters \citep{esl}. It is used for probability density estimation, classification and \textit{unsupervised} learning e.g. clustering, anomaly/outliers detection.}. One model, $l(\lambda)$, is fitted to the hyperparameter values
associated with the smallest (best) values of the loss function and the other,
$g(\lambda)$, is fitted to the hyperparameter values for all other values of the loss function. 
The suggested new set of hyperparameters, $\lambda$, at each $TPE$ iteration is the one resulting in the lowest value of $g(\lambda)/l(\lambda)$.

As noted in Section~\ref{sec:technique}, hyperparameters should not be learned from training data. We use 4-fold \textit{Cross-Validation} ($CV$; \citealt{esl}) during the hyperparameter search to prevent overfitting. For each trial with the values of hyperparameters proposed by \textit{TPE} the data were split into 4 non-overlapping parts (usually called \textit{folds}). The split was made by preserving the proportion of classes in folds (``stratified'' split).
Three of the four folds were combined into a training sample where the classifier with trial hyperparameters values was fitted and one fold became the evaluation sample that was used to evaluate the $F_1$-score. This combination of folds in training/evaluation samples was done 4 times in such a way that each of the 4 folds was used as the evaluation sample once. 
We use the following procedure to combine individual $F_1$-scores of the 4 splits into one value. ${\rm TP}$ and $\rm FP$ obtained for each split are summed and these summed ${\rm TP}$ and $\rm FP$ numbers are used to calculate the $F_1$-score.
Unlike direct averaging of $F_1$-scores of each split, this procedure is nearly free
of bias due to highly imbalanced data sets \citep{forman2010apples}. 
The cross-validation estimate of the $F_1$-score, $F_1^{CV}$, is an estimate of the algorithm's prediction performance on an unseen data set.
$F_1^{CV}$ is the quantity that was subject to maximization using the $TPE$ algorithm.

We performed a few thousand iterations of \textit{TPE} on classifiers that have
many hyperparameters and several hundred on the classifiers with few hyperparameters. 
It takes a couple of days of computing time on a Core~i5 desktop to find the best hyperparameters 
for $RF$, $SGB$ and $NN$. The computing time was less for the other algorithms.
For $NN$ and $SGB$ we first fixed the learning rate to some default values
(0.2 and 0.1) and searched for the best hyperparameters. 
We then searched for the best learning rate keeping other hyperparameters fixed.
For hyperparameters that were set and not fitted, we tried a few other choices manually. 
We tried $kNN$ with different distance metrics and $NN$ with more than two hidden layers.
We also tried $L1$-regularization for $LR$ (with poor performance that could be attributed to correlation of the features).

\section{Results and Discussion}
\label{sec:disc}

\subsection{Comparison of algorithm performance}
\label{sec:cv_comparison}

\begin{figure}
	\includegraphics[width=0.47\textwidth,clip=true,trim=0.0cm 0cm 0cm 0.5cm]{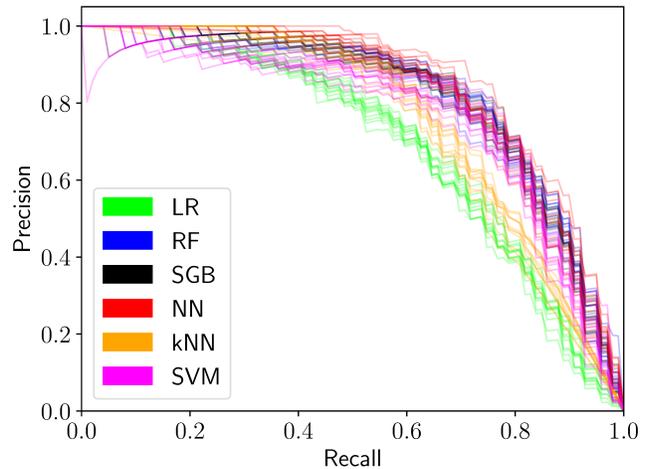}
        \caption{Precision-Recall curves for 6 algorithms with 12 different
                 splits of data set into folds during CV. Nearly identical performance is apparent for the four best algorithms.}
	\label{fig:4algo}
\end{figure}

The best values of $F_1^{CV}$ (Section~\ref{sec:hyperpartuning}) obtained for each algorithm along
with the corresponding values of tuned hyperparameters are presented in
Table~\ref{table:algos}. As expected for a small training data set, the
performance of classifiers depends on the way the data are split into folds during CV.
Figure~\ref{fig:4algo} shows the \textit{Precision-Recall} curves for each of the
6 algorithms. The hyperparameters used are the best for one (common
to all algorithms) of the CV splits, that was the result of the fixed random seed used. 
The different curves of the same color show the effect of different CV splits on the performance of each algorithm. 
$SVM$, $RF$, $GB$ and $NN$ show nearly equal performance.

$LR$ showed the worst performance as indicated by its Precision-Recall curve in Figure~\ref{fig:4algo} 
and the low value of $F_1^{CV}$. 
Note however, that the algorithm's $F_1$-score (as measured by $F_1^{CV}$) is still above the values reached by 
the traditional selection based on individual variability indices (Table~\ref{table:algos}).
Low performance can be understood as the $LR$ is a linear model that separates
classes with a linear decision surface that could result in high bias in case of classes that are not linearly separable (when nonlinear feature combinations better predict the data). 

$kNN$ also showed lower performance compared to the four best classifiers. 
This may result from the presence of class outliers (training objects surrounded by objects of a different class in feature space), 
which is especially pronounced in the case of highly imbalanced data sets. 
Moreover, the large number of features promote the \textit{curse of dimensionality} \citep{hughes} - a phenomenon that in high dimensional volume most of the points lie close to its boundary. Thus all vectors become remote from a given vector equally far and one needs exponentially more training data to represent density in a
highly dimensional space \citep{Beyer1999}. 
One has to mention that the classes of variable and non-variable stars are very inhomogeneous. 
The class of ``variables'' includes objects of various types (eclipsing binaries, pulsating variables) 
changing their brightness with different amplitudes and on various timescales (see Section~\ref{sec:bias}). 
The class of ``non-variables'' includes non-variable objects with properly measured brightness, 
as well as the few objects with corrupted measurements that have high values of variability indexes (features) but do not pass visual inspection of their light curves. Thus, the ``similarity hypothesis'' (see \ref{sec:kNN}) may fail in this case.
Finally, inclusion of some noisy features could also lead to degraded performance. 
We tested the latter possibility by adding an extra data preprocessing step: selecting the $K_{\rm best}$ best features as measured by $ANOVA$ $F$-value between features and class \citep{feature_selection_review,FANOVA} and found $K_{\rm best} = 16$, but only with a marginal (0.002) gain in $F_1^{CV}$.

Formally, the highest $F_1^{CV}$ was obtained by $NN$. The best $NN$
architecture consists of a fully-connected network structure with one input layer
with 18 neurons (determined by the number of features used), one hidden
layer with 13 neurons (both with Rectified Linear Units activation functions)
and an output layer with sigmoid activation function. No dropout and relaxed weights constrains are
preferred by the best model for the input layer.

In addition to the OGLE-II LMC light curves described in Section~\ref{sec:lightcurves}, 
we also compared classifiers on $Kr$ and $TF1$ data sets described in
\cite{2017MNRAS.464..274S}. After excluding the most correlated features 
(with $r >0.995$) we were left with 20 and 24 features, respectively. 
The performance of all considered algorithms on the $Kr$ data set
is nearly equal ($F_1^{CV} = 0.88$ for $kNN$, 0.90 for $LR$, $RF$ and $SVM$, 0.91 for
$SGB$ and 0.92 for $NN$). For $TF1$ the relative performance of the classifiers is
about the same as for the LMC\_SC20 data set, but at a lower overall level 
(resulting from a larger number of corrupted measurements in this data set)
with the best $F_1^{CV} \approxeq 0.78$ achieved again by the $NN$ classifier.

\begin{figure}
	\centering
	\includegraphics[width=0.48\textwidth]{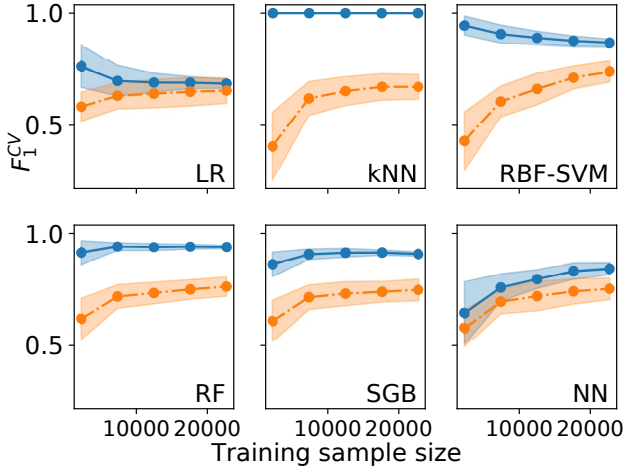}
    \caption{Learning curves for the LMC\_SC20 data set. Solid lines denote $F_1$-score on training sample, dot-dashed - cross-validation estimate of $F_1$-score on unseen data. Shaded regions show uncertainty estimated using 40 different splits of data set in training and validation sample.
    Two typical learning curve shapes are evident.
    $LR$ reveals comparable relatively low $F_1$ values on both training and validation sets
    that remain constant with growing training sample size. This is the sign of a bias of the classifier. The learning curves of the other classifiers show $F_1$ on the training data set higher than $F_1$ obtained on the independent validation data set (i.e. classifiers are overfitting) that is increasing with training sample size (that implies absence of a bias).}
	\label{fig:learningcurves}
\end{figure}

\subsection{Testing further modifications to the algorithms}
\label{sec:moretests}

\subsubsection{Learning curves and feature pre-conditioning for $LR$}
\label{sec:learningcurvesandprecondition}

To explore the possibilities of further increasing the performance of the algorithms we first considered \textit{learning curves} \citep{raschka2015python} 
-- the dependence of classifier performance (measured by the $F_1$-score)
on the amount of training data used (Figure~\ref{fig:learningcurves}).
For all the classifiers considered except for $LR$, the learning curves 
show that the $F_1$-score on the training data set is higher\footnote{Perfect classification of training data by $kNN$ is the result of \textit{weighted} distance metric used. Thus for prediction of training data it is equivalent to $kNN$ with $k$=1.} than the one obtained on the independent validation data set and the latter still increases at the maximum training sample size. This indicates that using a larger training set should further increase performance of these algorithms.
On the other hand, $LR$ shows comparable, relatively low $F_1$ values on training and validation sets.
These two characteristic types of learning curves correspond, respectively, to high-variance (in our case -- $kNN$, $SVM$, $RF$, $SGB$, $NN$) and high-bias ($LR$) algorithms (see Section~\ref{sec:technique}) for the used data set.

To improve the performance of $LR$ we tried to reduce its bias by accounting for non-linear feature interactions. 
First, in addition to the highly correlated ($r > 0.995$, Section~\ref{sec:prepocessing}) features we excluded the ones that show low $F_{\rm 1~max}$ in the original paper by \cite{2017MNRAS.464..274S}
-- $\sigma_{\rm NXS}^2$ and $v$ (Table~\ref{tab:indexsummary}).
We also excluded features with the lowest rank (as measured by feature coefficients in regression), which were lowering
the maximum achievable CV estimate of $F_1$-score using \textit{Recursive Feature
Elimination} method (kurtosis and $\sigma_{\rm NXS}^2$ again). Then, instead of raw features, we used their 
second order polynomial combinations and several first PCA-components of raw features
(the number was determined by the $TPE$ search optimizing $F_1^{CV}$).
This resulted in a performance ($F_1^{CV}=0.78$) comparable to that of the other classifiers.
We conclude that $LR$ may work as well as the other considered algorithms, but requires a special preparation of the input data.

\subsubsection{Exclude uninformative features}

We tried to exclude two features (kurtosis and skewness), which have the least relation to variability class (as reported by $ANOVA$ $F$-value between label/feature) from the input of the best classifier, $NN$, to check if the removal of these most noisy features increases the performance of the $NN$ classifier. After excluding the features, we repeated the $TPE$ search for optimal hyperparametes. The resulting $NN$ has marginally ($\Delta F_1^{CV} \approx 0.005$) degraded performance and simpler architecture (11 instead of 13 neurons on the hidden layer, stronger regularization via dropouts and weight constrains). Excluding kurtosis and skewness from the input of the third-best $SGB$ classifier also results in a slightly decreased performance ($\Delta F_1^{CV} \approx 0.01$). This suggests that even the least-important of the considered features contain some 
useful information that can be taken into account by the best classifiers $NN$ and $SGB$.

To test how many features are necessary to obtain high $F_1$-scores we used
the $SGB$ method as it is pretty straightforward to get the importance of features
using this algorithm \citep{esl}.
Although we used hyperparameters tuned for
18 features after successively excluding the least important
features, we found that with 9 features ($J$, $J_{\rm time}$, $I$,
$Magnitude$, ${\rm IQR}$, $1/\eta$, kurtosis, skewness, $I_{\rm sgn}$) we can still
obtain $F_1^{CV}$ as high as 0.77 and using only 3 ($J$, kurtosis, $I$)
results in $F_1^{CV}$ = 0.62.

We also tried to use several PCA-components (Figure~\ref{fig:pca}) as features instead of the original features listed in Table~\ref{tab:indexsummary}. 
The expectation was that by using several first PCA-components we may reduce the
noise introduced by a number of (nearly) uninformative features. 
For this test we used $RF$ classifier and added the number of used PCA-components 
to the list of optimized hyperparameters (see Table~\ref{table:algos}). 
We allowed $max\_features$ to vary from 3 to 5 and
the number of PCA-components from 5 to 18. The best value of $F_1^{CV}$ was 0.75 with
18 PCA-components and $max\_features$ = 4. Thus the classifier performs best
when using essentially all features. 
The degraded performance could be attributed to PCA keeping only linear combinations of features. 

\subsubsection{Ensemble combining multiple classifiers}
\label{sec:ensembling}

We have tried to combine individual algorithm predictions using ensembling. 
To approximate the case of classifying 
unseen data we used different random seeds when splitting the sample into training/test splits during cross-validation estimation of the $F_1$-score. This resulted in a slightly worse estimated performance of the algorithms that used HP optimized with different CV-splits.

First we used \textit{Hard Voting} of individual algorithms: the class that obtains the majority of votes of individual classifiers is chosen. 
We tried using all algorithms with weights equal to their $F_1^{CV}$ during HP optimization. We also tried to use only the four algorithms with the highest performance ($NN$, $SVM$, $SGB$ and $RF$). 
The voting resulted in $F_1^{CV}$ estimates slightly higher than the best values for individual algorithms. The corresponding gains in $F_1^{CV}$ were 0.007 and 0.004 for all and the four best algorithms, respectively.

As the predictions of individual algorithms are uncalibrated we tried \textit{rank averaging} of the probability outputs of individual learners\footnote{\url{https://mlwave.com/kaggle-ensembling-guide/}}. Predictions of individual models were turned into ranks, averaged and the result was normalized. Using all classifiers resulted in degraded performance (-0.018) relative to the best individual classifier. Averaging ranks of predictions of the four best-performing algorithms gives the same $F_1^{CV}$ (0.0006). At the same time using two of the worst performance algorithms in averaging brings some improvement relative to their individual score (0.032).

We also combined class and probabilistic predictions of individual
algorithms using higher-level (meta) algorithm -- $LR$ using the \textit{Stacking Generalization} or \textit{stacking} method \citep{Wolpert92stackedgeneralization} both alone and with original
(\textit{lower-level}) features\footnote{{We used \texttt{mlxtend} Python package \citep{raschkas_2016_594432}.}}. 
The largest improvement (0.007) was obtained when 
using only the class predictions of the four best classifiers. This could be the result of uncalibrated probabilistic outputs of the base algorithms.

\begin{figure}
	\centering
	\includegraphics[width=0.48\textwidth]{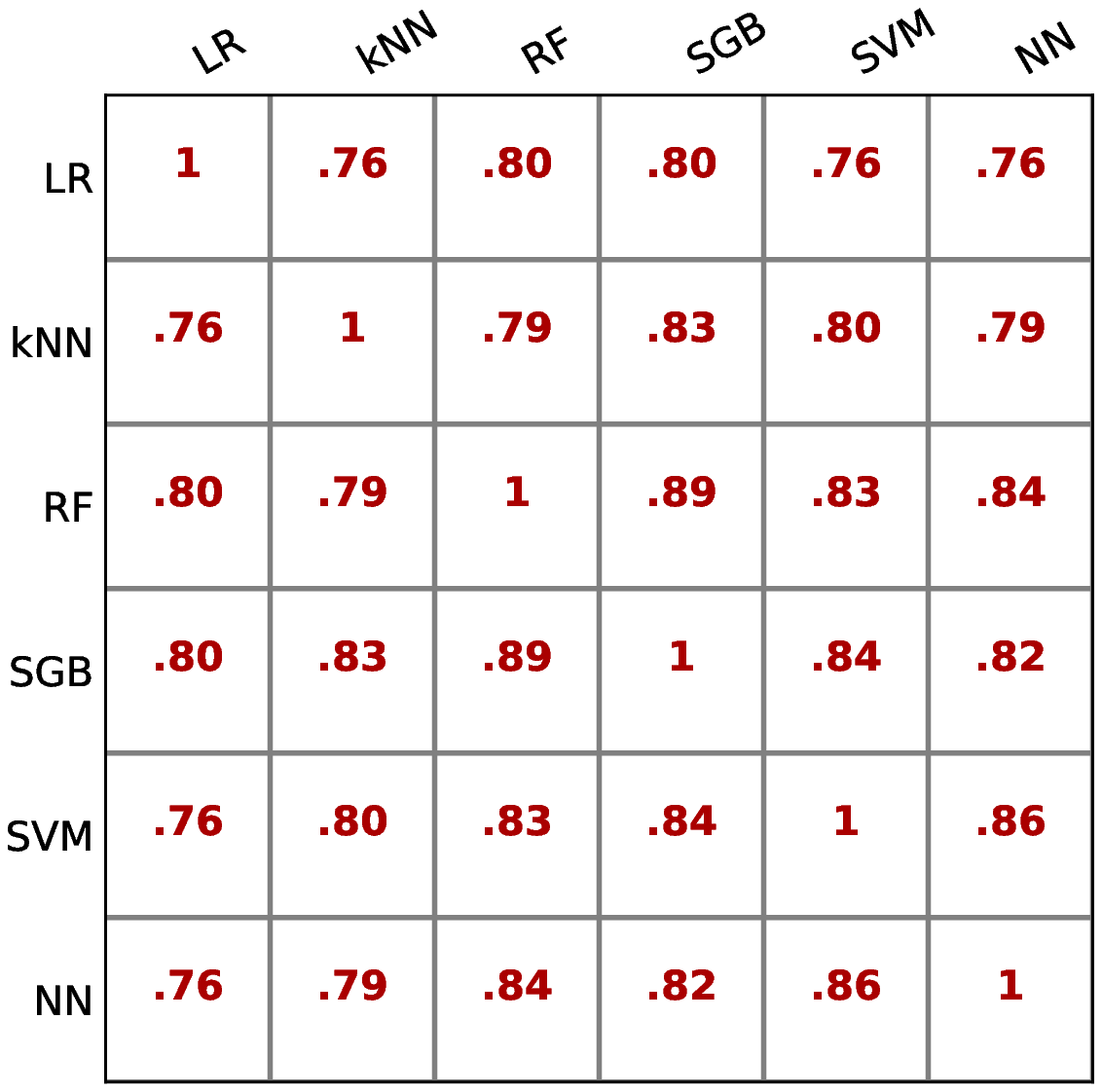}
    \caption{Pearson correlation coefficient between algorithm predictions estimated using CV on LMC\_SC20 data set. These  values are obtained with one fixed split of the data sample in training/test samples used in CV. Depending on the split, the presented values change with $\sigma \sim 0.01$-0.02 as estimated using 30 different splits.}
	\label{fig:corr_predictions}
\end{figure}

We attribute insignificant improvements of these ensembling methods to the high correlation between predictions of individual algorithms (see Figure~\ref{fig:corr_predictions}; \citealt{Sollich1996}). This is because all classifiers HP were tuned to have the highest $F_1^{CV}$ using the same CV splits of the training data. Using different CV splits during HP optimization for each algorithm or a larger training sample (that will allow calibration of the algorithms' probabilistic outputs) will make the ensembling methods more effective  \citep{Ting:1999:ISG:1622859.1622868,Sigletos:2005:CIE:1046920.1194903}.

\subsubsection{Possible future improvements}
\label{sec:future}
As can be seen from the learning curves presented in Figure~\ref{fig:learningcurves}, all high-performance classifiers would benefit from increasing the amount of training data. A larger sample of variables will also allow one to calibrate classifiers and combine probabilistic output  of multiple classifiers using stacking (Section~\ref{sec:ensembling}). Finally, as discussed in Section~\ref{sec:blind}, a larger sample size will help to avoid overfitting due to small-sized training samples, which could be unrepresentative of the general population.

A promising way to achieve a larger training set size could be the artificial enlargement
of training data (\textit{data augmentation}; see e.g. \citealt{Hoyle11062015})
by introducing possible variations to known constant/variable star light
curves (e.g.~changing variability amplitude, noise level, addition of
instrumental trends -- see Section~\ref{sec:blind}, etc.). 
According to Section~\ref{sec:technique} another promising way for improvement is engineering new features that quantify the object's image shape, profile and position on a CCD chip, proximity to other detected objects, correlation of magnitude measurements with external parameters such as seeing and airmass, periodicity in light variations, shape of the period-folded light curve, etc.

\subsection{Blind test on the new data set}
\label{sec:blind}

\begin{figure}
	\centering
	\includegraphics[width=0.48\textwidth]{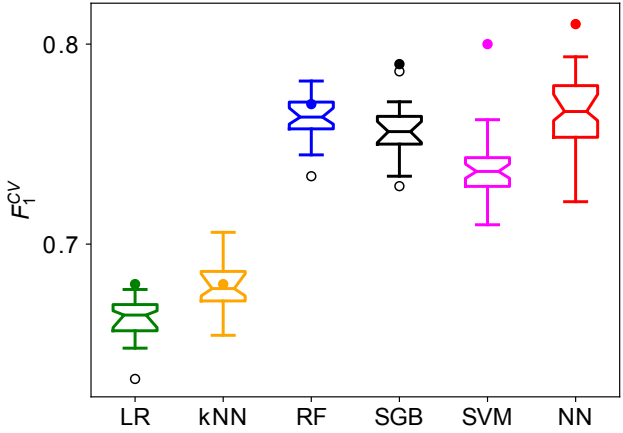}
    \caption{Boxplot of $F_1^{CV}$ values obtained by classifiers with optimized HP (see Table~\ref{table:algos}) for 30 different CV splits of LMC\_SC20 data on training/evaluation splits (not including the one used for HP tuning). The box extends from the lower to upper quartile values of the data, with a line at the median and narrowing of the box denotes the confidence band on the median. The upper error bar 
    extends from the box to last data point that is less than $Q_3 + 1.5\cdot IQR$ and the lower 
    error bar
    will extend to first data point that is greater than $Q_1 - 1.5\cdot IQR$, where $Q_1$ and $Q_3$ are first and third quartiles. Open circles 
    outside of the bars 
    are considered as outliers. Filled circles 
    represent the best values of $F_1^{CV}$ for CV split used in HP tuning (Table~\ref{table:algos}).}
	\label{fig:cv_boxplot}
\end{figure}

The actual performance on new (unseen) data is hard to estimate. 
As our data sample is quite small, we did not reserve 
some part of it for testing classifiers on the new data \citep{esl}. 
Performance on new data should be slightly lower than the estimations obtained using CV on the original data set (Table~\ref{table:algos}).
We estimate the effect of this by considering distributions of $F_1^{CV}$ values obtained by classifiers with best HP from Table~\ref{table:algos} for 30 different splits of LMC\_SC20 data in train/evaluation subsamples (not including the one used for HP tuning; Figure~\ref{fig:cv_boxplot}).
That is data set is splitted in 4 parts (for using in 4-fold CV, Section~\ref{sec:hyperpartuning}) differently 30 times. Thus each of the training/test samples used in 30 repeating runs of the 4-fold CV procedure contains some different (but overlapping) data.
As each of the 30 splits of the data sample in 4 folds was different from the one used for HP tuning (Table~\ref{table:algos}), the obtained $F_1^{CV}$ values are lower than the best $F_1^{CV}$ values presented in Table~\ref{table:algos}. The difference is 0.05 for $SVM$, while they are nearly the same for $LR$ (-0.01) and  
$kNN$. The typical error estimated using the variance of $F_1^{CV}$ between the different splits is 0.01.

On the other hand, the CV estimate of prediction performance  
is pessimistic because only some portion of data is used to fit the model (e.g. 75$\%$ in our case of 4-fold CV). 
Thus $F_1$ on the new data set with the size of LMC\_SC20 will be higher for high-variance algorithms (all except $LR$). 
The value of this bias can be estimated using learning curves (Figure~\ref{fig:learningcurves}; \citealt{esl}). 
It is interesting that $SVM$ which demonstrated the highest drop of performance on the new CV splits, should gain the most in performance from enlarging the training sample according to its learning curve (Fig.~\ref{fig:learningcurves}).

Finally, if LMC\_SC20 is not representative of the overall variable star population, then we expect a degraded performance of classifiers on new data sampled from that population (i.e. overfitting, see the discussion in Section~\ref{sec:bias}). This 
could be reduced with a larger sample size.

We have tested the $NN$ classifier with the best hyperparameters (derived in Section~\ref{sec:hyperpartuning}) on the new data
set, which consist of 31798 light curves (field LMC\_SC19, Section~\ref{sec:data}). $NN$ was fitted on the whole training data set (LMC\_SC20) with the derived best hyperparameters and its predictions
were evaluated. We used the default threshold (0.5) as this was the value used for hyperparameter optimization.
The predicted variables were checked in existing catalogues (Section~\ref{sec:lightcurves}) and by
visual inspection. Among the 205 candidates classified as variable stars,
178 occurred to be real variables (TP), 27 were considered FP.

\begin{figure}
	\includegraphics[width=0.24\textwidth]{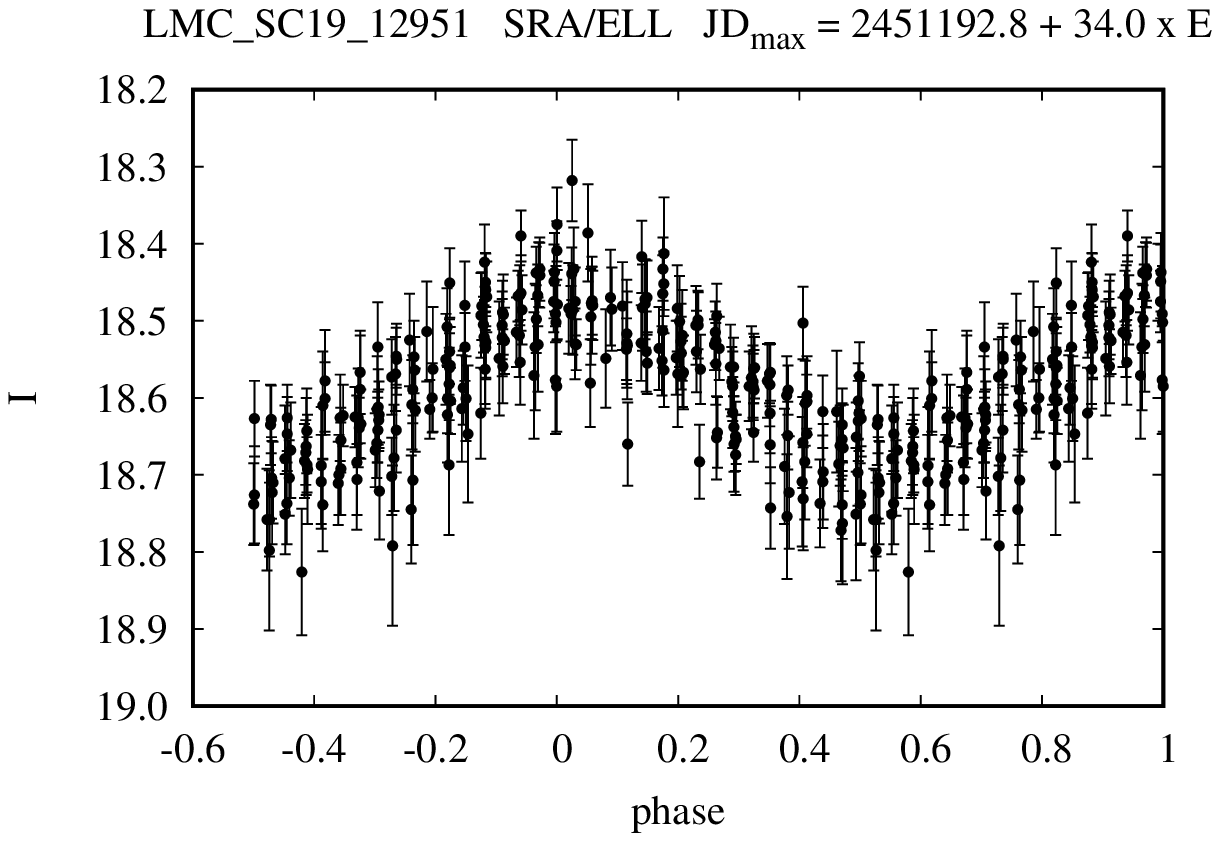}
	\includegraphics[width=0.24\textwidth]{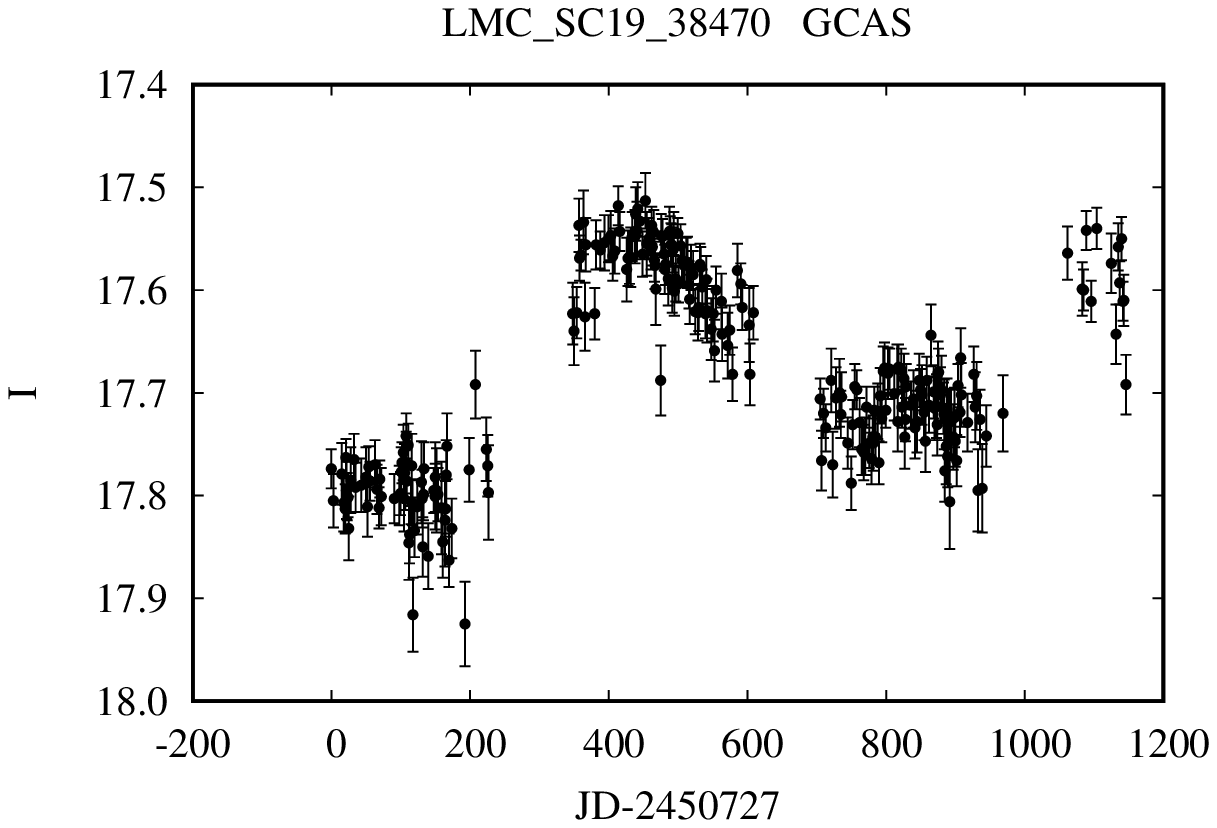}
	\includegraphics[width=0.24\textwidth]{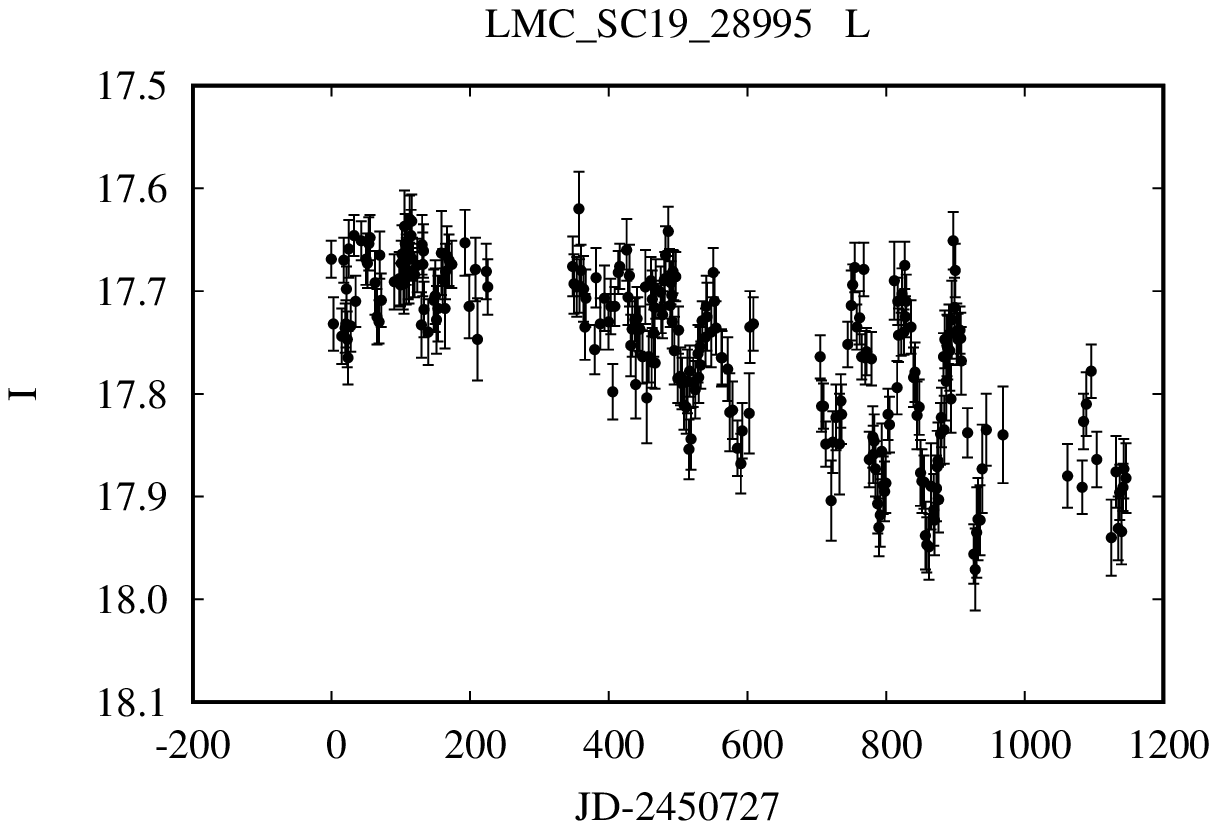}
	\includegraphics[width=0.24\textwidth]{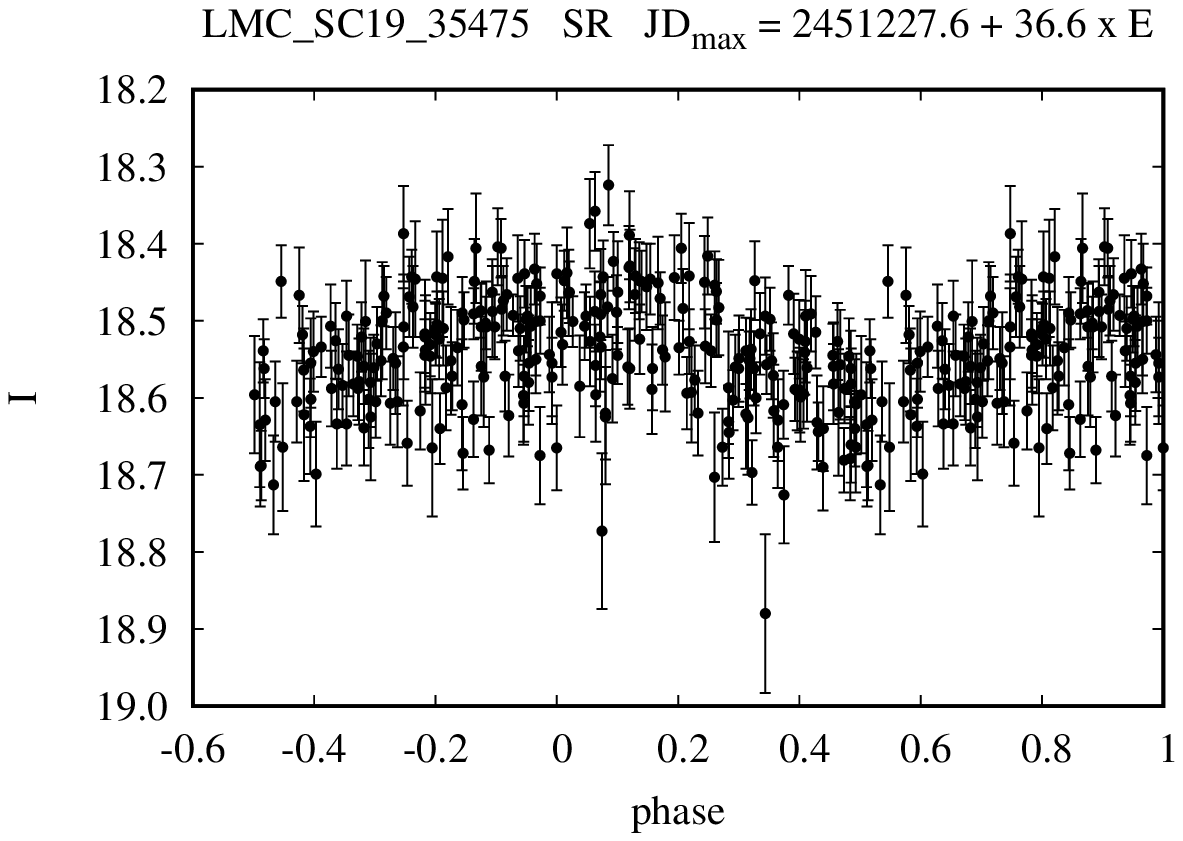}
	\includegraphics[width=0.24\textwidth]{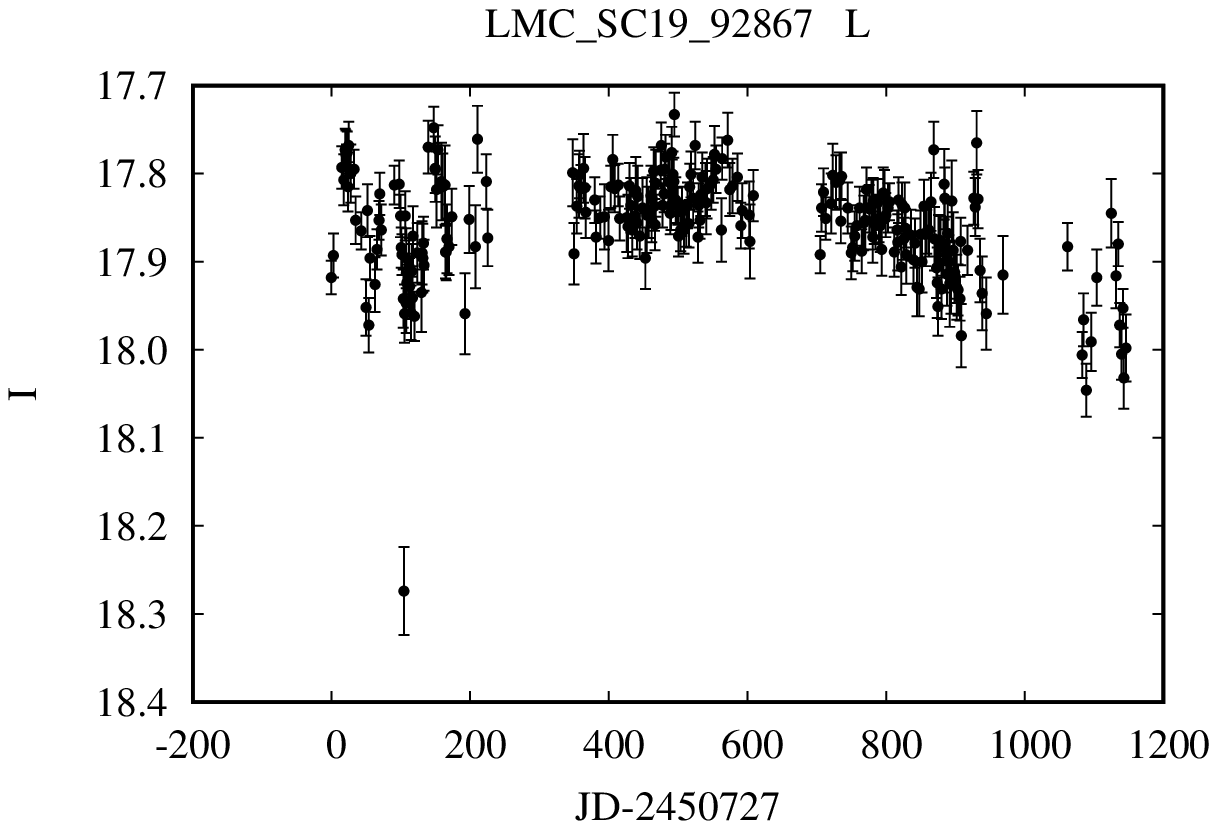}
	\includegraphics[width=0.24\textwidth]{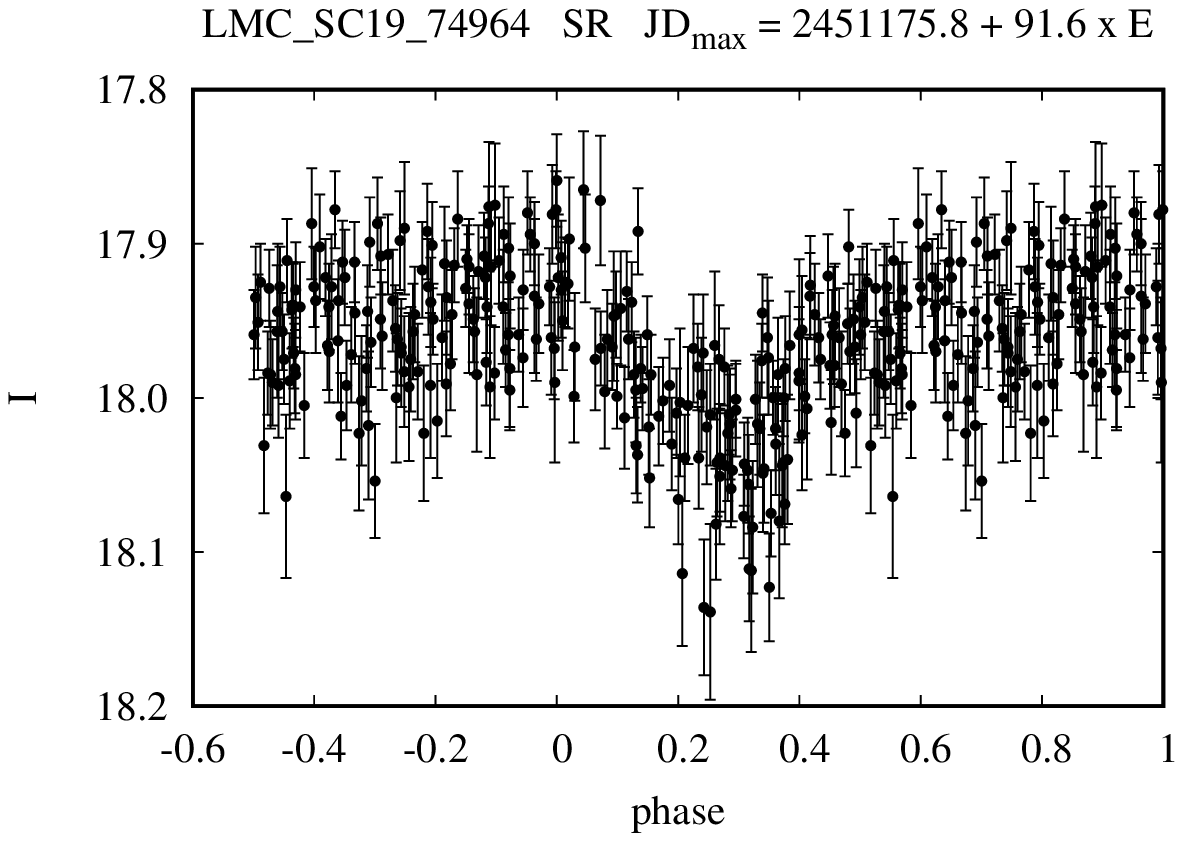}
	\includegraphics[width=0.24\textwidth]{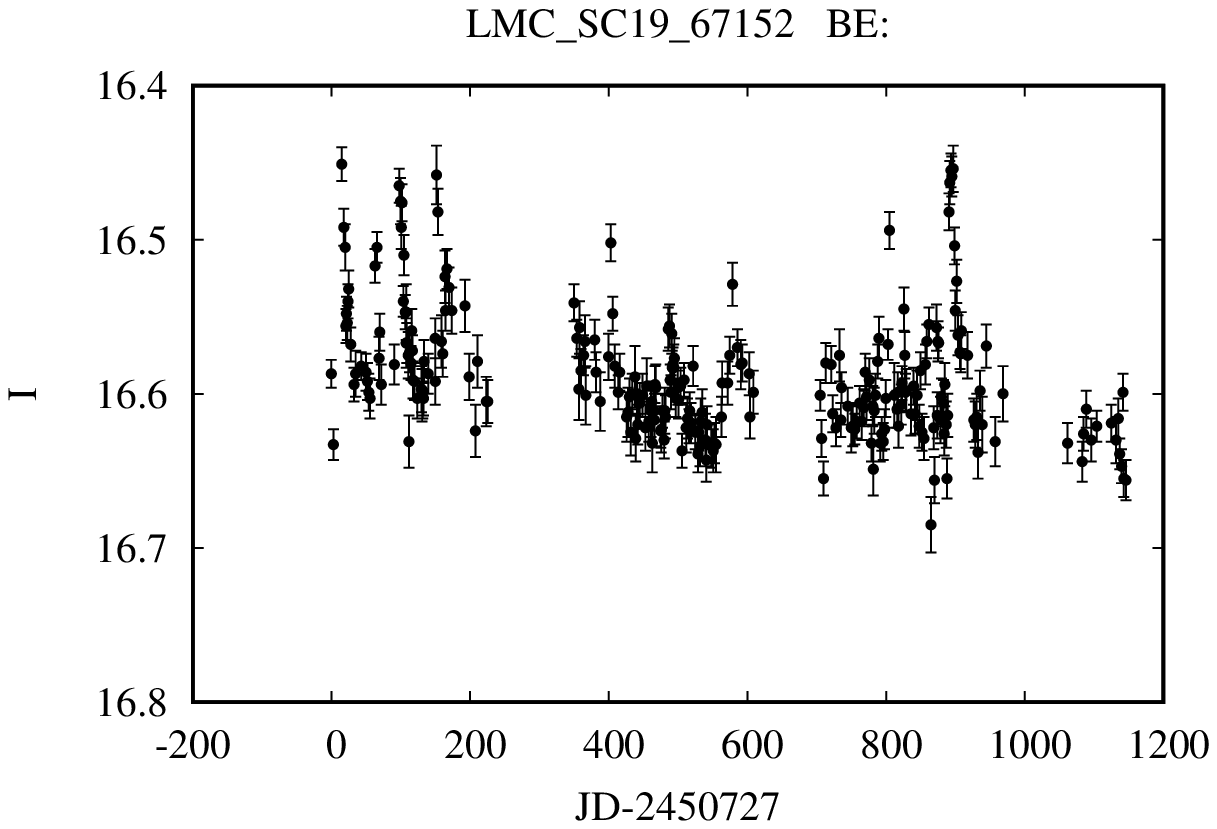}
	\includegraphics[width=0.24\textwidth]{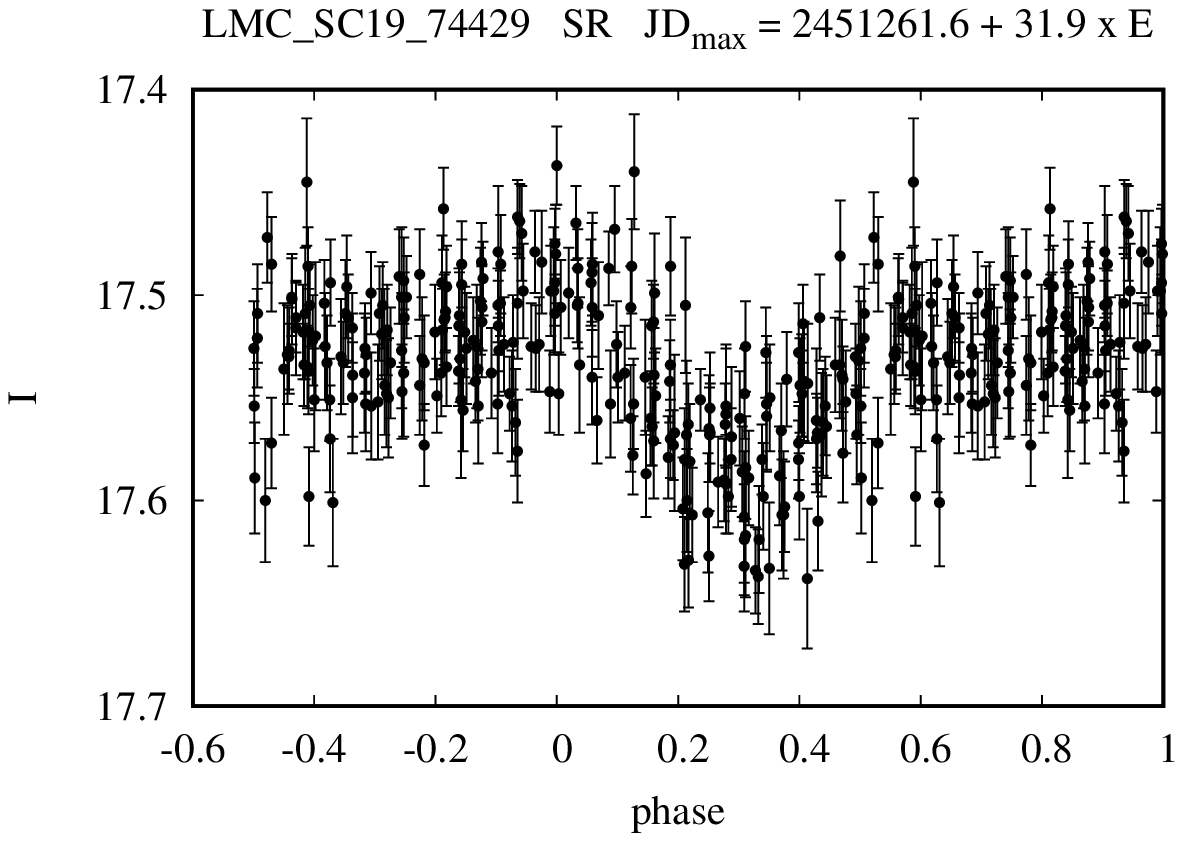}
	\includegraphics[width=0.24\textwidth]{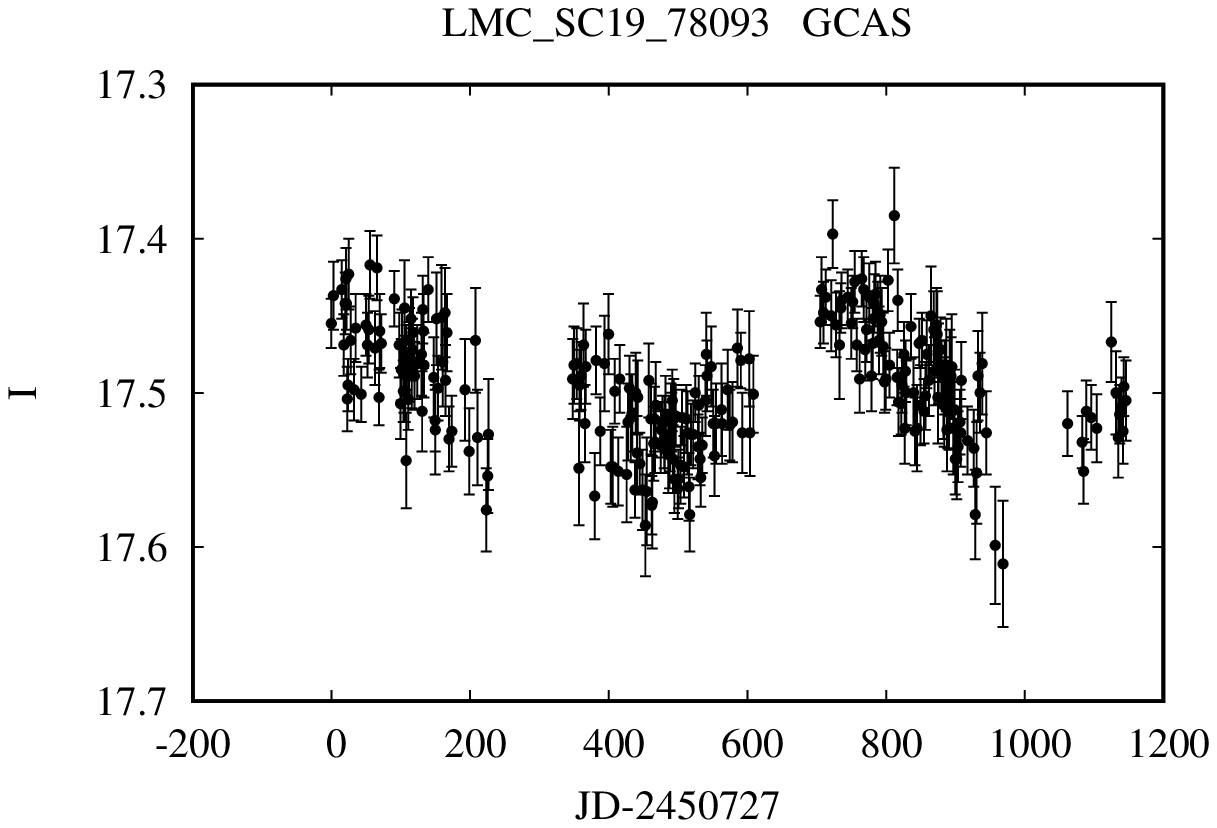}
	\includegraphics[width=0.24\textwidth]{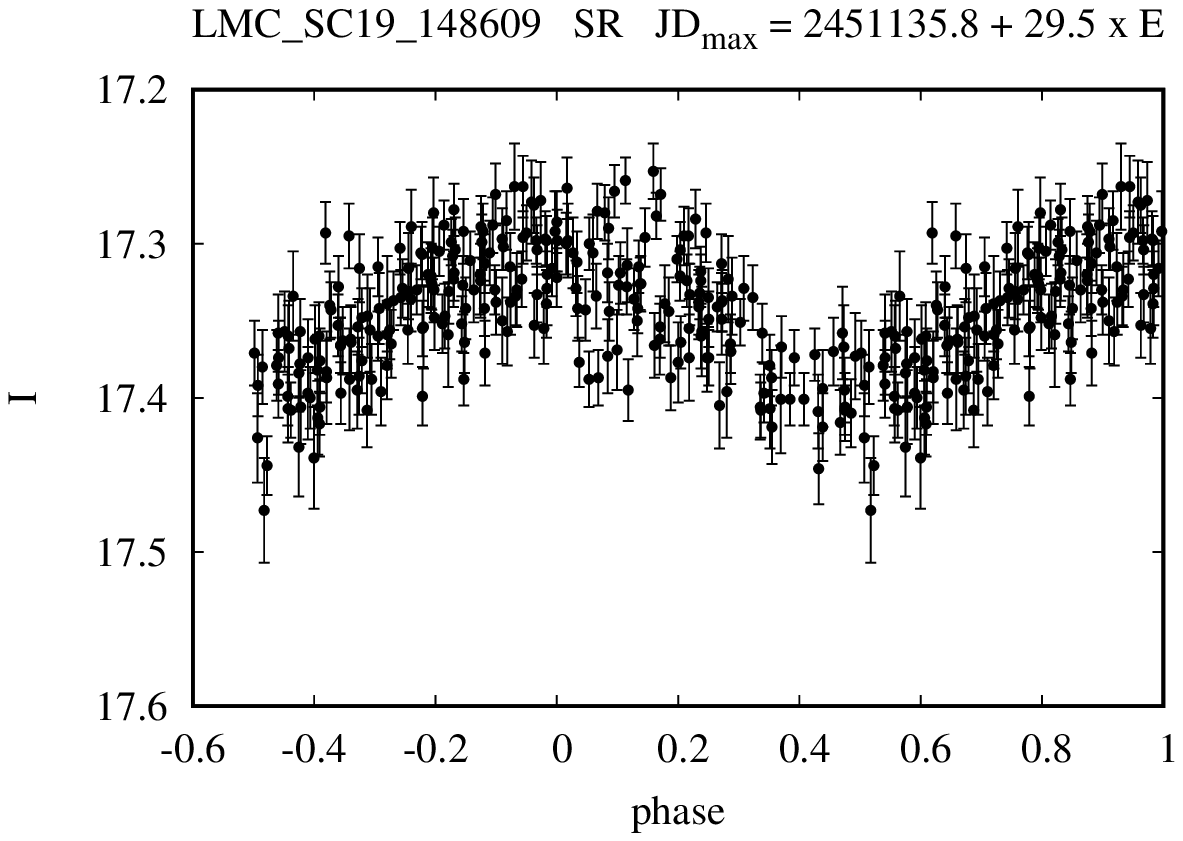}
	\includegraphics[width=0.24\textwidth]{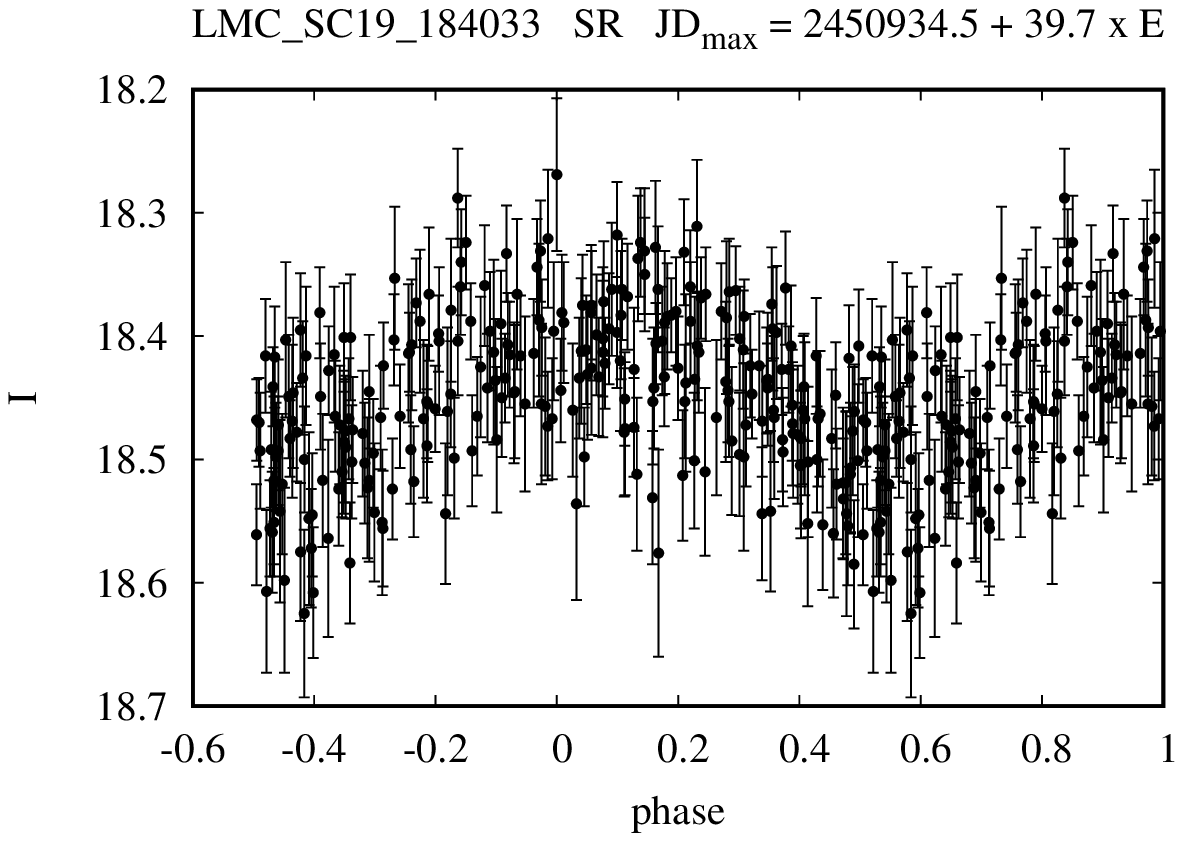}
	\includegraphics[width=0.24\textwidth]{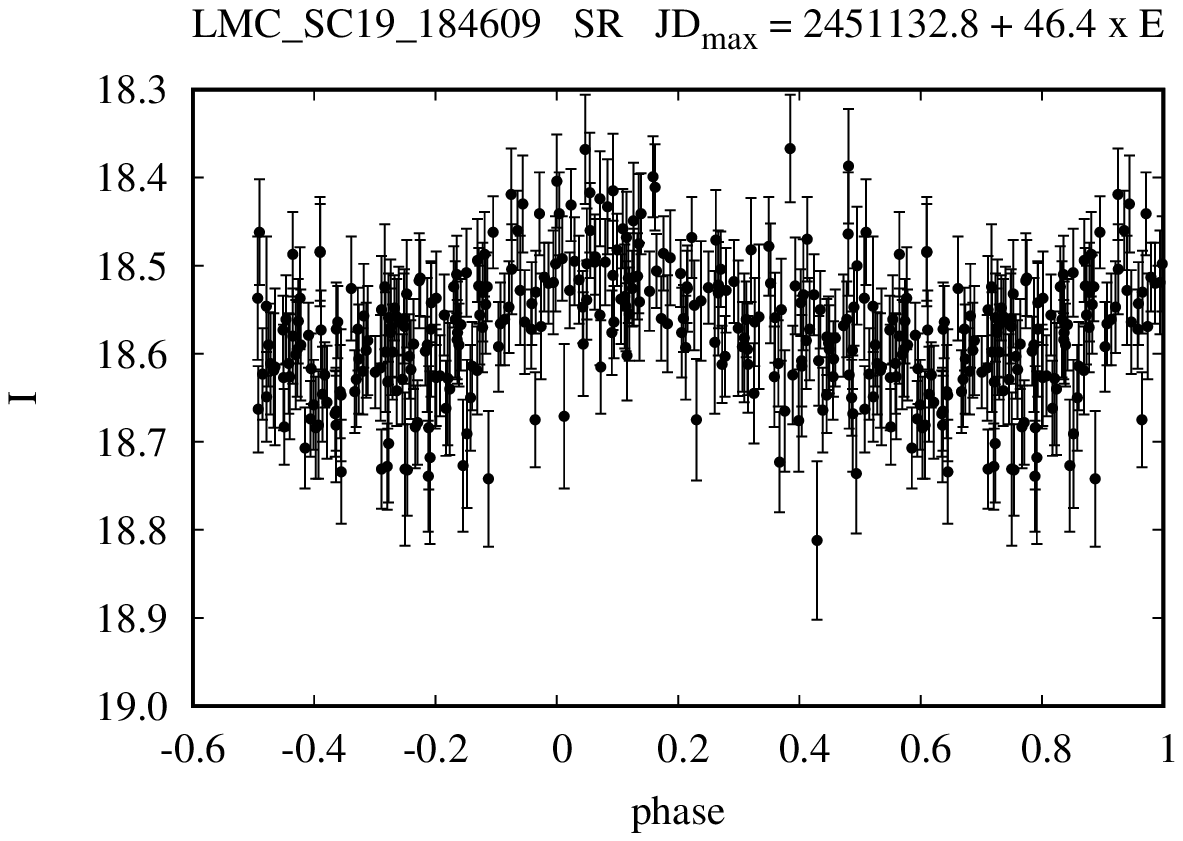}
	\includegraphics[width=0.24\textwidth]{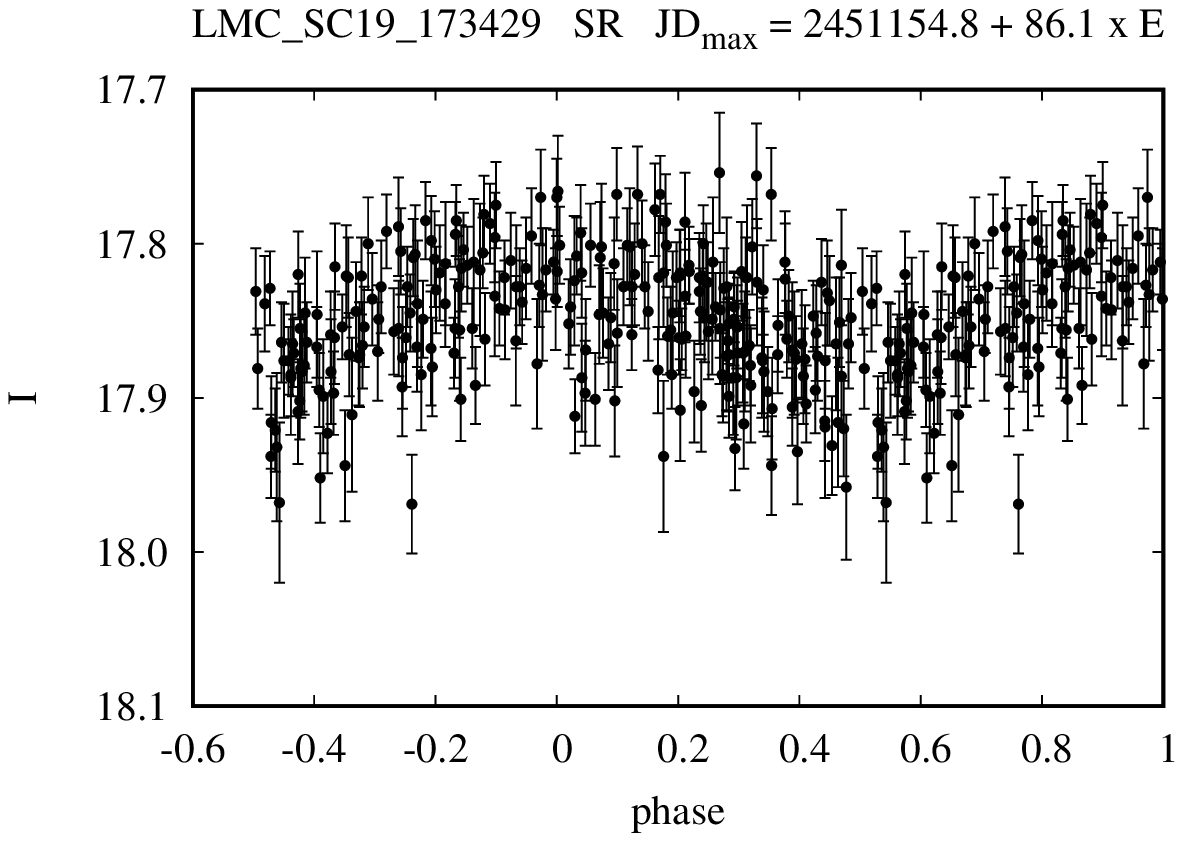}
        \caption{Light curves of the newly identified variable stars listed in Table~\ref{tab:newvars}. These are examples of true positives (TP): candidate variables identified by the $NN$ classifier that passed visual inspection. Light curves of periodic variables are phase folded with the indicated light elements. For non-periodic variables the light curves are plotted as a function of time.}
	\label{fig:newvarslightcurves}
\end{figure}

\begin{figure}
	%
	\includegraphics[width=0.24\textwidth]{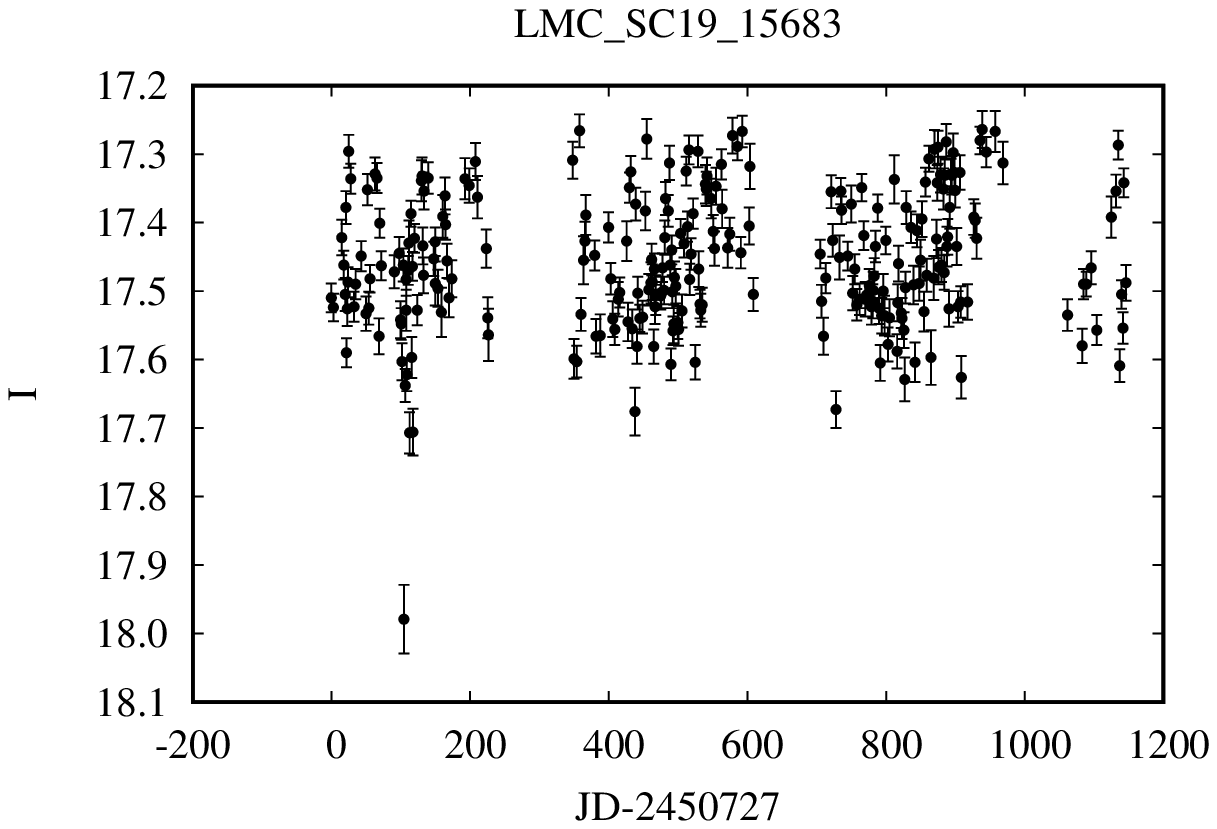}
	\includegraphics[width=0.24\textwidth]{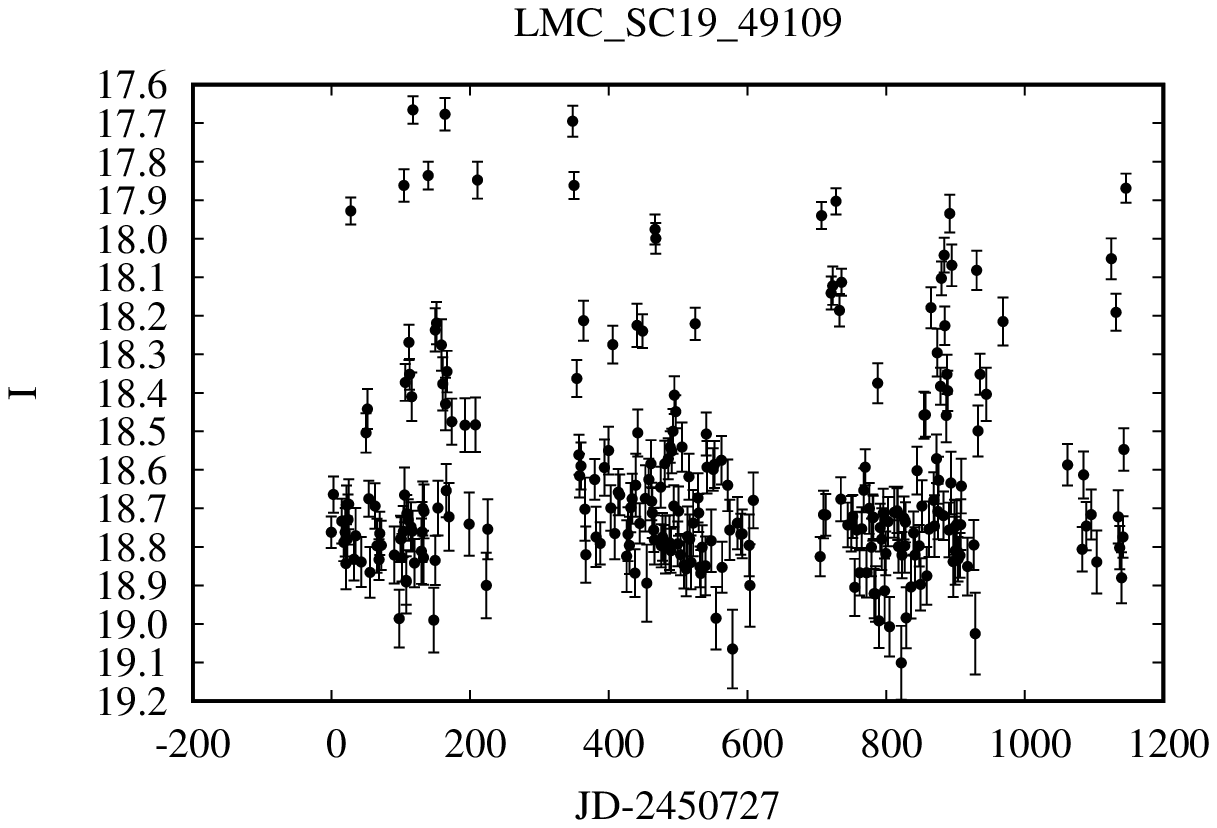}
	\includegraphics[width=0.24\textwidth]{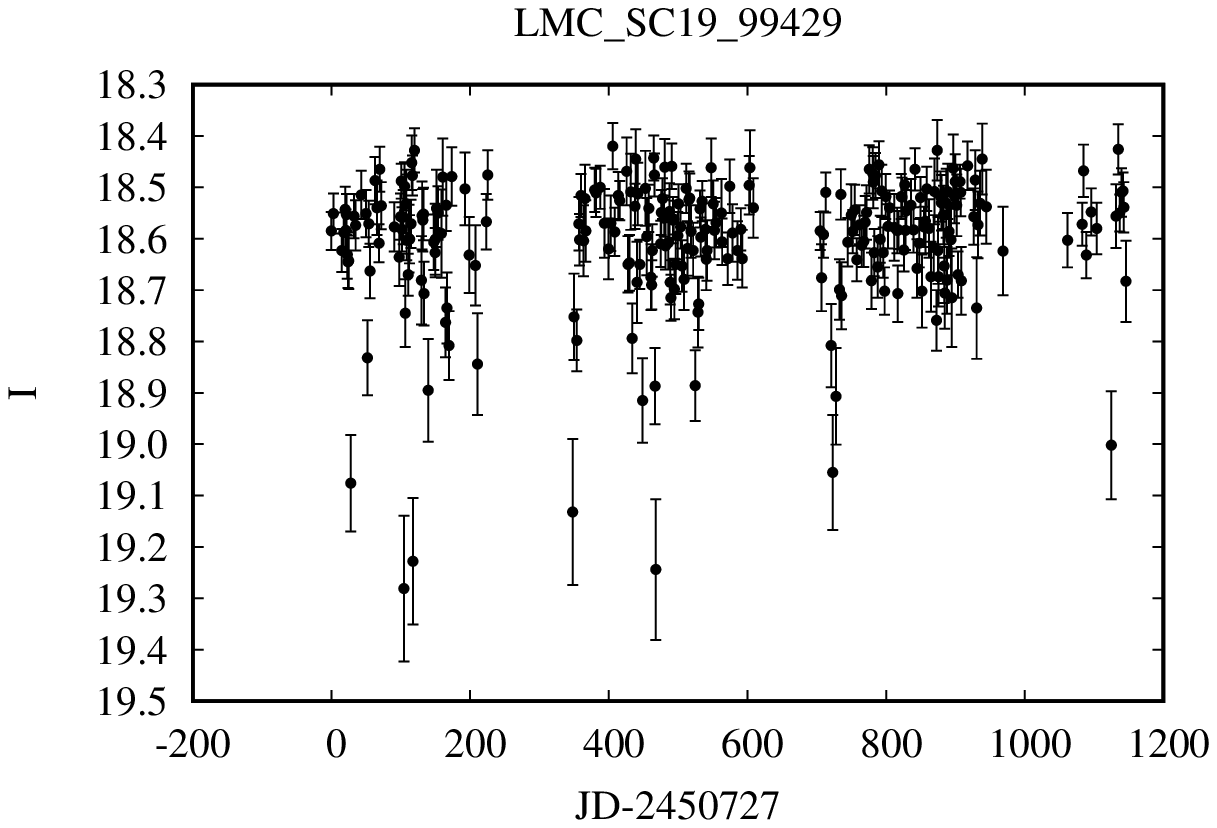}
	\includegraphics[width=0.24\textwidth]{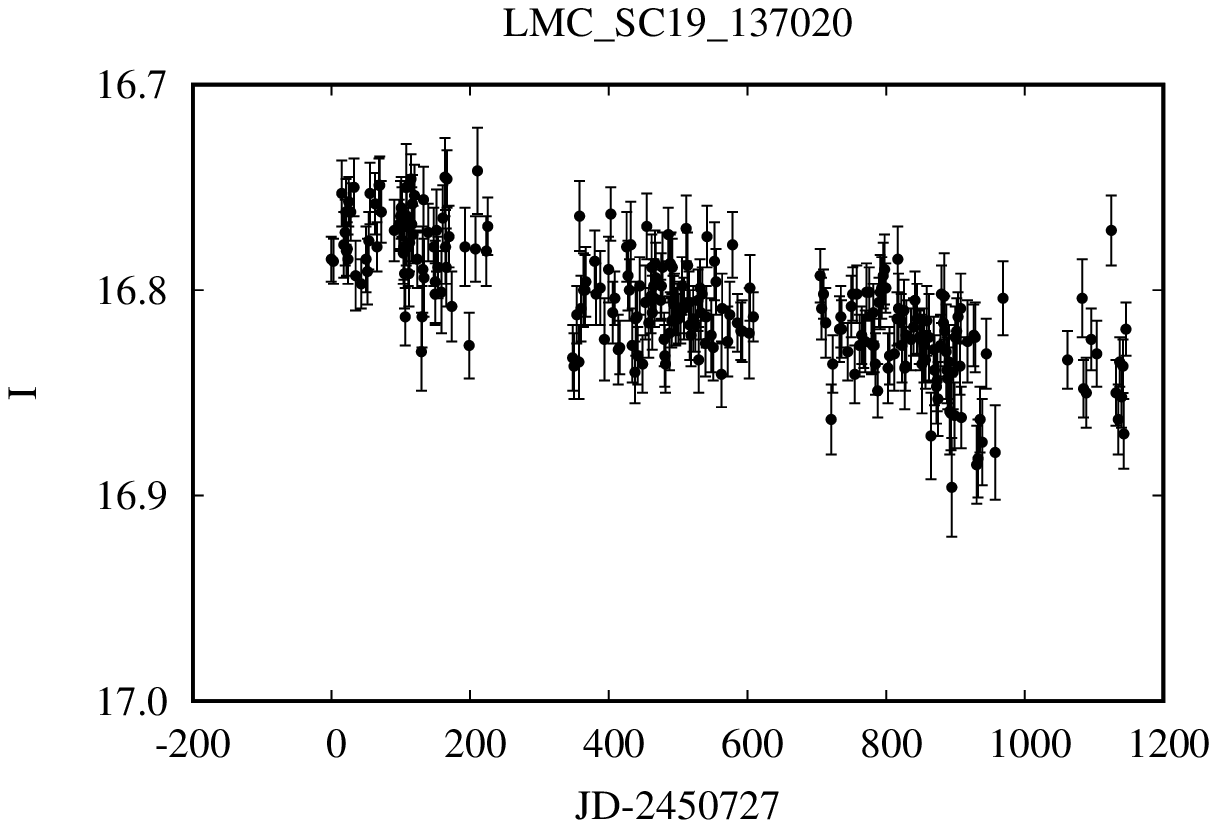}
        \caption{Example light curves of candidate variables rejected during visual inspection (FP).}
	\label{fig:falsepositivelc}
\end{figure}

The separation of true variables from false candidates division may not be perfect, it involves the following assumptions:
\begin{itemize}
 \item If a candidate variable is matched with a catalog, it is considered a 
TP. We ignore the possibility that an object may have no detectable
variations in OGLE-II data while being detected as variable by another
survey.
 \item We consider as TP candidate objects that are not matched to the catalogs of known
     variables, but upon visual inspection are identified as variable stars of
     a known type (Figure~\ref{fig:newvarslightcurves}).
 \item We consider as FP all candidates showing a continuous brightness increase
or decline if they are not matched with known variables from the catalogs (lower right panel of Figure~\ref{fig:falsepositivelc}).
This is done to exclude possible long term instrumental trends and apparent
variations caused by proper motion \citep{2001MNRAS.327..601E}.
It is possible that some true variables showing long-term brightness changes
may be misattributed to instrumental trends and mislabeled as FP.
 \item We consider as FP candidates showing an increased scatter in their light curves (compared to other objects of similar brightness), while showing no detectable periodicity in these variations (Figure~\ref{fig:falsepositivelc}). Specifically, we consider as FP those objects showing non-periodic dimming or brightening on a timescale shorter then the typical observing cadence. Young stellar objects and flare stars may show this type of behaviour. Hot/cold pixels underneath the star image may also produce light curves of these shapes. The inspection of images associated with individual measurements (that are not available to us) is necessary to judge if the measurements of a given object are reliable. We choose to exclude candidates showing this type of behaviour from the list of confirmed variables.
\end{itemize}

Among the 178 confirmed variable objects in LMC\_SC19, 12 have never been reported as variable before.
Table~\ref{tab:newvars} presents the list of newly identified variables, their colors from \cite{2000AcA....50..307U} and the suggested classification according to the GCVS scheme \citep{2017ARep...61...80S}. 
Table~\ref{tab:newvars} also lists one new variable, LMC\_SC19\_184609, which was not selected as a candidate variable by the final run of the $NN$ classifier. This variable was identified by us during a test run with hyperparameters of the $NN$ classifier differing from the ones listed in Table~\ref{table:algos} (but some other variables were missed in this run). In order to obtain a more exhaustive list of variables one needs to lower the classifier's threshold or optimize its hyperparameters using a different performance metric (as discussed in Section~\ref{sec:performancemetric}). This will come at the price of an increased number of false candidates, which have to be rejected during visual inspection. The need to find an optimal trade-off between the rate of false candidates and 
completeness is common to all variability detection techniques.
Machine learning techniques considered here provide a more favorable ratio of true variables to false detections compared to the traditional methods (Table~\ref{table:algos}).

The light curves of the new variables are presented in Figure~\ref{fig:newvarslightcurves}. The period search was performed using the  \cite{1975Ap&SS..36..137D} discrete Fourier transform method implemented in an online period search tool\footnote{\url{http://scan.sai.msu.ru/lk/}}.
These newly identified variables give an idea of what kind of variables are missed by previous variability searches in the LMC (Section~\ref{sec:lightcurves}): they have low amplitudes $\Delta I \lesssim 0.25$ and many are periodic with long periods $\gtrsim 30^d$.

\begin{figure}
	\includegraphics[width=0.24\textwidth]{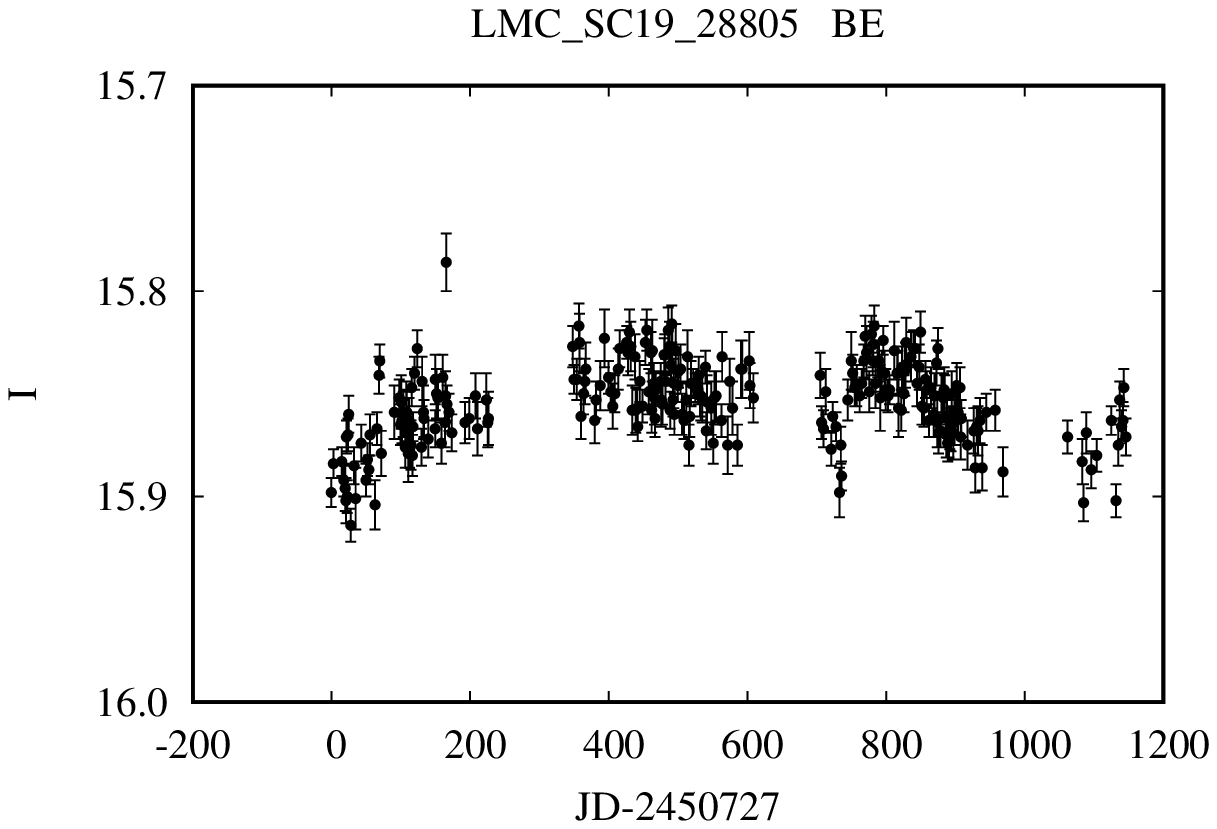}
    \includegraphics[width=0.24\textwidth]{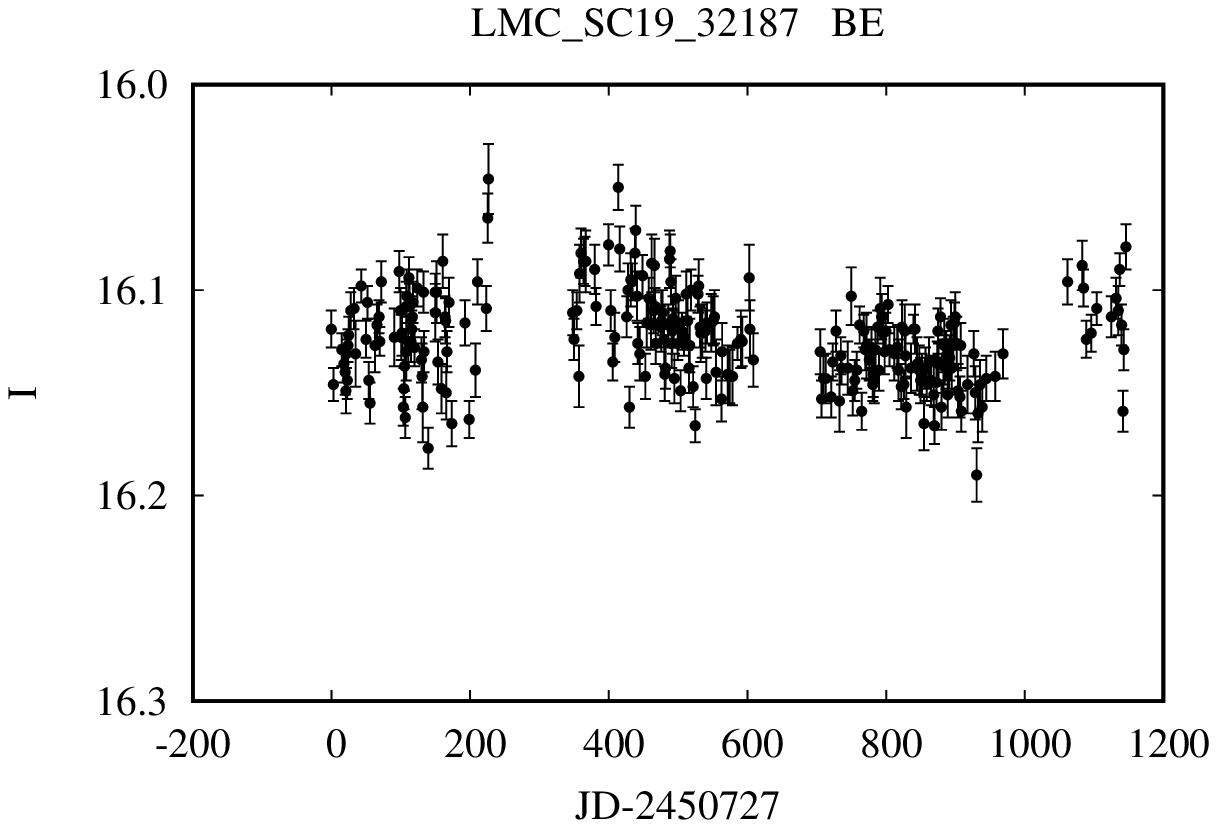}
	\includegraphics[width=0.24\textwidth]{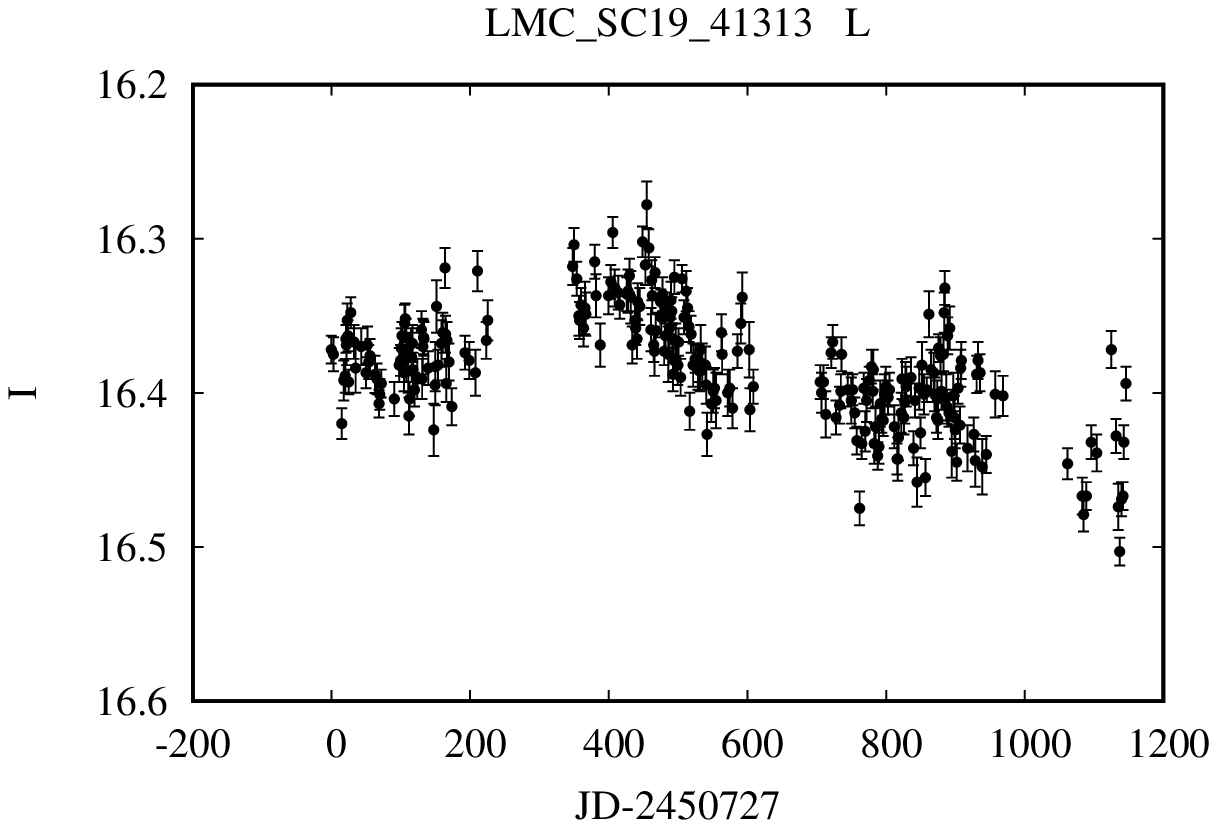}
	\includegraphics[width=0.24\textwidth]{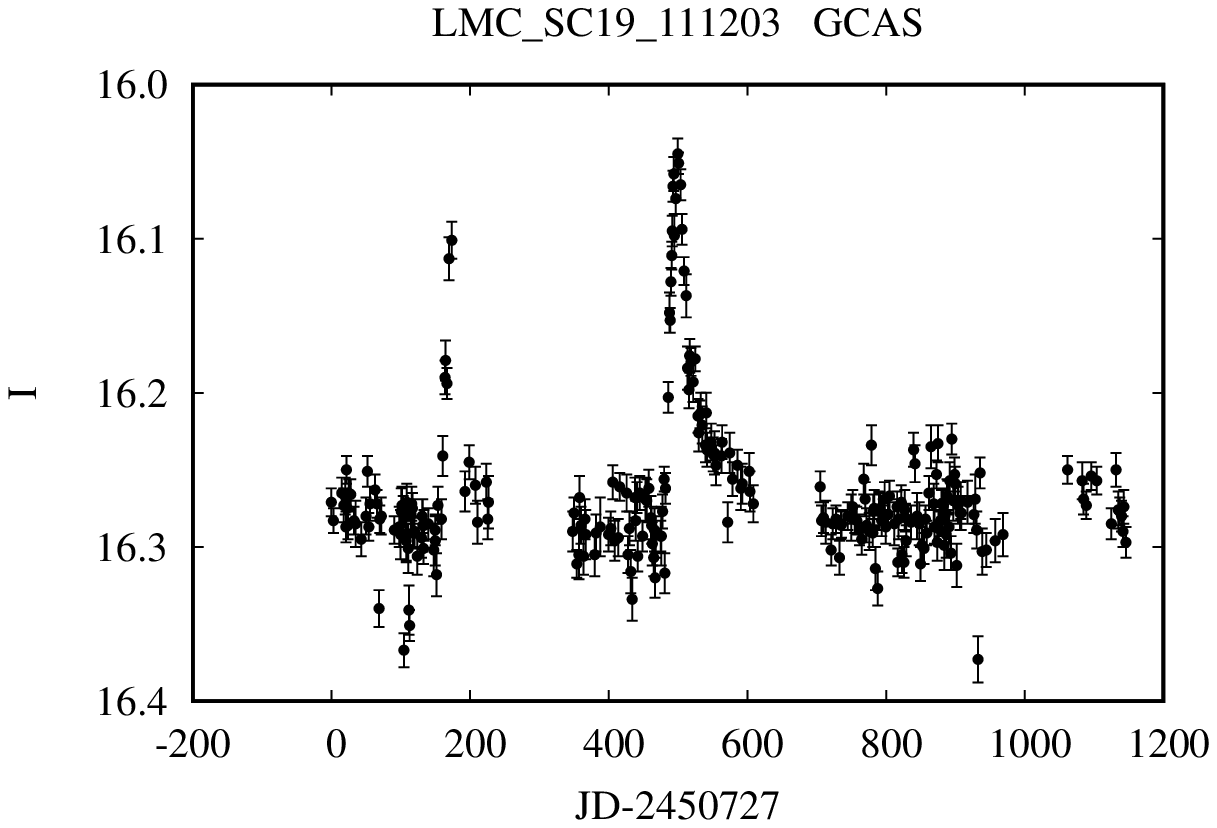}
	\includegraphics[width=0.24\textwidth]{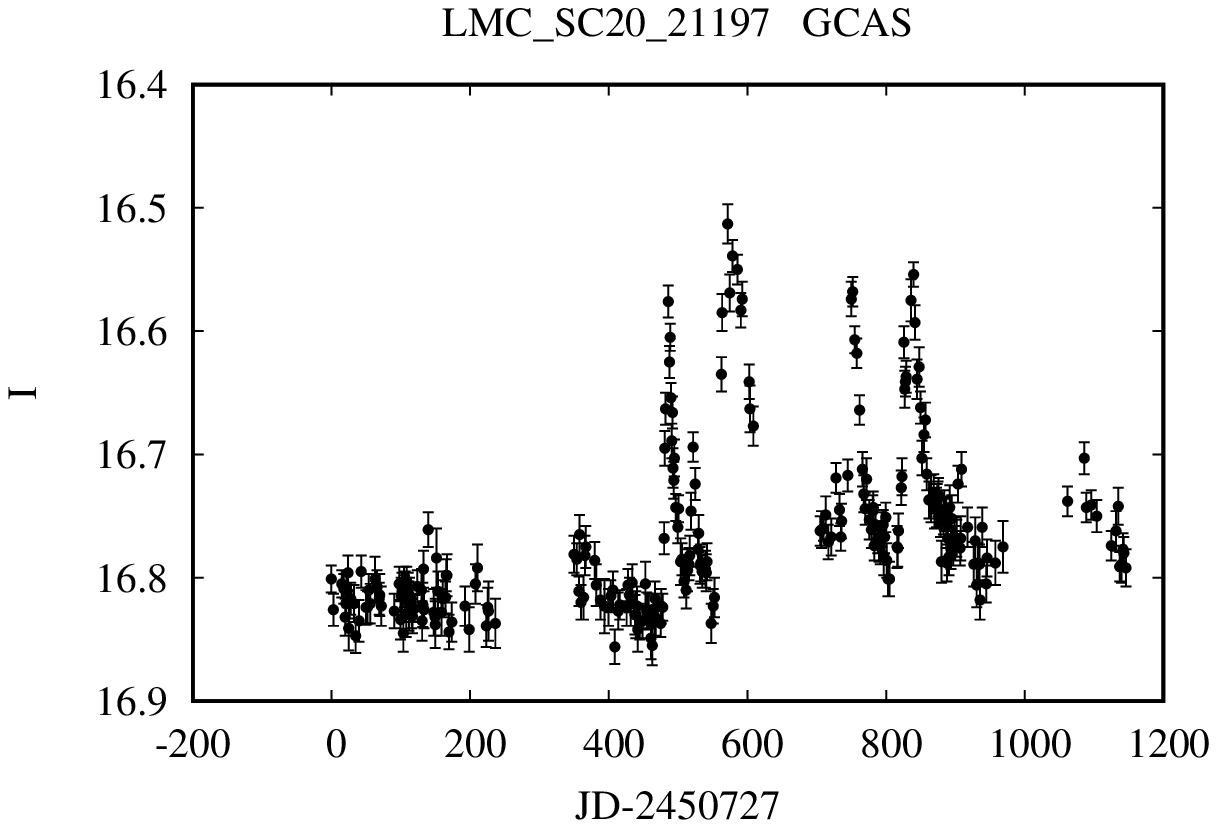}
	\includegraphics[width=0.24\textwidth]{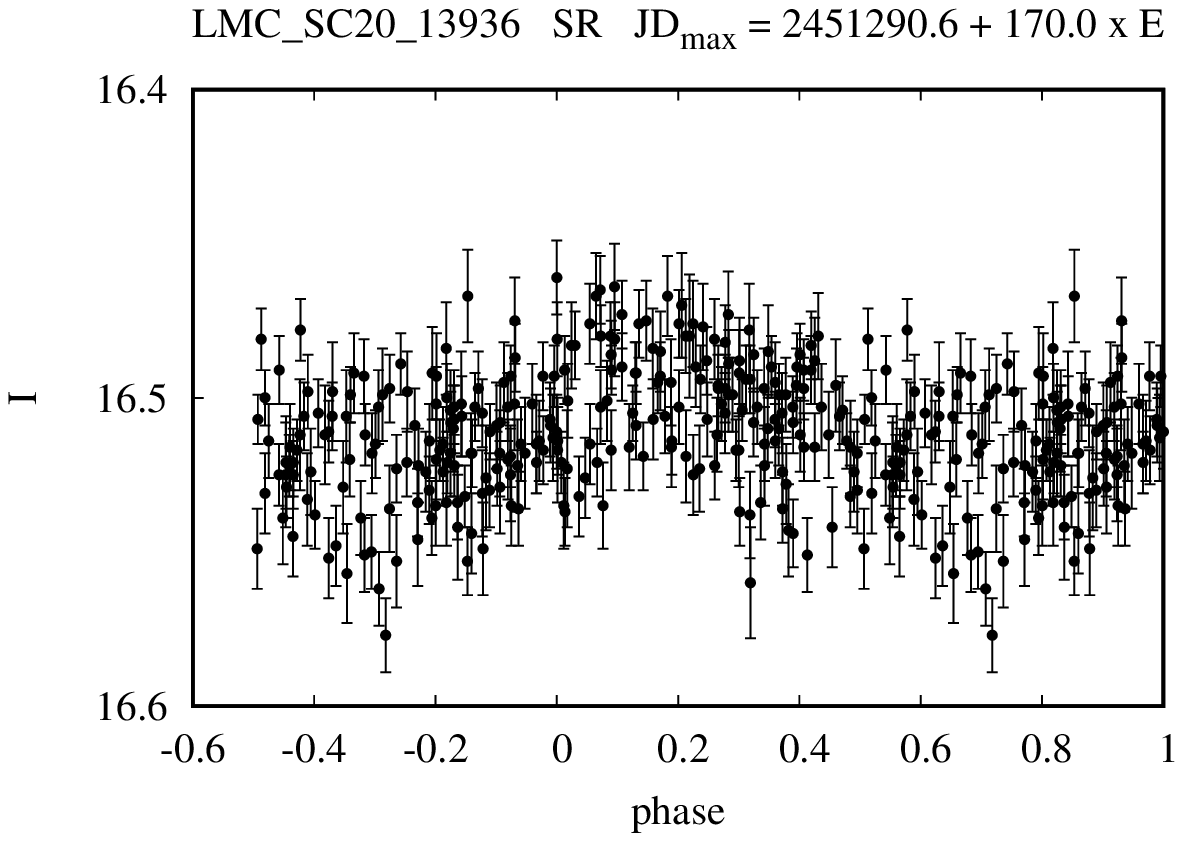}
	\includegraphics[width=0.24\textwidth]{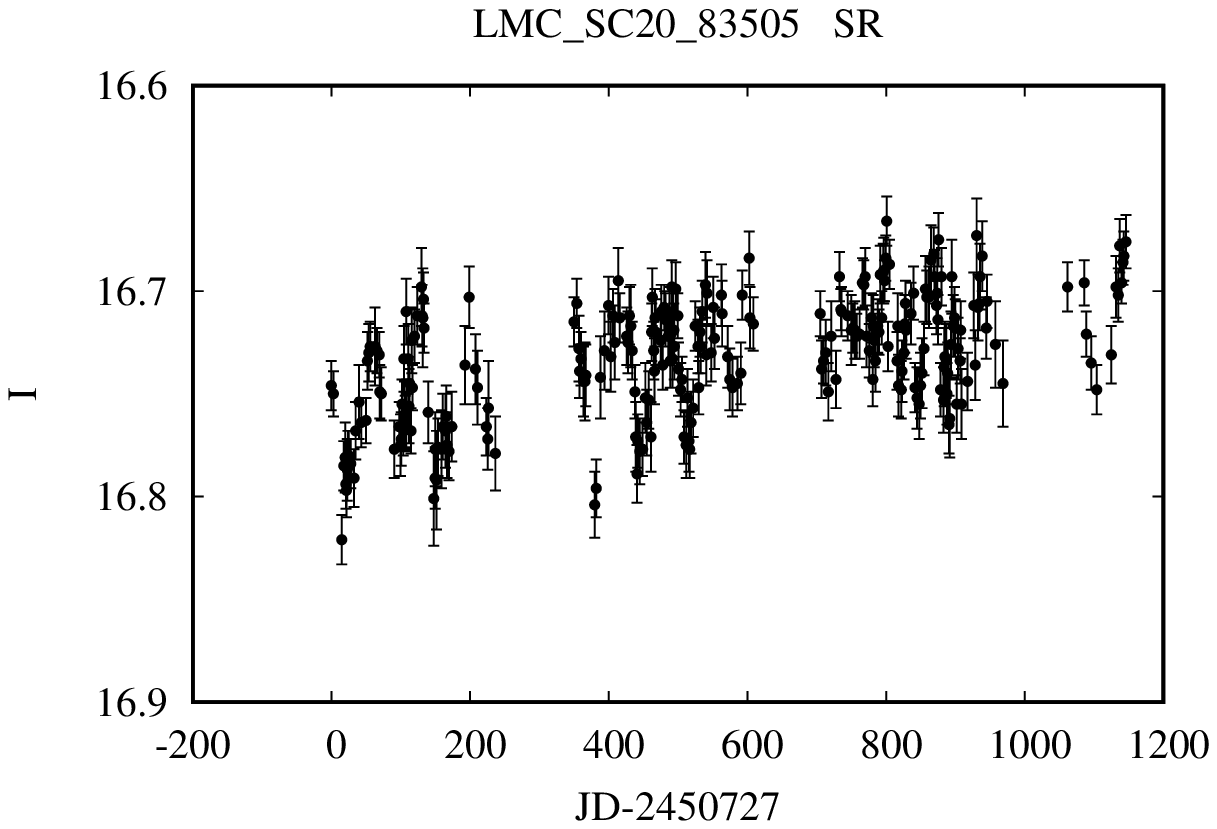}
	\includegraphics[width=0.24\textwidth]{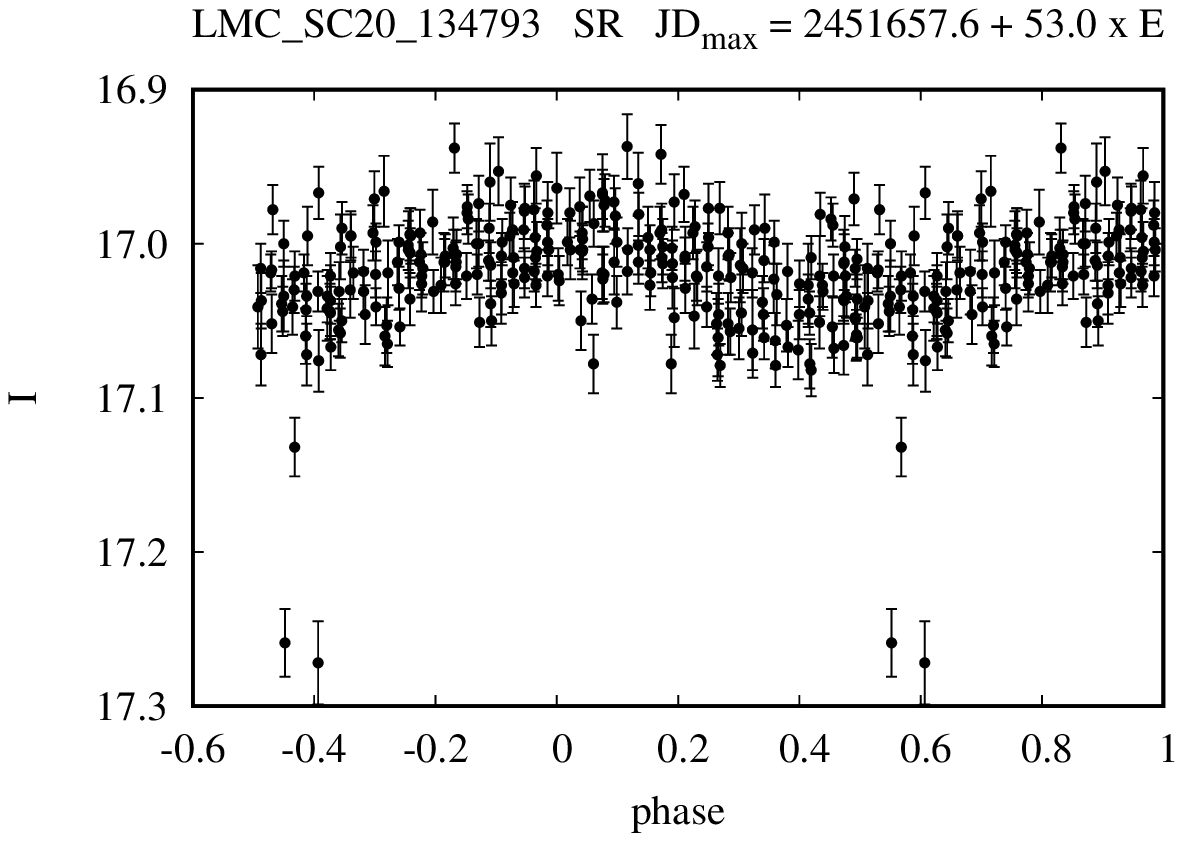}
	\includegraphics[width=0.24\textwidth]{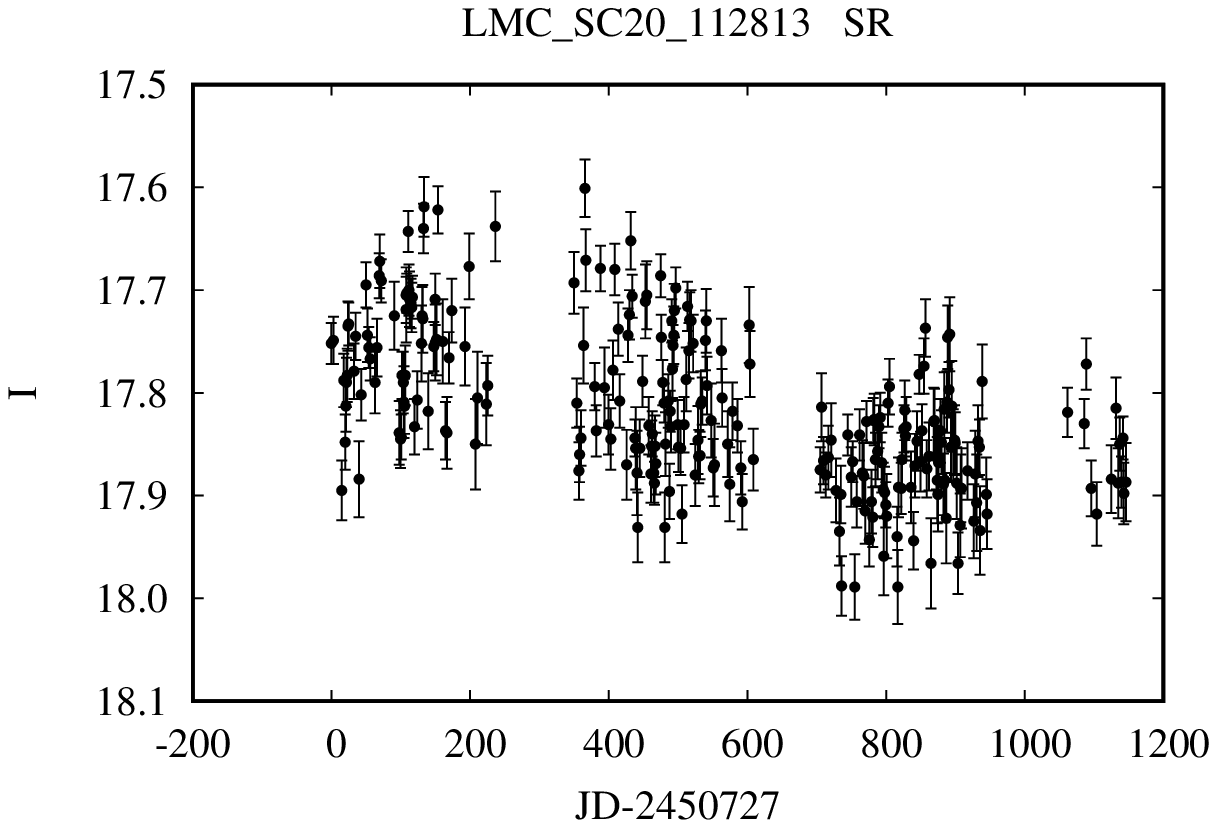}
	\includegraphics[width=0.24\textwidth]{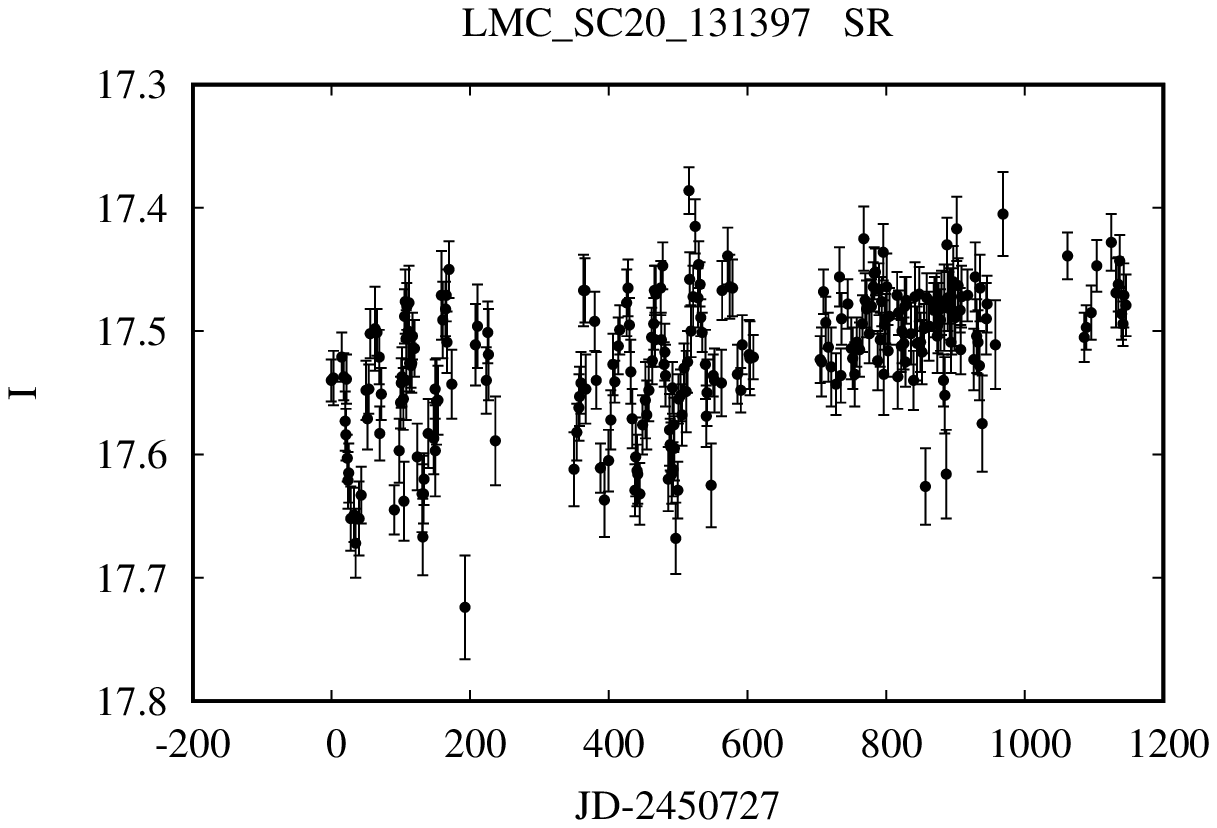}
	\includegraphics[width=0.24\textwidth]{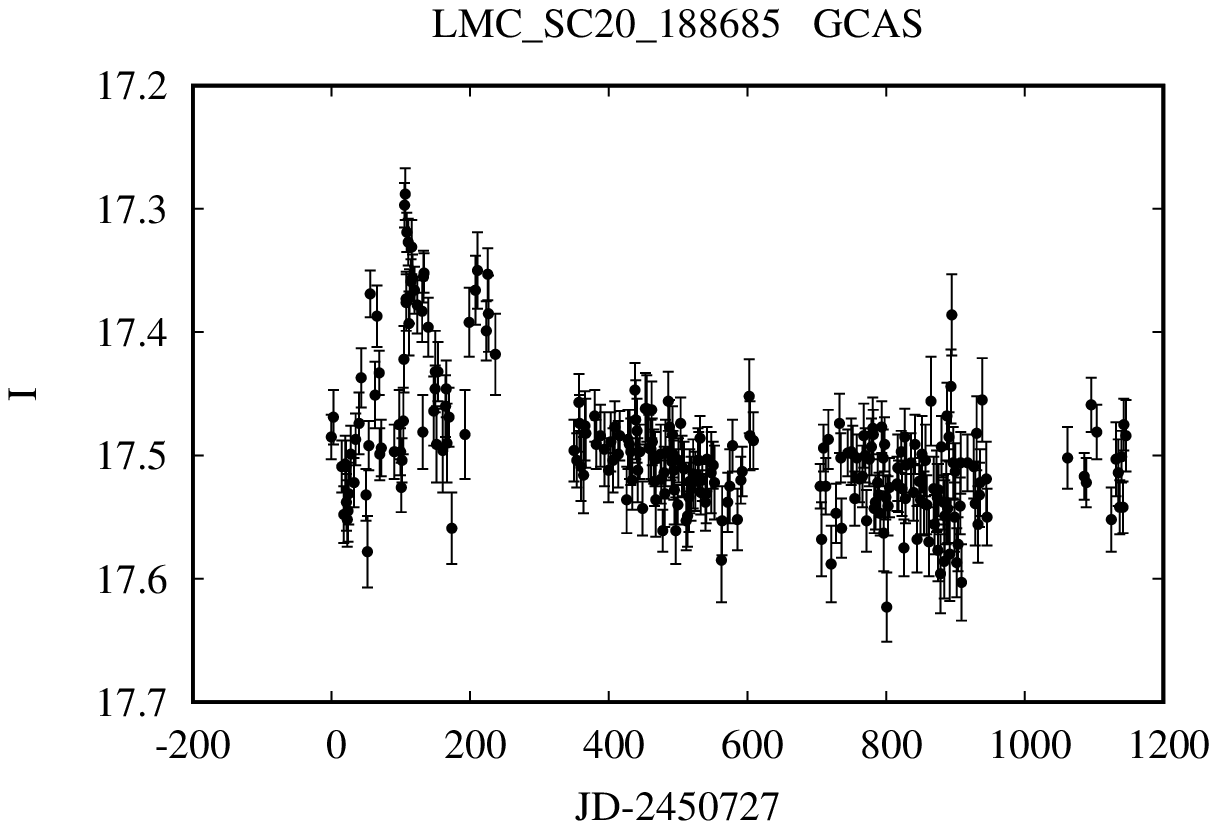}
        \caption{Light curves of variable stars with no previous reported classification (Table~\ref{tab:knownvars}). Variability of these stars was discovered with DIA. 
        The light curves are phased with the indicated light elements for  LMC\_SC20\_13936 and LMC\_SC20\_134793 and plotted as a function of time for the remaining stars.}
	\label{fig:knownvarslightcurves}
\end{figure}

Eleven variable sources discovered with DIA by \cite{2001AcA....51..317Z} had no classification suggested in the literature.
In order to account for these variables in Table~\ref{tab:vartypes}, we classify them (Table~\ref{tab:knownvars}) based on their light curves (Figure~\ref{fig:knownvarslightcurves}) and colors measured by \cite{2000AcA....50..307U}.


\begin{table*}
 \caption{New variable stars identified in the field LMC\_SC19 using the $NN$ classifier with hyperparameters resulting in the best $F_1$-score for LMC\_SC20.}
 \label{tab:newvars}
 \begin{tabular}{r c c c c r r c}
    \hline
Name & Position (J2000) & $I$-band range & Type & Light elements & $B-V$ & $V-I$ & Remarks\\
     &                  & (mag)          &      &                & (mag) & (mag) &        \\
    \hline

LMC\_SC19\_12951  & 05:42:40.86 $-$70:47:08.7 & 18.50--18.70 & SRA/ELL & $   JD_{\rm max} = 2451192.8 + 34.0 \times E$ &  $1.047$ &  $1.120$ & (1)\\
LMC\_SC19\_38470  & 05:42:41.10 $-$70:18:07.2 & 17.55--17.80 & GCAS    &                                           &  $0.039$ &  $0.040$ & \\
LMC\_SC19\_28995  & 05:42:42.43 $-$70:28:34.8 & 17.70--17.90 & SR      & $   JD_{\rm max} = 2451623.7 + 70.2 \times E$ &  $1.364$ &  $1.500$ & (2)\\
LMC\_SC19\_18475  & 05:42:54.55 $-$70:23:19.7 & 18.50--18.60 & SR      & $   JD_{\rm max} = 2451227.6 + 36.6 \times E$ &  $1.120$ &  $1.297$ & \\
LMC\_SC19\_92867  & 05:43:13.34 $-$70:15:23.2 & 17.80--18.00 & L       &                                           &  $1.203$ &  $1.285$ & (3)\\
LMC\_SC19\_74964  & 05:43:17.78 $-$70:36:02.7 & 17.95--18.10 & SR      & $   JD_{\rm max} = 2451175.8 + 91.6 \times E$ &  $1.134$ &  $1.171$ & \\
LMC\_SC19\_67152  & 05:43:24.27 $-$70:44:16.3 & 16.45--16.65 & BE:     &                                           & $-0.007$ & $-0.002$ & (4)\\
LMC\_SC19\_74429  & 05:43:37.06 $-$70:37:03.3 & 17.50--17.60 & SR      & $   JD_{\rm max} = 2451261.6 + 31.9 \times E$ &  $1.271$ &  $1.364$ & \\
LMC\_SC19\_78093  & 05:43:41.45 $-$70:32:19.9 & 17.45--17.60 & GCAS    &                                           &  $0.065$ &  $0.124$ & \\
LMC\_SC19\_184033 & 05:44:54.88 $-$70:18:02.3 & 18.40--18.50 & SR      & $   JD_{\rm max} = 2450934.5 + 39.7 \times E$ &  $0.979$ &  $1.096$ & \\
LMC\_SC19\_148609 & 05:44:52.60 $-$71:01:38.1 & 17.30--17.40 & SR      & $   JD_{\rm max} = 2451135.8 + 29.5 \times E$ &  $1.104$ &  $1.206$ & \\
LMC\_SC19\_184609 & 05:45:00.36 $-$70:17:26.8 & 18.50--18.60 & SR      & $   JD_{\rm max} = 2451132.8 + 46.4 \times E$ &  $0.444$ &  $1.349$ & (5)\\
LMC\_SC19\_173429 & 05:45:01.34 $-$70:31:23.1 & 17.80--17.90 & SR      & $   JD_{\rm max} = 2451154.8 + 86.1 \times E$ &  $0.967$ &  $1.041$ & \\
    \hline
    \end{tabular}
\begin{flushleft}
(1)\,$2^{\prime\prime}$ from an X-ray source 1WGA\,J0542.6$-$7047. 
(2)\,Periodic variations with changing amplitude are superimposed on a long-term declining trend.
(3)\,The faint outlier point in the light curve (Figure~\ref{fig:newvarslightcurves}) is likely not real.
(4)\,Irregular flares lasting 10--20$^d$ superimposed on a slow declining trend.
(5)\,Found in one of the test run with hyperparameter values different from the ones listed in Table~\ref{table:algos}.
\end{flushleft}
\end{table*}

\begin{table*}
 \caption{Classification of the variable stars discovered with DIA.}
 \label{tab:knownvars}
 \begin{tabular}{r c c c c c c c}
    \hline
Name & Position (J2000) & $I$-band range & Type & Light elements & $B-V$ & $V-I$ & Remarks \\
     &                  & (mag)          &      &                & (mag) & (mag) &        \\
    \hline
LMC\_SC19\_28805  & 05:42:47.47 $-$70:28:49.6 & 15.85--15.90 & BE   &                                            & $0.125$  & $0.355$ & (1)\\
LMC\_SC19\_32187  & 05:42:59.07 $-$70:26:01.0 & 16.10--16.15 & BE   &                                            & $0.028$  & $0.021$ & \\
LMC\_SC19\_41313  & 05:43:00.57 $-$70:15:45.8 & 16.35--16.45 & L    &                                            & $0.551$  & $0.912$ &     \\
LMC\_SC19\_111203 & 05:43:53.34 $-$70:49:07.4 & 16.05--16.30 & GCAS &                                            &          & $0.004$ &     \\
LMC\_SC20\_21197  & 05:45:21.69 $-$70:50:21.3 & 16.50--16.80 & GCAS &                                            & $0.020$  & $0.017$ &     \\
LMC\_SC20\_13936  & 05:45:22.51 $-$70:57:24.2 & 16.50--16.55 & SR   & $   JD_{\rm max} = 2451290.6 + 170.0 \times E$ & $1.572$  & $1.464$ &     \\
LMC\_SC20\_83505  & 05:45:49.71 $-$70:43:18.3 & 16.70--16.80 & SR   & $   JD_{\rm max} = 2450856.8 + 70.3 \times E$  & $0.927$  & $1.108$ & (2) \\
LMC\_SC20\_134793 & 05:46:29.70 $-$70:43:56.8 & 17.00--17.05 & SR   & $   JD_{\rm max} = 2451657.6 + 53.0 \times E$  & $1.423$  & $1.173$ & (3) \\
LMC\_SC20\_112813 & 05:46:31.25 $-$71:09:13.6 & 17.65--17.90 & SR   & $   JD_{\rm max} = 2451092.8 + 21.1 \times E$  & $0.926$  & $1.175$ & (4)\\
LMC\_SC20\_131397 & 05:46:54.52 $-$70:45:01.4 & 17.50--17.65 & SR   & $   JD_{\rm max} = 2451256.6 + 51.5 \times E$  & $1.498$  & $1.172$ & (2,5) \\
LMC\_SC20\_188685 & 05:47:02.33 $-$70:40:37.2 & 17.30--17.55 & GCAS &                                            & $-0.090$ & $-0.032$ &    \\
    \hline
    \end{tabular}
\begin{flushleft}
{These variables were originally identified by \cite{2001AcA....51..317Z}, but no classification was published.}
(1)\,B0IIIe spectral type according to \cite{2012MNRAS.425..355R}.
(2)\,Periodic brightness variations superimposed on a rising trend. 
(3)\,Three faint outliers are likely not real.
(4)\,Periodic variations superimposed on a long-term wave.
(5)\,Periodic variations stop around JD2450900 and reappear around JD2451800.
\end{flushleft}
\end{table*}

\begin{figure}
    %
	\includegraphics[width=0.24\textwidth]{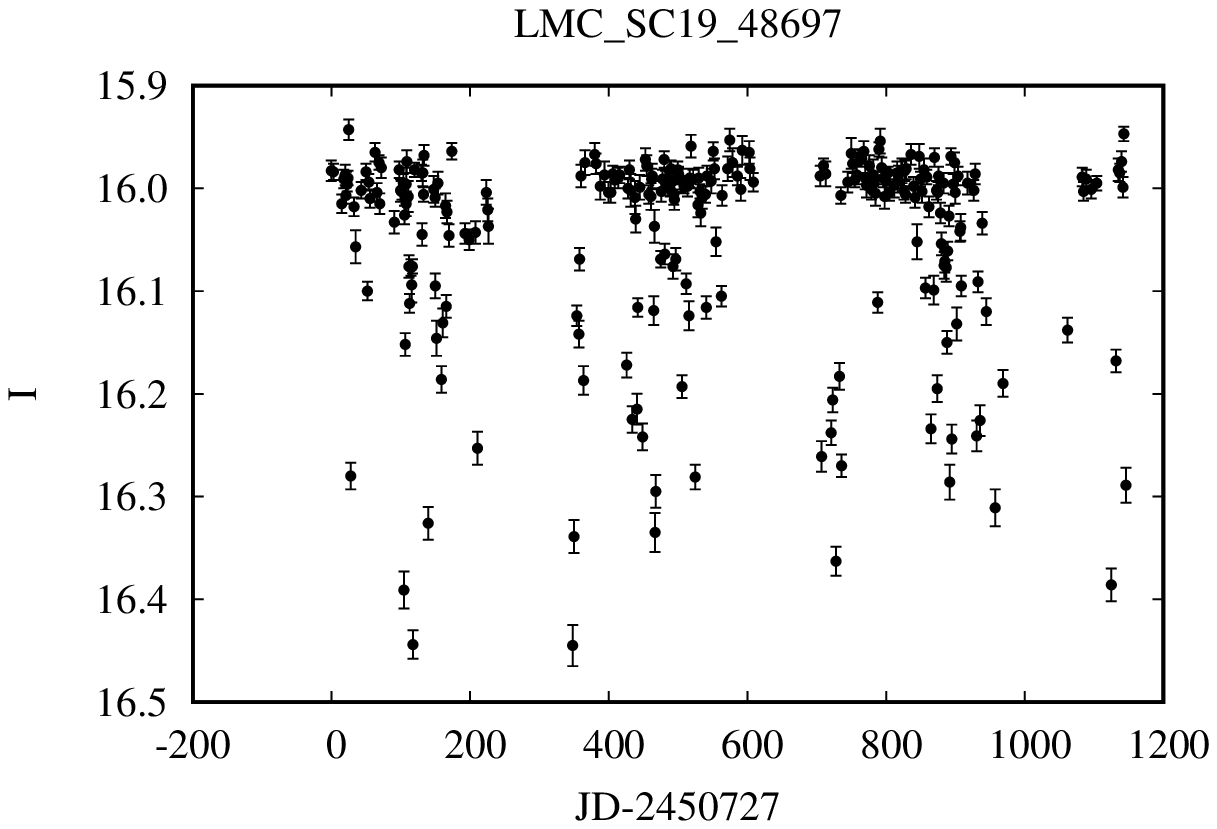}	\includegraphics[width=0.24\textwidth]{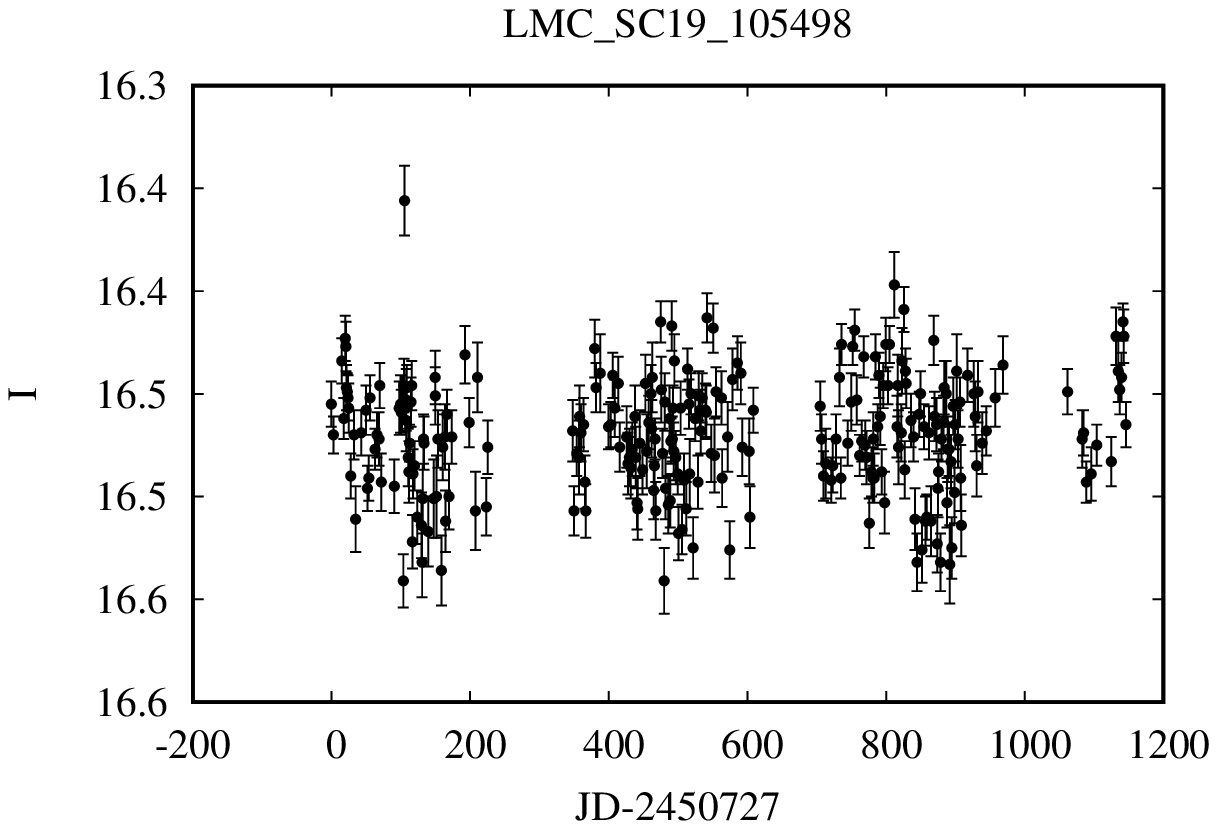}
	\includegraphics[width=0.24\textwidth]{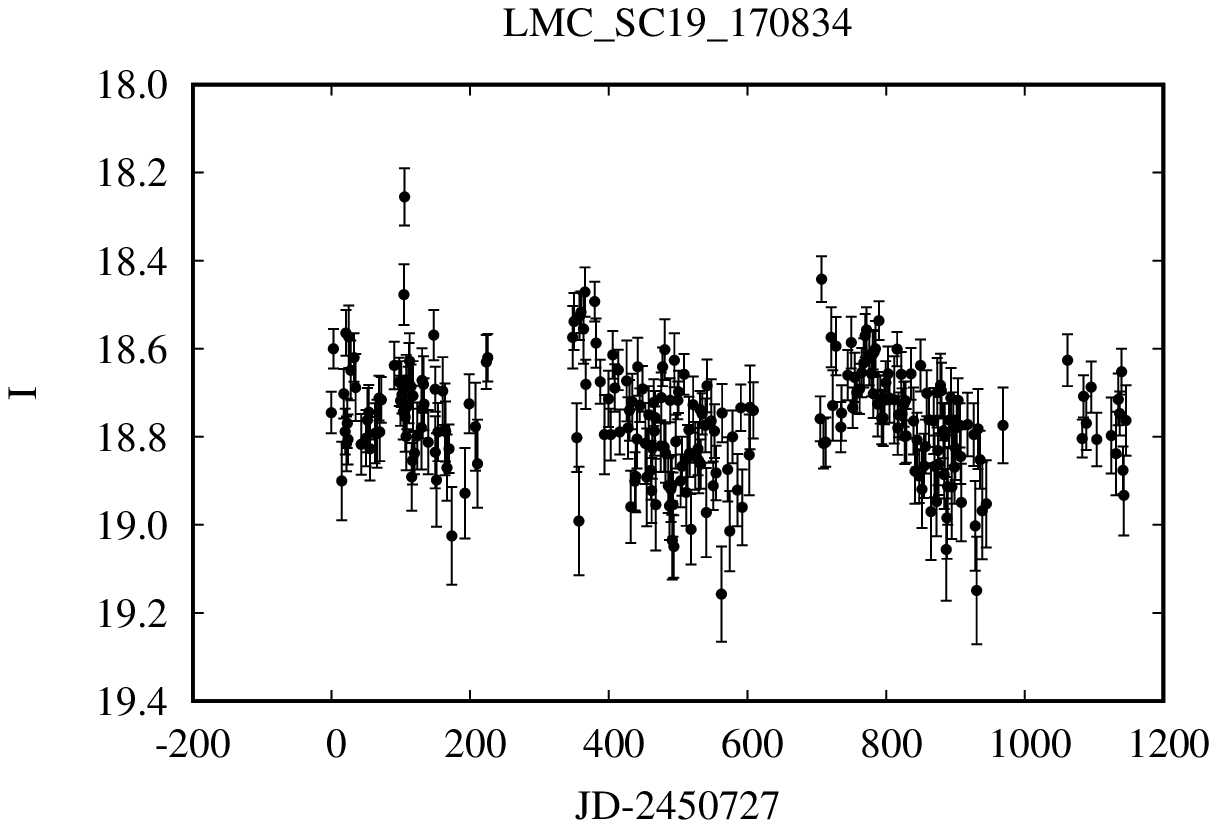}
	\includegraphics[width=0.24\textwidth]{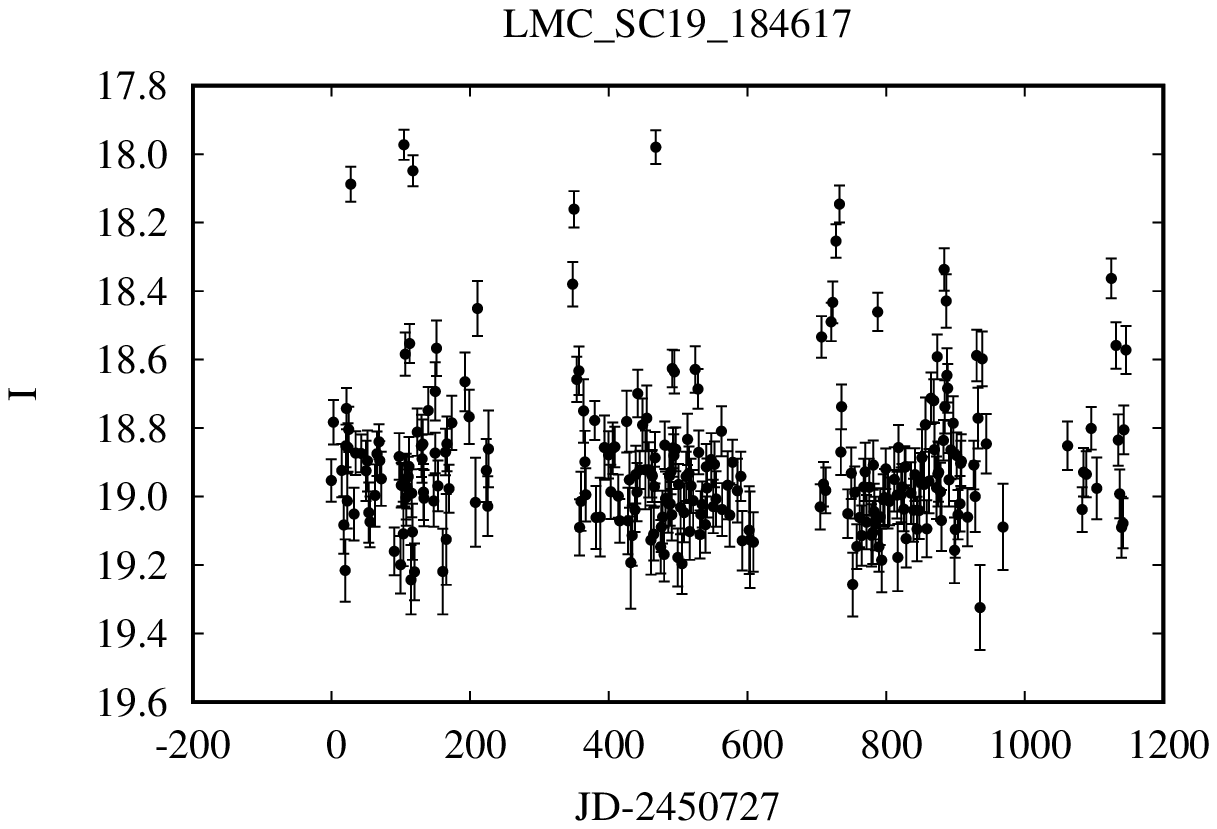}
        \caption{Example light curves having elevated values of individual variability indexes that were {\it correctly rejected} by the $NN$ classifier (TN).}
	\label{fig:truenegativelc}
\end{figure}

\begin{figure}
	%
	%
	\includegraphics[width=0.24\textwidth]{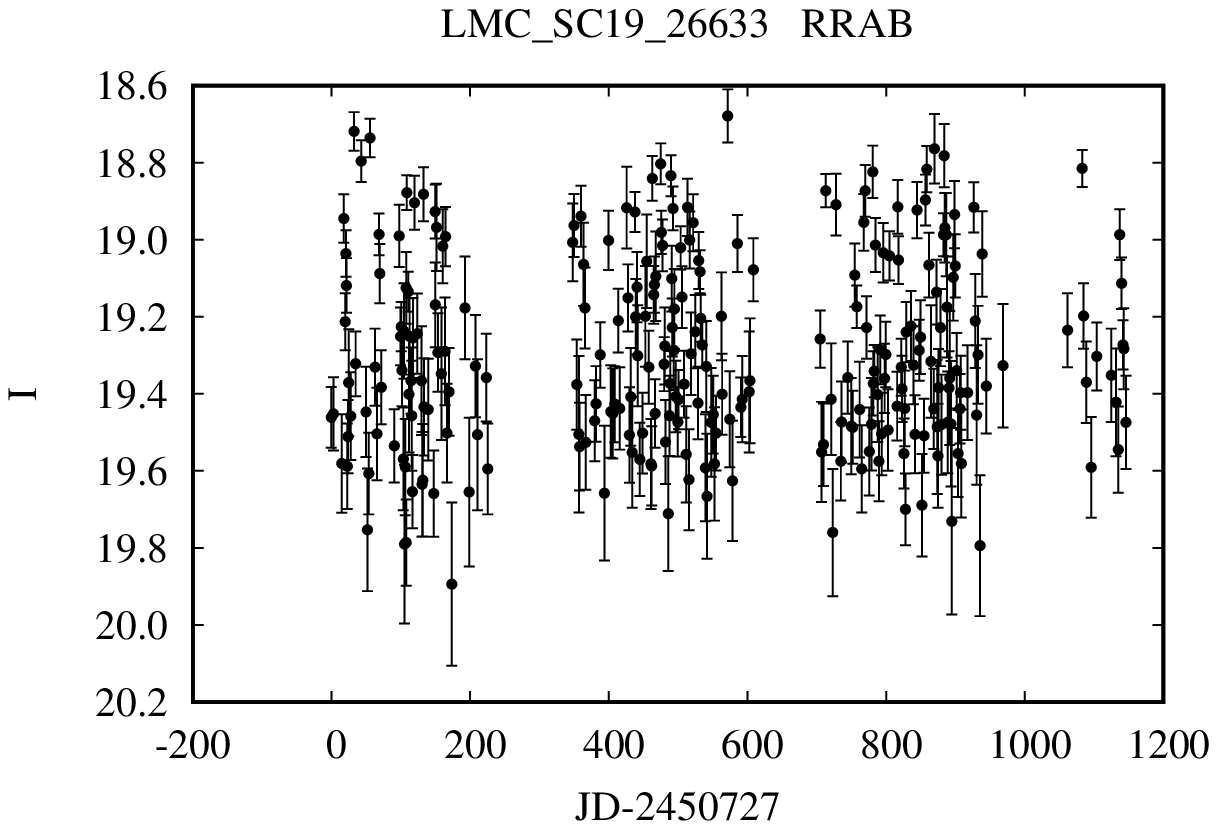}
	\includegraphics[width=0.24\textwidth]{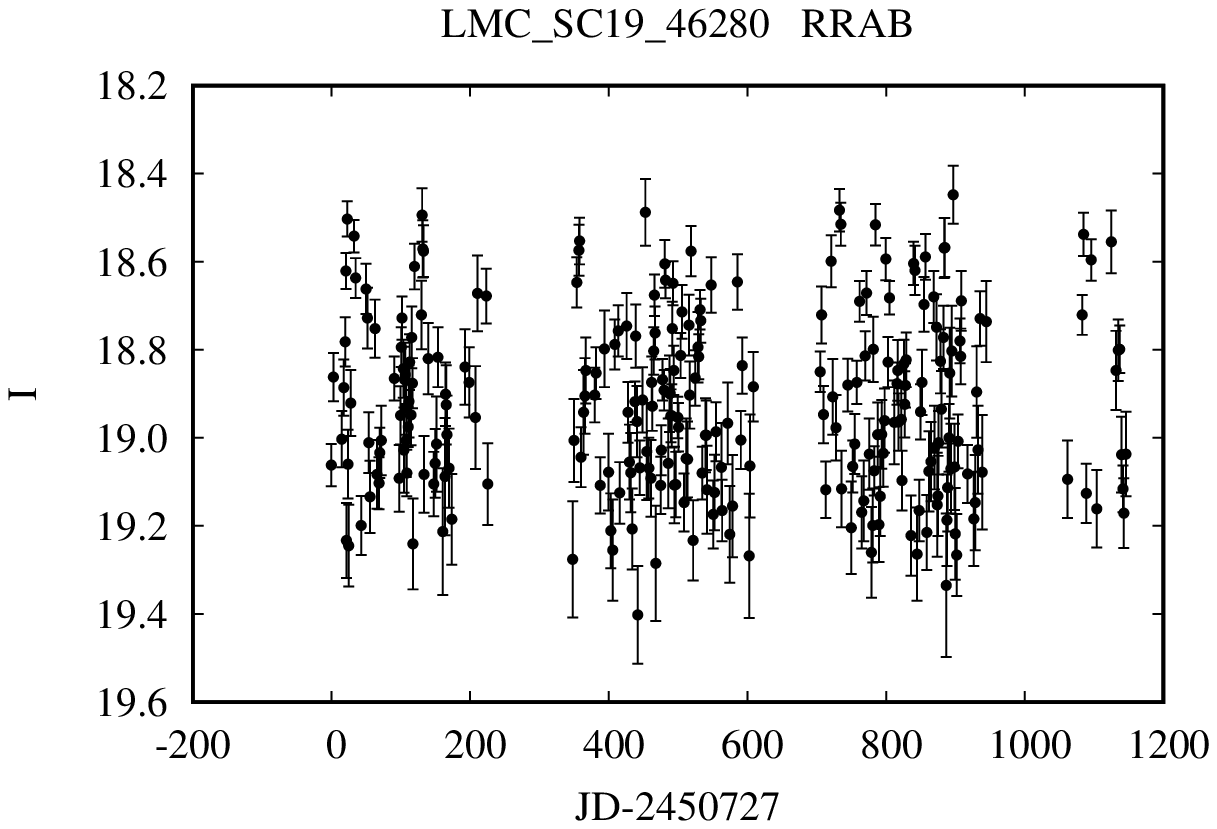}
	\includegraphics[width=0.24\textwidth]{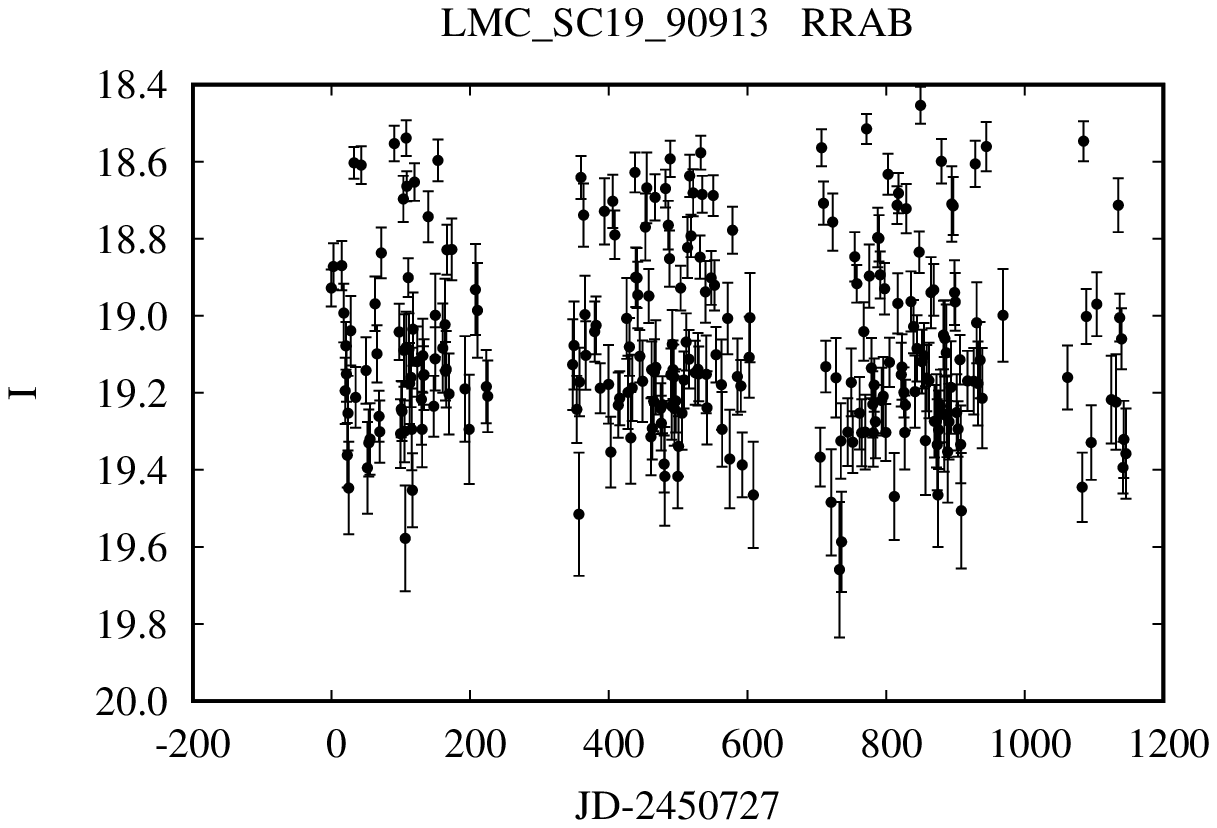}
	\includegraphics[width=0.24\textwidth]{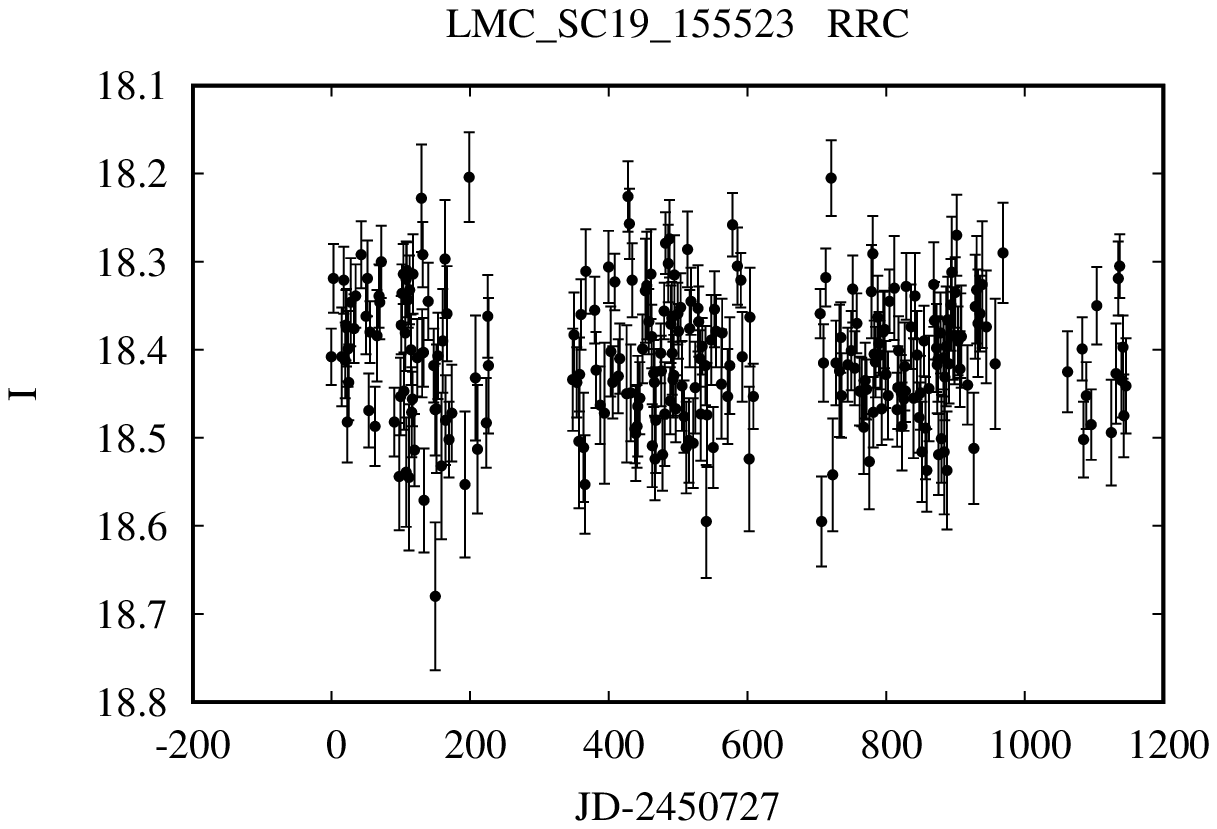}
	\includegraphics[width=0.24\textwidth]{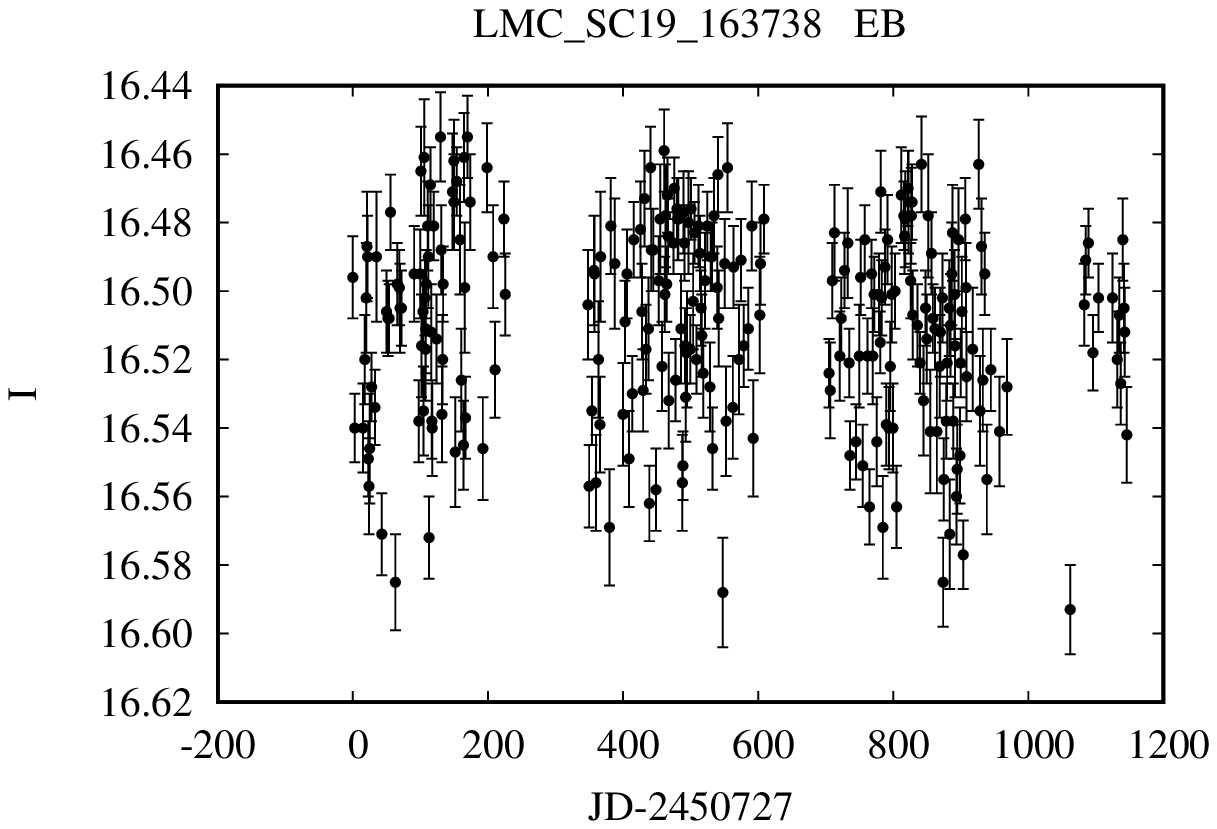}
	\includegraphics[width=0.24\textwidth]{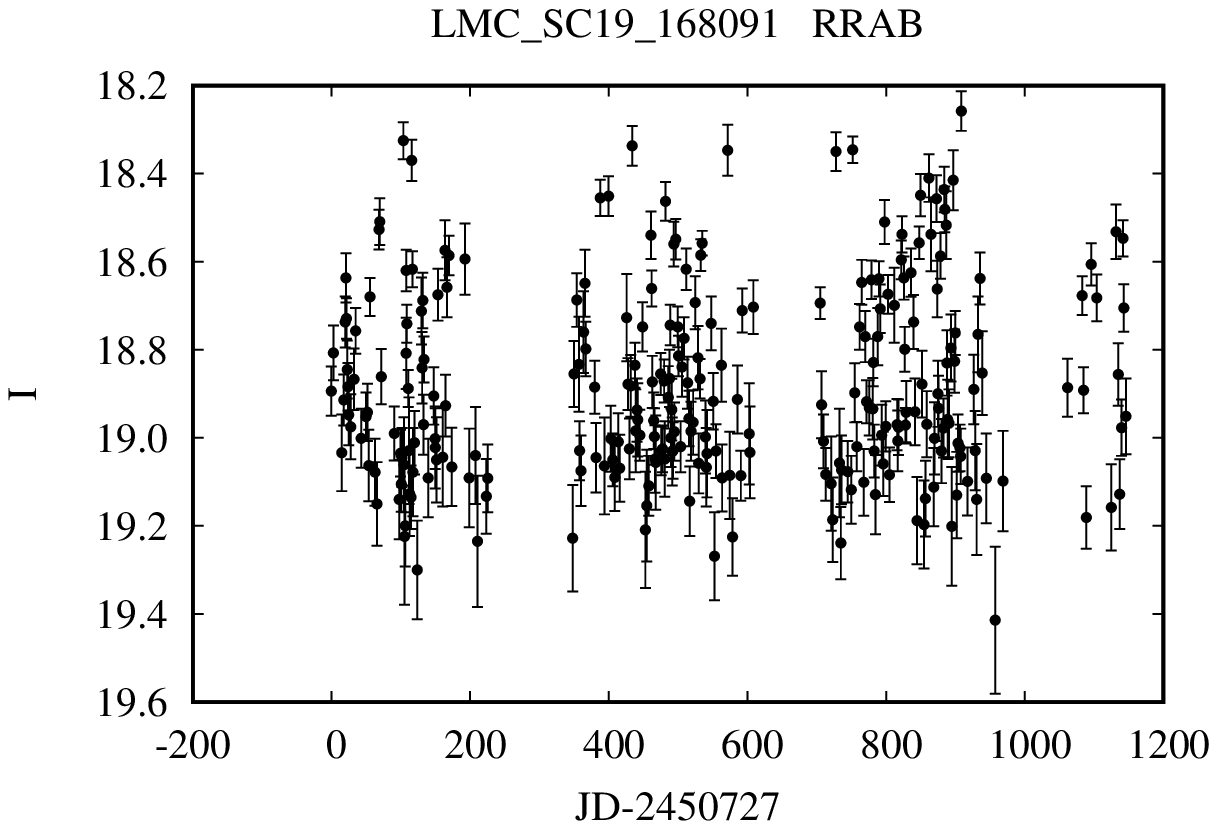}
	\includegraphics[width=0.24\textwidth]{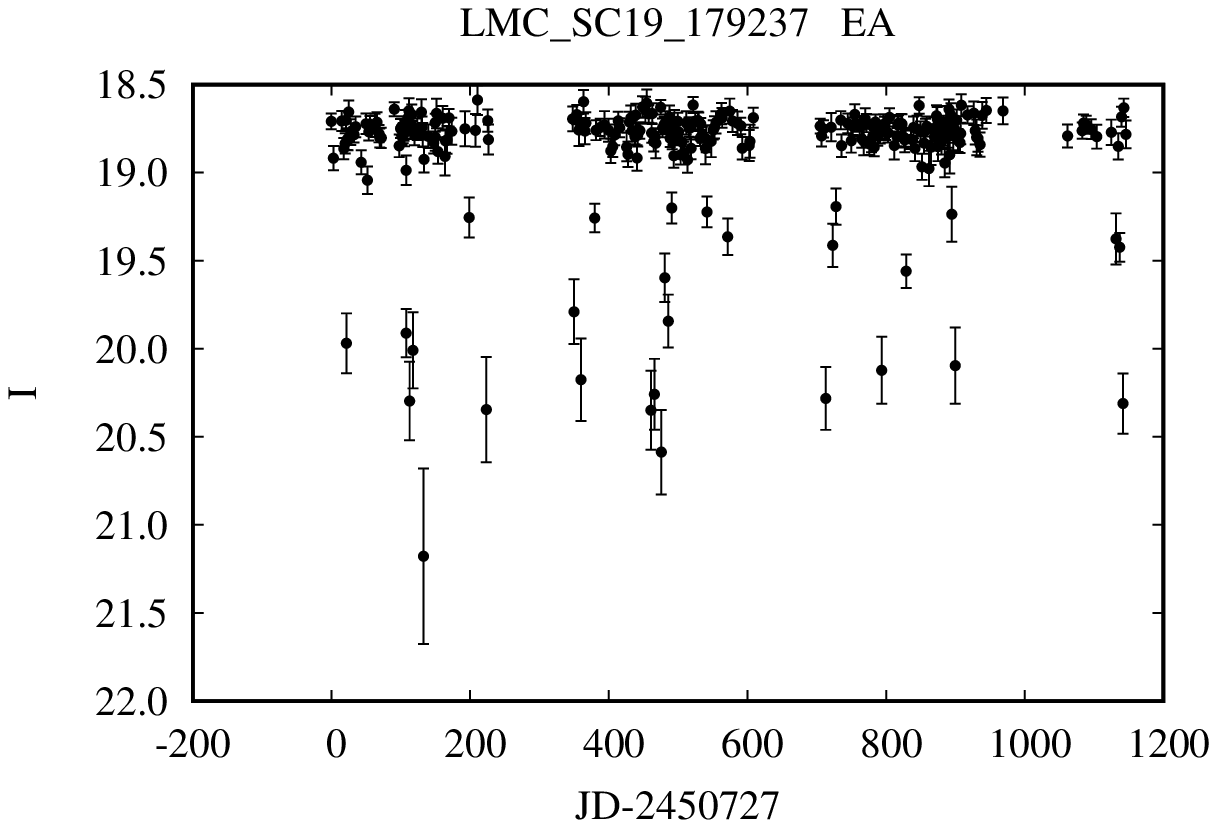}
	\includegraphics[width=0.24\textwidth]{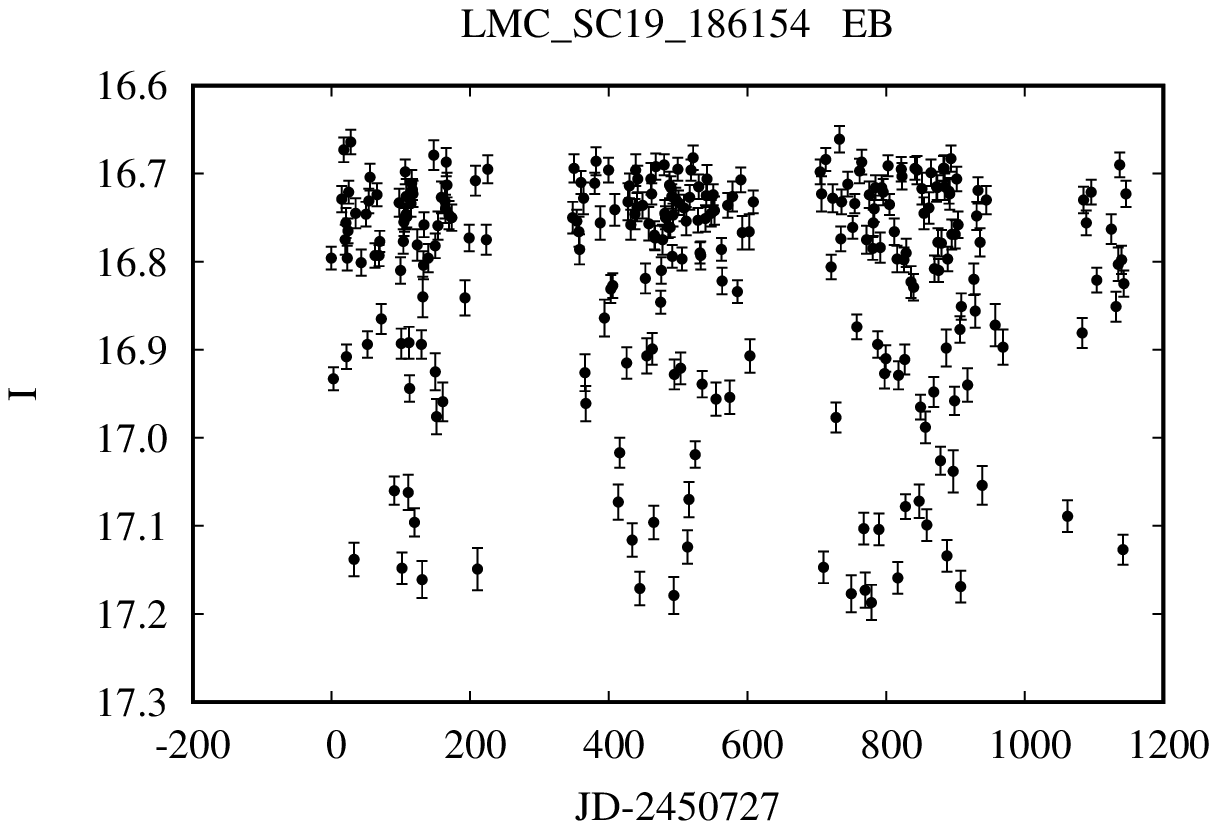}
        \caption{Light curves of known variables missclassified as non-variable by the $NN$ (FN). While all these variables are periodic, we plot here their light curves as a function of time rather than phase to highlight similarities with some FP (Fig.~\ref{fig:falsepositivelc}) and TN (Fig.~\ref{fig:truenegativelc}). Recall that none of the utilized variability features (Table~\ref{tab:indexsummary}) captures information about the period or the phased light curve shape.}
	\label{fig:falsenegativelc}
\end{figure}

Figures~\ref{fig:newvarslightcurves} and \ref{fig:knownvarslightcurves} present light curves of some of the variables correctly identified by the $NN$ classifier (TP). Figure~\ref{fig:falsepositivelc} illustrates light curves of objects that we believe were incorrectly selected by the $NN$ classifier as candidate variables (FP).
 Eight known variables were not detected by the $NN$ classifier (FN; Figure~\ref{fig:falsenegativelc}), three of them are eclipsing binaries identified by \cite{2003AcA....53....1W,2011AcA....61..103G} and the rest are RR~Lyrae stars \citep{2003AcA....53...93S,2009AcA....59....1S}. Figure~\ref{fig:truenegativelc} presents example light curves that were correctly identified by the classifier as non-variable (TN) while these objects have high values of some variability features and therefore would appear as false candidates in a variability search based on individual features (rather than their ML-based combination used here). As the light curves of FP and TN show high scatter of brightness measurements while showing no periodicity, it is most likely that the measurements are corrupted and do not reflect true brightness variations of these objects. Additional information, such as visual inspection of the images is required to identify the effects corrupting the measurements of these objects.

\subsection{Applicability to other photometric data sets}
\label{sec:otherdatasets}

The suggested approach to variability detection should be applicable to any large set of light curves given that:
\begin{enumerate}
\item a subset of these light curves is {\em a~priori} classified into variable and non-variable ones,
\item both classes include hundreds of examples or more,
\item the examples are representative of variability types and measurement artifacts found in the studied set of light curves.
\end{enumerate} 
These requirements are easily satisfied for surveys covering a large fraction of the sky as they include many previously known variable stars of various types listed in the GCVS and the AAVSO International Variable Star Index (VSX\footnote{\url{https://www.aavso.org/vsx/}}; \citealt{2006SASS...25...47W}). 
The photometric data suitable for the ML-based variability search are collected by a number of surveys including 
ASAS \citep{2002AcA....52..397P} and ASAS-SN \citep{2014ApJ...788...48S,2017PASP..129j4502K}, 
CRTS \citep{2009ApJ...696..870D}, 
DES \citep{2016MNRAS.460.1270D},
Gaia \citep{2017arXiv170203295E},
HATNet \citep{2004PASP..116..266B}, 
KELT \citep{2007PASP..119..923P}, 
MASCARA \citep{2017A&A...601A..11T},
NMW \citep{2014ASPC..490..395S},
NSVS \citep{2004AJ....127.2436W},
Pan-STARRS \citep{2010SPIE.7733E..0EK,2016arXiv161205560C},
PTF \citep{2009PASP..121.1395L}, 
SuperWASP \citep{2010A&A...520L..10B}, 
TrES \citep{2007ASPC..366...13A},
VVV \citep{2010NewA...15..433M}
with even more ambitious surveys being developed, among them 
LSST \citep{2008arXiv0805.2366I},
NGTS \citep{2017arXiv171011100W},
PLATO \citep{2014ExA....38..249R},
TESS \citep{2014SPIE.9143E..20R},
ZTF \citep{2017arXiv170801584L}.
The survey parameters such as photometric accuracy, observing cadence, single or multi-color observations, number of measurements per object in a single filter and magnitude range have an impact on the ability to discover various types of variable objects. The suggested ML-based variability detection approach is applicable regardless of the specifics of the survey's observing strategy.

Space photometry surveys such as Kepler \citep{2010Sci...327..977B} and CoRoT \citep{2009A&A...506..411A}
are capable of detecting brightness variations caused by magnetic activity (faculae, star spots; e.g. \citealt{2016A&A...589A..46S})
in Sun-like stars \citep{2013ApJ...769...37B} blurring the boundary between ``variable'' and ``non-variable'' stars. 
The question ``is there any detectable variability'' may still be relevant for the
fainter stars observed in these surveys. One may be interested in identifying stars more variable than the Sun \citep{2012A&A...539A.137M} or the ones showing periodic variability
\citep{2009A&A...506..519D,2011A&A...529A..89D} -- these problems require a different set of light curve features than the ones
considered here. The variability detection approach presented here
will likely not be useful for space astroseismology missions like 
MOST \citep{2003PASP..115.1023W}, BRITE \citep{2014PASP..126..573W,2016PASP..128l5001P,2017A&A...605A..26P} and the upcoming transit photometry mission CHEOPS \citep{2013EPJWC..4703005B} as they observe (with superior accuracy) only one or few stars at a time.

When applying the ML-based variability detection to new data sets, some light curve features listed in Table~\ref{tab:indexsummary} may lose their predictive power while some that are found to be the least informative for the OGLE-II data set could become useful. When designing a variability detection procedure for a new set of photometric observations, it is desirable to go through the full process (Section~\ref{sec:sum}) of feature selection/filtering, choosing multiple ML-algorithms, tuning their HP, checking for possible over/underfitting using learning curves before choosing the best algorithm and its HP values. The resulting classification performance will be different from the one reported in Table~\ref{table:algos} and could be either better or worse depending on the sample size, light curve quality and the exact set of features used for classification.

\section{Conclusions}
\label{sec:sum}

We explore a novel approach for selecting variable objects from a set of light curves.
The basic idea is to treat variability detection as a two-class classification problem (variable vs. non-variable objects) despite the intrinsic inhomogeneity of these classes and solve it with machine learning.
The procedure may be summarized as follows:
\begin{enumerate}
 \item Search a representative subset of all light curves for variability using traditional methods, e.g. by visually inspecting the light curves of all outliers in variability feature -- magnitude plots (Figure~\ref{fig:indexmagplots}). 
 It is important to get reasonable confidence that the variability search in the subset is exhaustive.
 This will be our training subset.
 \item For each light curve compute a set of features (Table~\ref{tab:indexsummary}) that highlight some or all types of 
 variability while hiding unimportant differences between the light curves (like the difference in the number of measurements).
 \item Choose a machine learning algorithm and tune its hyperparameters on the training subset using cross-validation as described in Section~\ref{sec:hyperpartuning}. Table~\ref{table:algos} presents an example of 
 optimal hyperparameter values.
 One may control the trade-off between the completeness of the variability search and the rate of false detections by selecting performance metrics (e.g. $F_{\beta}$ instead of $F_1$, Section~\ref{sec:performancemetric}) maximized during the 
 search for optimal hyperparameters
 \item Train the algorithm with the optimized hyperparameters on the whole training subset.
 \item Apply the algorithm to the full set of light curves and inspect the ones classified as variable. One may control the false detection rate at this stage by changing the classifier threshold.
\end{enumerate}
This procedure works even with a highly imbalanced training subsample of a modest size: 168 variables among 30265 OGLE-II light curves (Section~\ref{sec:lightcurves}; see also the cross-validation scores in Figure~\ref{fig:learningcurves}). Application to an independent set of 31798 OGLE-II light curves resulted in the selection of 205 candidate variables, 
27 of which turned out to be false detections and 178 real variables (12 of them new, Table~\ref{tab:newvars}, Figure~\ref{fig:newvarslightcurves}). 

To directly compare traditional variability search methods to the machine learning algorithms considered here, 
we restricted ourselves to the data sets used by \cite{2017MNRAS.464..274S} who compared the effectiveness of various variability indices (features).
In terms of the $F_1$-score (Table~\ref{table:algos}), all machine learning algorithms tested
here outperform each individual variability index as well as their linear
combination. 
The $NN$, $SVM$, $SGB$ and $RF$ algorithms show the best performance (Figure~\ref{fig:4algo}).
In addition to the OGLE-II data discussed in detail here, these conclusions are confirmed with two other data sets from \cite{2017MNRAS.464..274S}, which were collected with different telescopes and processed using different source extraction and photometry software (Section~\ref{sec:cv_comparison}).
To improve the variable object selection results even further, one needs to use a larger training sample and engineer additional
features that would quantify the object's image shape, its proximity to other
detected objects and periodicity in light variations.
The suggested ML-based variability detection technique should be applicable to any large ($\gtrsim 10^4$) set of light curves given that a representative sub-sample of these light curves is a~priori classified as ``constant'' or ``variable'' by other means (Section~\ref{sec:otherdatasets}).

\section*{Acknowledgments}

We thank the anonymous referees for helpful comments.
We thank Dr.~Laurent Eyer for pointing out the hypothesis-testing approach
to the problem of variability detection, Dr.~Alceste Bonanos, Dr.~Antonios Karampelas, Dr.~Nikolay Samus, Dr.~Maria~Ida Moretti for critically reading this manuscript.
KVS and PG are supported by the European Space Agency (ESA) under
the ``Hubble Catalog of Variables'' program, contract No.\,4000112940.
This research has made use of the International Variable Star Index (VSX)
database, operated at AAVSO, Cambridge, Massachusetts, USA.
This research has made use of the \texttt{VizieR} catalogue access tool, CDS,
Strasbourg, France. The original description of the VizieR service is
presented by \cite{2000A&AS..143...23O}.
We also relied on the catalog matching capabilities of \texttt{TOPCAT} \citep{2005ASPC..347...29T} and catalog and image visualization with \texttt{Aladin} sky atlas \citep{2000A&AS..143...33B}.
This research has made use of NASA's Astrophysics Data System.

\footnotesize{
 \bibliographystyle{mn2e}
 \bibliography{mlvid}
}

\appendix

\section{Clipped light curve features}
\label{sec:clip}

Corrupted photometric measurements result in outlier points in a light curve (Sec.~\ref{sec:data}, see for example LMC\_SC19\_92867 in Fig.~\ref{fig:newvarslightcurves} and LMC\_SC20\_134793 in Fig~\ref{fig:knownvarslightcurves}) that may alter the light curve feature values while having no relation to the object's variability. One way to minimize this problem is to apply clipping to the light curve before computing the feature values. \cite{2016A&A...587A..18K} perform $\sigma$-clipping before computing all the light curve features used for classification of periodic variable stars. As we are concerned with detection of non-periodic stars (as well as periodic ones) that may show variability only occasionally, we do not apply $\sigma$-clipping. Instead, for a few features that are most sensitive to outlier light curve points we compute both their unclipped and clipped versions (Table~\ref{tab:indexsummary}) as outlined below. 

\subsection{{\scshape VaST}-style clipped $\sigma$ -- $\sigma_{\rm clip}$}
\label{sec:sigmaclip}

This clipped statistic was used for variability detection in the early versions of the {\scshape VaST} code. From each light curve we drop 5\,per~cent of brightest and 5\,per~cent of faintest points, but not more than 5 points from each side and compute the unweighted standard deviation
$$
\sigma_{\rm clip} = \sqrt{ \frac{1}{N-1} \sum\limits_{i=1}^N (m_i-\bar{m})^2 }
$$
where $N$ is the number of points in the clipped light curve, $\bar{m}$ is the mean magnitude of the set ${m_i}$ of magnitude measurements remaining after clipping. In many data sets $\sigma_{\rm clip}$ proved to be a more useful variability indicator than $\sigma$ computed over the non-clipped light curve. It is also more sensitive than MAD and IQR  (Table~\ref{tab:indexsummary}) to rare variability events (flares, eclipses). Similar clipping schemes based on removing a predefined percentage or number of the brightest and faintest points were applied by \cite{2013AJ....146..101P,2013PASP..125..857T}.

\subsection{Clipped Stetson's indices $J_{\rm clip}$ and $L_{\rm clip}$}
\label{sec:clipstetson}

\cite{1996PASP..108..851S} suggested variability detection statistics $J$ and $L$ that rely on observations taken close in time being grouped into pairs. If both observations in a pair deviate in the same direction from the mean brightness, this indicates the light curve is smooth (as expected for an object varying on a timescale longer than the time difference between the observations in the pair). \cite{2017MNRAS.464..274S} suggested a modified version of these variability indices, $J_{\rm clip}$ and $L_{\rm clip}$, which does not form a pair if the magnitude difference between the two observations is larger than a predefined limit (indicating that one of the observations in the pair might be corrupted). The clipping in these indices is done on the magnitude difference in pairs, not on the original light curve. This modification however did not result in a considerable performance improvement compared to the original $J$ and $L$ when tested on real data \citep{2017MNRAS.464..274S}.

\cite{1996PASP..108..851S} advocates for iterative re-weighting as an alternative to clipping. This avoids having a sharp boundary between the observations that are ``in'' or ``out''. In the original $J$ and $L$ definitions, iterative re-weighting is applied only to the mean magnitude calculation, but not to the observations that form pairs.

\section{Variability feature -- magnitude plots}
\label{sec:indexmagplots}

\begin{figure*}
	\centering
	\includegraphics[width=0.45\textwidth]{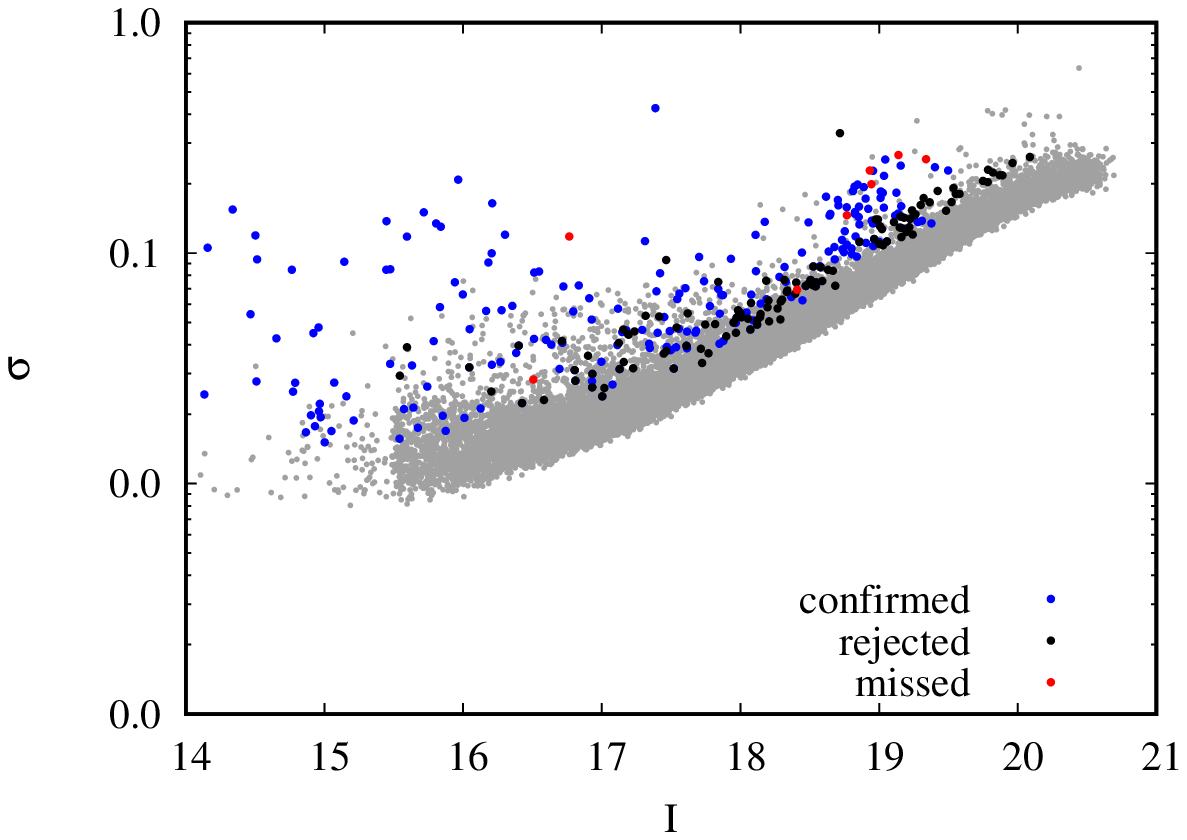}
	\includegraphics[width=0.45\textwidth]{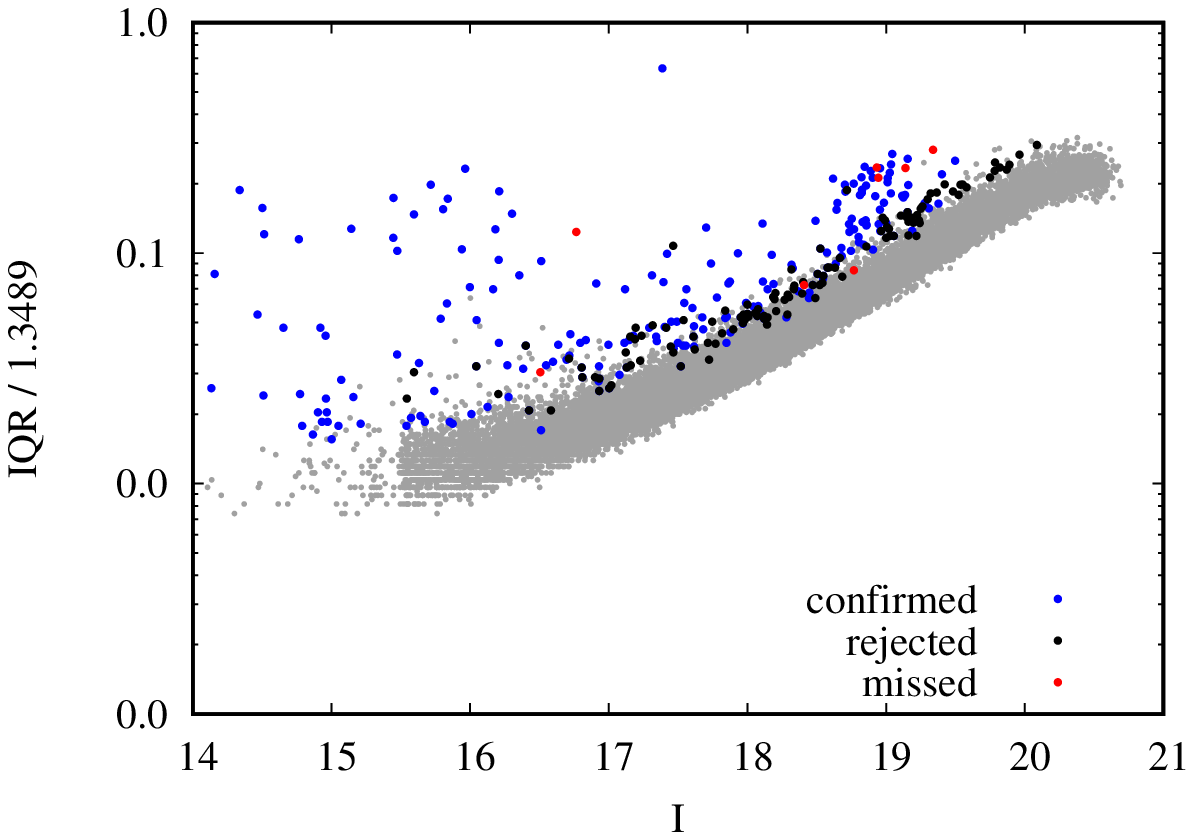}
	\includegraphics[width=0.45\textwidth]{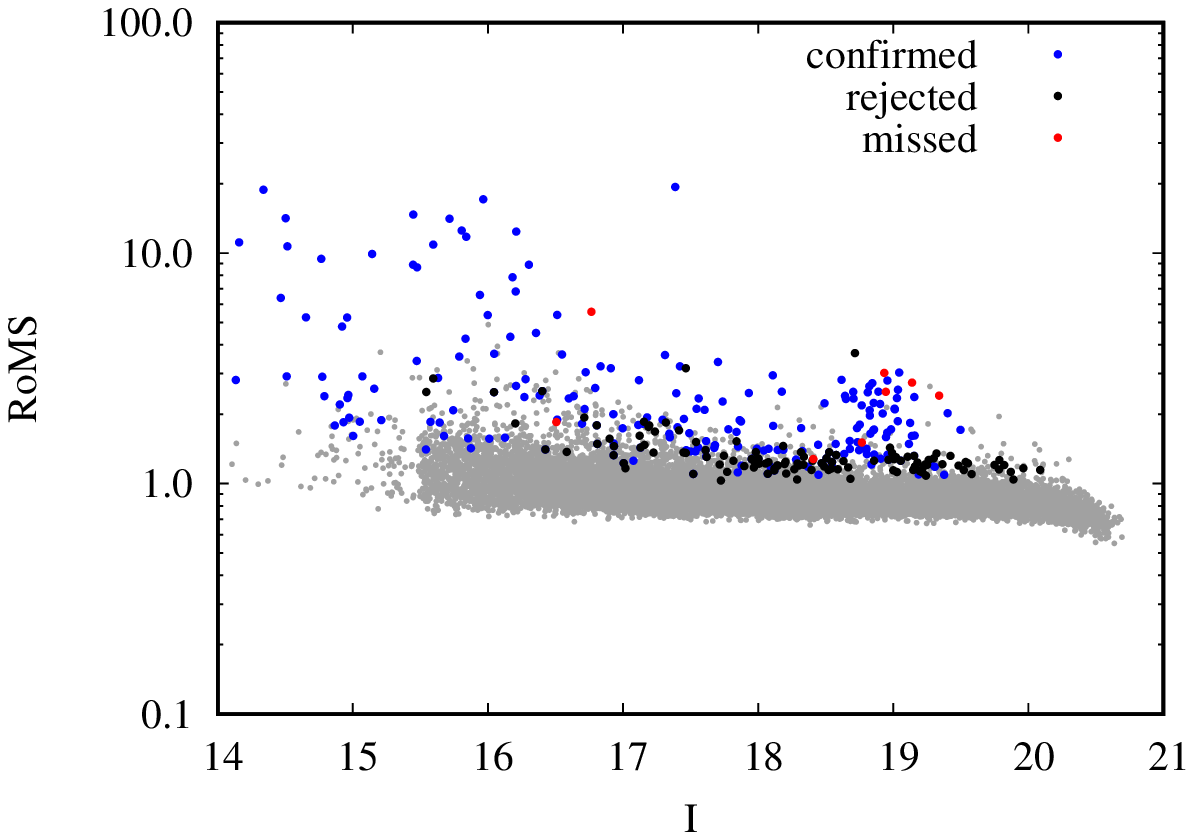}
	\includegraphics[width=0.45\textwidth]{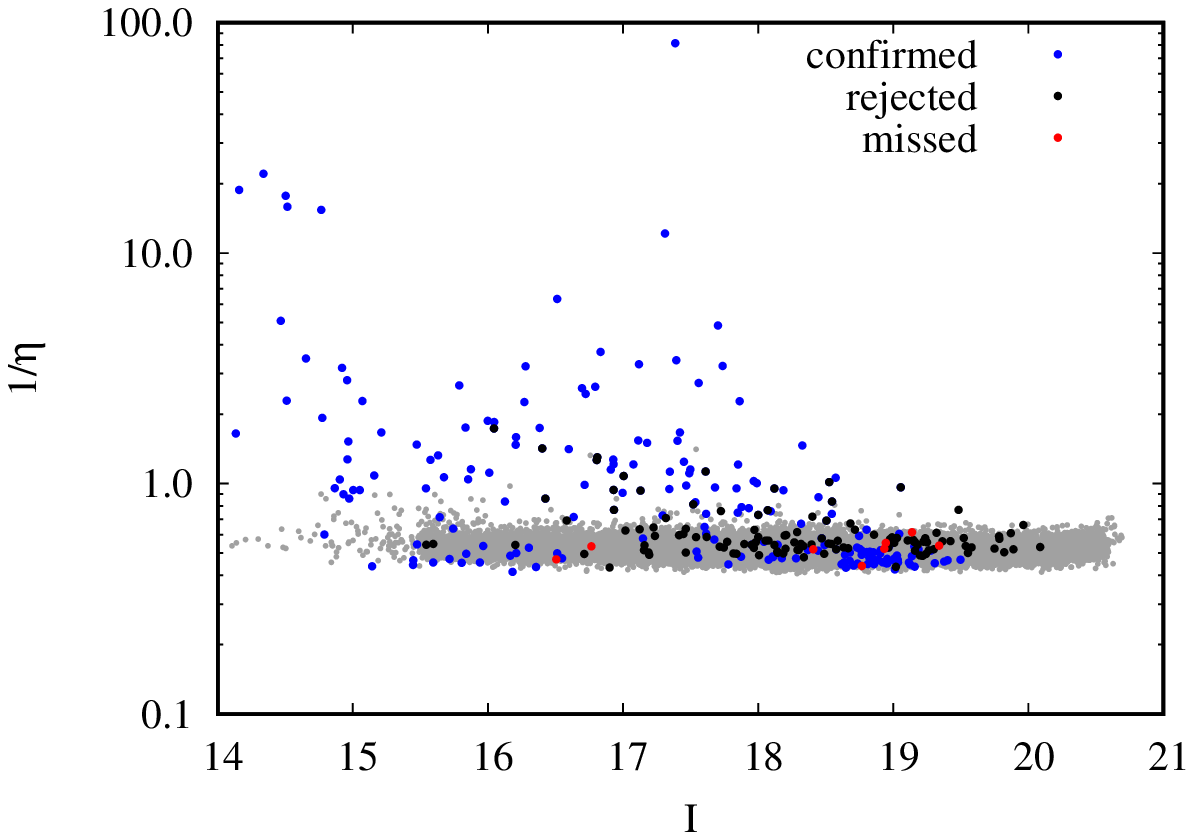}
        \caption{Variability feature vs. $I$ magnitude plots showing all objects in grey and highlighting candidate variables selected by the $NN$ classifier and confirmed by visual inspection (see example light curves in Figures~\ref{fig:newvarslightcurves} and \ref{fig:knownvarslightcurves}), rejected after visual inspection (Figure~\ref{fig:falsepositivelc}) as well as the known variable stars missed by the $NN$ classifier (Figure~\ref{fig:falsenegativelc}). The IQR is scaled to $\sigma$ of the Gaussian distribution so the numerical values of the two upper plots may be compared directly.}
	\label{fig:indexmagplots}
\end{figure*}

Figure~\ref{fig:indexmagplots} presents plots of selected individual variability features (Table~\ref{tab:indexsummary}) as a function of OGLE $I$ magnitude. Such plots are typically used to identify variable objects by selecting a magnitude-dependent cut-off for an individual feature (usually referred to in this context as ``variability index'') and visually inspecting light curves of all objects above the cut-off \cite[e.g.][]{2017MNRAS.464..274S}.

\label{lastpage}

\end{document}